%% file: draft.tex
\documentclass[twocolumn]{aastex62}
\shortauthors{Moskowitz \& Walker}
\usepackage{amsmath}

\usepackage{graphicx}
\abovedisplayshortskip
\belowdisplayshortskip
\begin{document}

\title{Stellar Density Profiles of Dwarf Spheroidal Galaxies}

\author{A. G. Moskowitz}
\affiliation{McWilliams Center for Cosmology, Department of Physics, Carnegie Mellon University, 5000 Forbes Avenue,
Pittsburgh, PA 15213, USA}
\email{amoskowitz@cmu.edu}

\author{M. G. Walker}
\affiliation{McWilliams Center for Cosmology, Department of Physics, Carnegie Mellon University, 5000 Forbes Avenue,
Pittsburgh, PA 15213, USA}

\begin{abstract}
  We apply a flexible parametric model, a combination of generalized Plummer profiles, to infer the shapes of the stellar density profiles of the Milky Way's satellite dwarf spheroidal galaxies (dSphs). We apply this model to 40 dSphs using star counts from the Sloan Digital Sky Survey, PanStarrs-1 Survey, Dark Energy Survey, and Dark Energy Camera Legacy Survey. Using mock data, we examine systematic errors associated with modelling assumptions and identify conditions under which our model can identify `non-standard' stellar density profiles that have central cusps and/or steepened outer slopes.  Applying our model to real dwarf spheroidals, we do not find evidence for centrally cusped density profiles among the fifteen Milky Way satellites for which our tests with mock data indicate there would be sufficient detectability.  We do detect steepened (with respect to a standard Plummer model) outer profiles in several dSphs---Fornax, Leo I, Leo II, and Reticulum II---which may point to distinct evolutionary pathways for these objects.  However, the outer slope of the stellar density profile does not yet obviously correlate with other observed galaxy properties. 
\end{abstract}


\section{INTRODUCTION}
\label{introduction}

The advent of deep, wide-field sky surveys has quadrupled the number of known dwarf spheroidal galaxies (dSphs) surrounding the Milky Way (e.g. \citet{sdss-dsphs}, \citet{desy2-dsphs}, and \citet{beasts}). These extremely faint systems contain old ($>$10 Gyr) stellar populations. They are dark matter dominated, with mass to light ratios reaching into the hundreds or even thousands (see \citet{mateo1998}, \citet{mcc}, and \citet{simon2019}). Because of their diminutive masses, and subsequent sensitivity to dark matter distributions and baryonic processes, it is challenging for galaxy formation theories to generate both the currently observed number and internal structure of dSphs \citep{bullock-bk-2018}. The dSphs also provide excellent targets for indirect dark matter detection experiments in the Local Group, as their old, quiescent stellar populations and lack of gas minimize astrophysical backgrounds \citep{indirect-detection}. For these reasons, dSphs provide an important small-scale constraint to theories of dark matter and cosmology. In order to test these theories, however, dSph dynamical masses must be accurately measured.

The mass profile of a dSph is usually inferred by measuring the line-of-sight velocity distribution of stars in the galaxy and comparing to a model via the spherical Jeans equation \citep[e.g.][]{jeans,intro_ags,intro_read,intro_strigari}, which relates the enclosed dynamical mass profile to the stellar velocity dispersion and stellar density profiles. While much effort has been devoted to measuring the magnitudes and shapes of stellar velocity dispersion profiles \citep{Mateo1991,Kleyna2002,Battaglia2006,MW2007}, relatively little attention has been paid to measuring the shapes of the stellar density profiles.  

Most studies of dwarf spheroidal galaxies usually parameterize stellar density profiles, $\nu(r)$, using simple analytic models of fixed functional form, such as the exponential, \citet{King62}, or \citet{plummer1911} profiles \citep[e.g.][]{irwin_1995,M18b,sextans-new,MYfornax}. Such models characterize a dSph's size using only one or two parameters, and assume a fixed slope at $r=0$, either a cusp ($\frac{\mathrm{d} log(\nu(r))}{\mathrm{d} log(r)}|_{r=0} < 0$) or a core ($\frac{\mathrm{d} log(\nu(r))}{\mathrm{d} log(r)}|_{r=0} = 0$).   Therefore, adoption of such models does not provide actual measurements of the shapes of stellar density profiles.  This is especially relevant given the diagnostic implications of stellar density profiles at both small and large radius.  For example, as shown by \citet{corescuspsdSphs}, for dSph-like systems with flat velocity dispersion profiles, the central slope of the dark matter density profile relates directly to that of the stellar density profile. Moreover, since the result of \citet{tidessim} indicate that tidal stripping can alter the shape of the stellar profile, the light profile is an important clue to the dynamical history of a galaxy. Thus efforts to infer dark matter distributions and study the dynamical state of dwarf galaxies can benefit from careful measurements of the shapes of stellar density profiles.  

Some well-known parametric models do provide sufficient flexibility to fit a broad range of stellar density profile shapes, such as the $\alpha \beta \gamma$ model that allows different power-law indices at both small and large radius \citep{abgmodel}. However, the $\alpha\beta\gamma$ model lacks analytic integrals for the 2D projected stellar density.  Performing the requisite  numerical integrations then consumes significant computational resources during fits to star-count data and/or subsequent dynamical modeling.  

Here we develop a model for fitting dSph stellar density profiles that is both flexible and economical, with analytic expressions for its 3D and projected versions, as well as the number of stars enclosed within a given radius.  Following \citet{Read2017}, we use a generalized version of the Plummer model as a basis function, expressing the overall stellar density as a sum of Plummer profiles (and/or Plummer-like profiles with steepened gradients at large radius).  Using mock data, we demonstrate that our models have sufficient flexibility to distinguish among input models that follow cusped and cored central stellar density profiles, and/or have Plummer-like or steeper outer profiles.  We apply this model to known dSphs within the footprints of major sky surveys, providing a uniform analysis that returns inferences about the stellar density profiles of these objects.

\section{METHODOLOGY}
\label{sec:data preparation}

We consider that the number of stars contained in a particular square bin of the sky follows a poission distribution. Following \citet{pandas}, we adopt the following likehood function:
\begin{center}
\begin{equation}
L=e^{-N_{predicted}}  \Pi_{i=1}^{N} \Sigma (\vec{R}_{i}) \hspace{10mm} \label{eq:llf}
\end{equation}
\end{center}

where $\Sigma (\vec{R}_{i})$ is the projected stellar density at the position of a particular star $i$ specified by its coordinates $\vec{R}_{i}$ relative to the center of the dSph\footnote{For clarity, we specify the position of a star on the sky by $R$. We specify the 3-dimensional spherical radial coordinate as $r$. We use $R_{h}$ for the projected half light radius---i.e., the radius of the circle that encloses half of the galaxy's stars}. $N_{predicted}=\int \Sigma (\vec{R}) dA_{field}$ is the number of stars in the entire field as predicted by the model stellar profile; $N$ refers to the number of stars observed in the field. 
 In modeling the projected stellar density $\Sigma(R)$, we assume that the 3-dimensional stellar density of member stars (as opposed to foreground/background contamination) can be expressed as the weighted sum of individual components that each follow a simple analytic profile.  

We base our profile on the generalized version of a \citet{plummer1911} profile, for which the 3D stellar density of $N_{\rm tot}$ member stars is
\begin{equation}
\nu_{mem}(r)=N_{\rm tot}k_{\nu}\frac{b^{n-3}}{(b^2+r^2)^{n/2}}
 \end{equation}

where $b$ is the scale radius of the profile and $k_{\nu}=\frac{\Gamma(n/2)}{\Gamma(\frac{n-3}{2})\pi^{3/2}}$ is a constant that normalizes the profile. We use this function as a basis function, expressing the overall 3D density profile of stars in the galaxy as

\begin{equation}
\nu_{mem}(r)=N_{\rm tot}k_{\nu} \sum_{i=1}^{N_{components}} \dfrac{w_{i}b_{i}^{n-3}}{(b_{i}^{2}+r^{2})^{n/2}},\label{eq:multi_nu}
\end{equation}

where $w_i$ is the weight assigned to the $i^{\rm th}$ component, $b_i$ is the scale radius of that component, and $N_{\rm tot}$ is the total number of member stars, integrated over all components. The projected density of member stars on the sky is 
\begin{eqnarray}
\Sigma_{mem}(R)=2 \int_{R}^{\infty} \dfrac{\nu_{mem}(r)rdr}{\sqrt{r^{2}-R^{2}}}\\
    =k_{\Sigma} N_{\rm tot}\sum_{i=1}^{\rm N_{\rm components}}\frac{w_i b_i^{n-3}}{(b_i^2+R^2)^{\frac{n-1}{2}}},
\end{eqnarray}
with normalization constant $k_{\Sigma}=\frac{n-3}{2\pi}$. The number of members enclosed within a circle of radius $R$ on the sky is 
\begin{eqnarray}
N_{\rm mem}(R)=2 \pi \int_{0}^{R} \Sigma_{mem} (S) SdS\\
=N_{\rm tot} \sum_{i=1}^{N_{\rm components}}w_i\left[1-\frac{ b_i^{n-3}}{(b_i^2+R^2)^{\frac{n-3}{2}}}\right],
\end{eqnarray}

In theory, $n$ could have an arbitrary value, but the integral for $N_{mem}(\vec{R})$ is not analytic for all values of $n$ when an elliptical morphology is allowed within a circular field. Therefore, in this work, we consider two values of $n$: $n=5$, the commonly used \citet{plummer1911} value, and $n=9$, which allows for a steeper outer profile. As Section \ref{sec:mock_data} demonstrates, these choices give sufficient flexibility to distinguish astrophysically interesting differences between centrally cored and cusped profiles, as well as Plummer-like outer profiles and steeper gradients.

In practice, the weights $w_{i}$ and scale radii $b_{i}$ are not the actual free parameters used in the fit. The sum of the weights is degenerate with the overall normalization of the fit, so only $N_{components}-1$ of the weights are free parameters. For our choice of $N_{components}=3$, we use two free parameters $z_{2}$ and $z_{3}$ defined as: 
\begin{center}
\begin{equation}
\begin{split}
w_{1}\equiv \dfrac{1}{1+z_{2} +z_{3}}
\\
w_{2}\equiv\dfrac{z_{2}}{1+z_{2} +z_{3}} \\
w_{3}\equiv\dfrac{z_{3}}{1+z_{2} +z_{3}} \label{eq:weights}
\end{split}
\end{equation}
\end{center}

For the scale radii of each component, we enforce $b_{1}>b_{2}>b_{3}$, to avoid multi-modalities that would arise due to swapping labels among the three components. Furthermore, supposing that the available data span a sufficiently large fraction of the galaxy's area, we enforce the condition that $b_{1} < R_{field}$. Therefore, the free parameters we use are  $m_{1}$, $m_{2}$ and $m_{3}$, where $m_{i}$ is a number between 0 and 1:
\begin{center}
\begin{equation}
\begin{split}
b_{1}=m_{1}*R_{field} \\
b_{2}=m_{2}*b_{1} \\
b_{3}=m_{3}*b_{2} \label{eq:bs}
\end{split}
\end{equation}
\end{center}

\begin{deluxetable*}{cccc}
\tablecaption{Free Parameters\label{params}}
\tablehead{
\colhead{Parameter} & \colhead{Description} & \colhead{Prior Range} & \colhead{Equation Reference}}

\startdata
$\log_{10}(f_{mem})$ & Fraction of stars in the field that are members of the dSph & $-5$ - $0$ & \ref{eq:multi_nu} (part of $N_{0}$) \\
$\log_{10}(\Sigma_{MW})$ & Projected density of Milky Way contaminant stars in the field & $-2$ - $8$ ($ \mathrm{deg} ^{-2}$) & \ref{eq:llf},\ref{eq:multi_nu} (part of $N_{stars,model}$ and $N_{0}$) \\
$\log_{10}(z_{2})$ & Second component unnormalized weight & $-3$ - $3$& \ref{eq:weights} \\
$\log_{10}(z_{3})$ & Third component unnormalized weight & $-3$ - $3$ &  \ref{eq:weights} \\
$\log_{10}(m_{1})$ & First component scale radius factor & $-3$ - $0$ & \ref{eq:bs} \\
$\log_{10}(m_{2})$ & Second component scale radius factor & $-3$ - $0$ & \ref{eq:bs} \\
$\log_{10}(m_{3})$ & Third component scale radius factor & $-3$ - $0$ & \ref{eq:bs} \\
$\epsilon$ & Ellipticity & $ 10^{-4}$ - $0.9$ & \ref{eq:ell_rad} \\
$\theta_{0}$ & Fit position angle (East of North) & $0$ - $\pi$ & \ref{eq:ell_rad} \\
\enddata
\tablecomments{Free parameters used to fit our 3-Component model. }

\end{deluxetable*}

In order to allow for flattened morphologies, we take the magnitude of the projected position vector $\vec{R}$ in Equation \ref{eq:llf} to be an `elliptical radius'
\begin{center}
\begin{equation}
R_{e} = \sqrt{\xi ^{2}+\eta^{2}} \Big(\cos(\theta-\theta_{0})^{2}+\dfrac{\sin(\theta-\theta_{0})^{2}}{(1-\epsilon)^{2}}\Big) \label{eq:ell_rad}
\end{equation}
\end{center}
where $\xi$ and $\eta$ are the standard coordinates of a star's RA and DEC with the origin at the center of the galaxy in question (centers listed in Table \ref{galaxiesall}). 
In Equation \ref{eq:ell_rad}, $\theta=\tan^{-1} \big(\dfrac{\xi}{\eta} \big)$ is the angle of the star with respect to the $\xi$ axis (which points toward East), $\theta_{0}$ is the position angle of the ellipse (increasing East of North), and $\epsilon\equiv 1-b/a$ is the ellipticity of the projected stellar density profile, where $a$ and $b$ are semi-major and semi-minor axes, respectively. Even with elliptical symmetry and allowing for a circular field of finite radius, the integral for $N_{predicted}$ in Equation \ref{eq:llf} remains analytic for the Plummer-like ($n=5$) profiles that we consider. 

Henceforth, we refer to our model based on the Plummer profile and $N_{components}=$1 or $3$ as ``1-Plummer'' and ``3-Plummer'' profiles, and to our model with steeper outer slope ($n=9$) as ``1-Steeper'' and ``3-Steeper'' profiles.  In total our models have nine free parameters.  Table \ref{params} lists these parameters along with the ranges over which their tophat priors are nonzero.  We use MultiNest \citep{multinest}, a nested sampling software package, to perform the fits. Unlike the maximum-likelihood methods used by \citet{martin2008} and \citet{M18b}, nested sampling via MultiNest calculates the Bayesian evidence, allowing for quantitative model comparison (discussed in section \ref{sec:mock_data}).  MultiNest also returns a random sampling of the posterior probability distribution function.  

\section{Validation with Mock Data}
\label{sec:mock_data}

In order to gauge the reliability of our methodology, we first test our fitting procedure against mock data sets drawn from known stellar density profiles. We generate mock data sets by randomly sampling the radial coordinate $R$ from the projection of an $\alpha\beta\gamma$ \citep{abgmodel} model, $\nu(r)\propto (r/r_s)^{-\gamma}(1+(r/r_s)^\alpha)^{(\gamma-\beta)/\alpha}$, with the transition parameter held fixed at $\alpha$=2.  In order to test for recovery of a variety of central and outer slopes, we draw from input models having either central cores ($\gamma=0$) or cusps ($\gamma=1$), and outer profiles that follow either the standard Plummer ($\beta=5$) or steeper ($\beta=9$) forms.  We assign $\theta$ coordinates according to an assumed ellipticity  $\epsilon_{mock}=0.6$. Parameter values for scale radii $r_{s}$ ($1'-8'$) and total number of member stars ($10-10000$) are chosen to be similar to the real dSph data sets we analyze. We also add a background of uniformly-distributed contaminant stars based on membership fractions ranging between 0.003 and 0.1. All mock data sets have a field of view of $1^{\circ}$.  We fit each of 2000 mock data sets four different ways, allowing $N_{components}=1$ or $N_{components}=3$ for both Plummer and Steeper basis functions. 

\subsection{Systematic Errors}
\begin{figure*}
\begin{centering}
    \begin{tabular}{@{}ccc@{}}
        \includegraphics[scale=0.75]{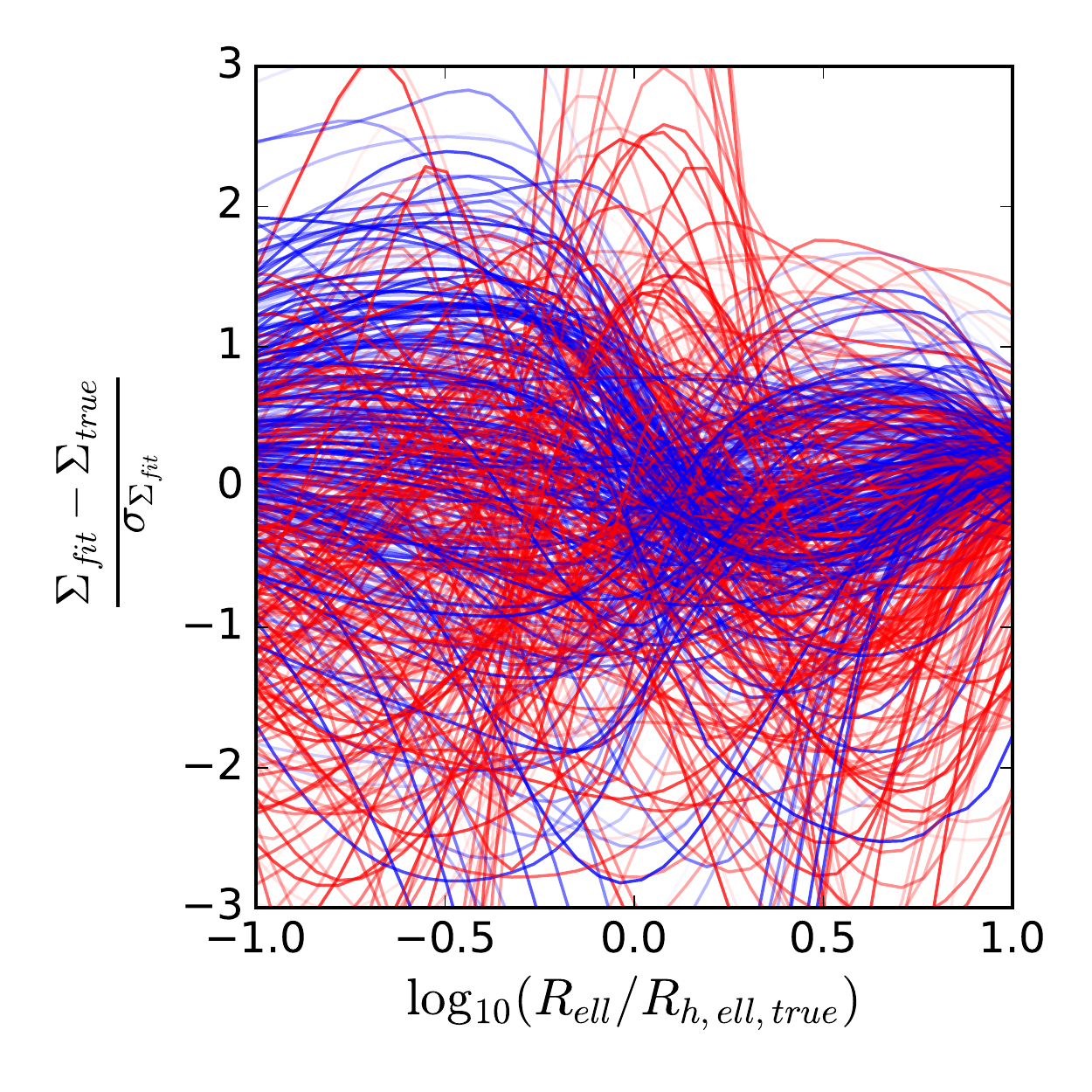}&\hspace{-0in}\includegraphics[scale=0.75]{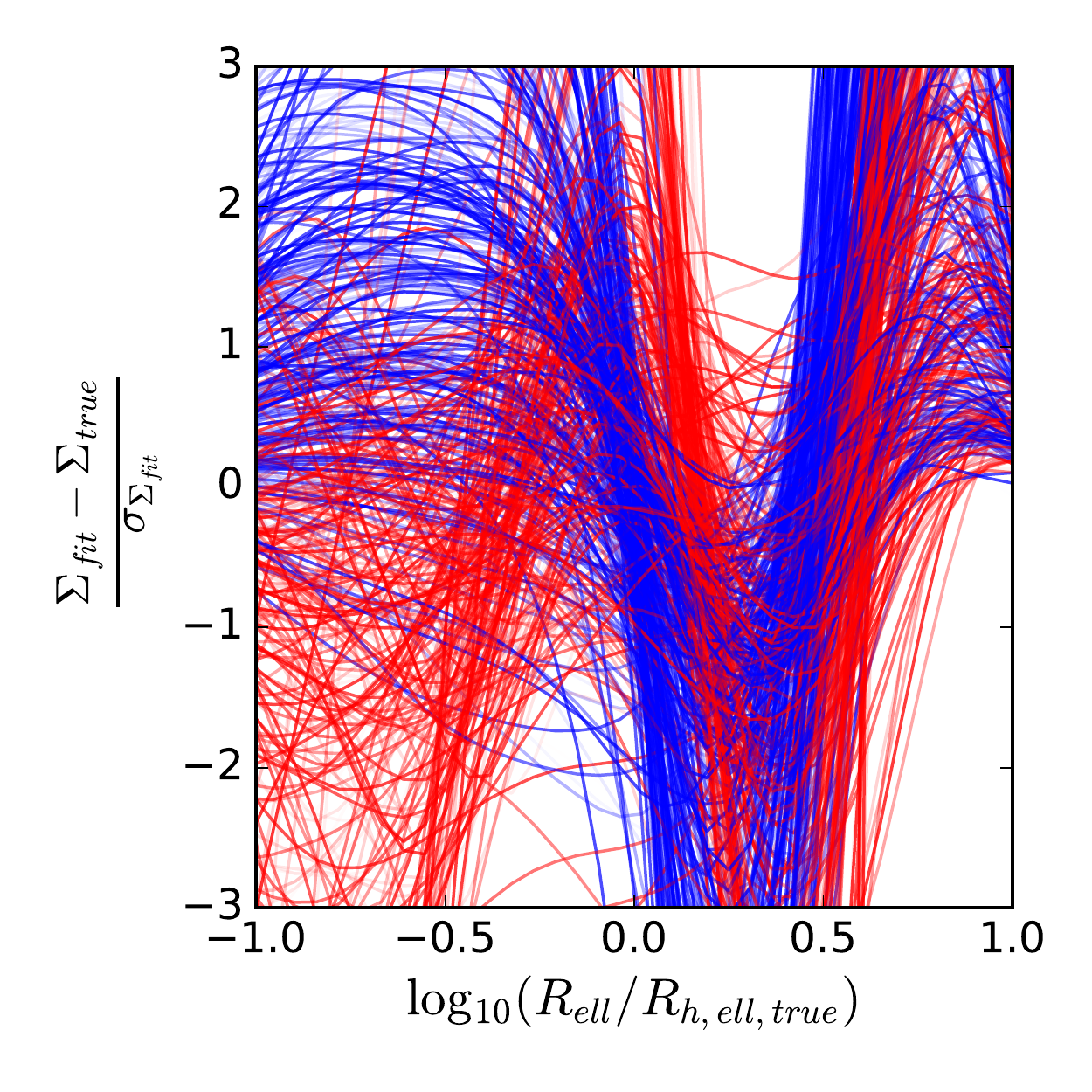}
        \\
        \includegraphics[scale=0.75]{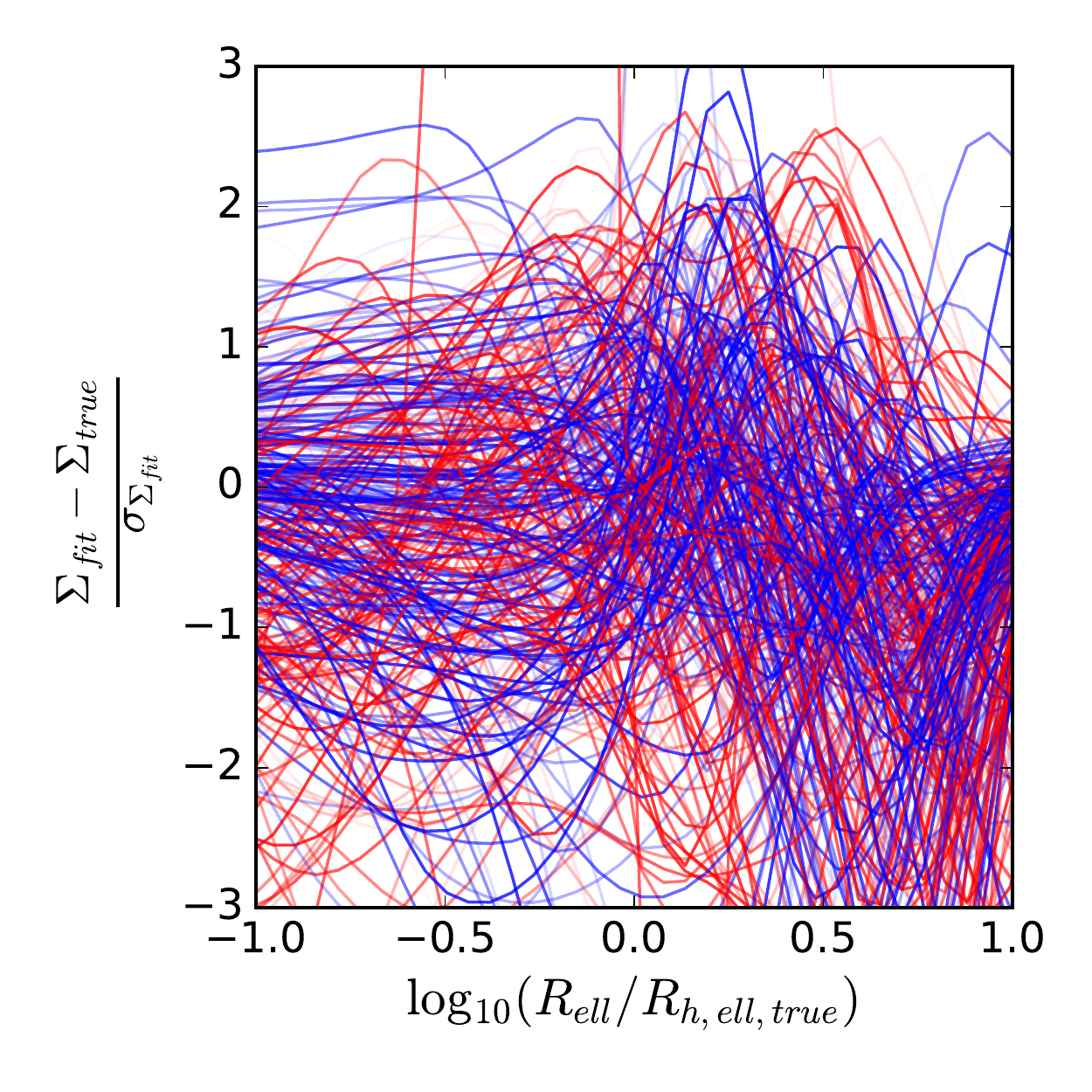}&\hspace{-0in}\includegraphics[scale=0.75]{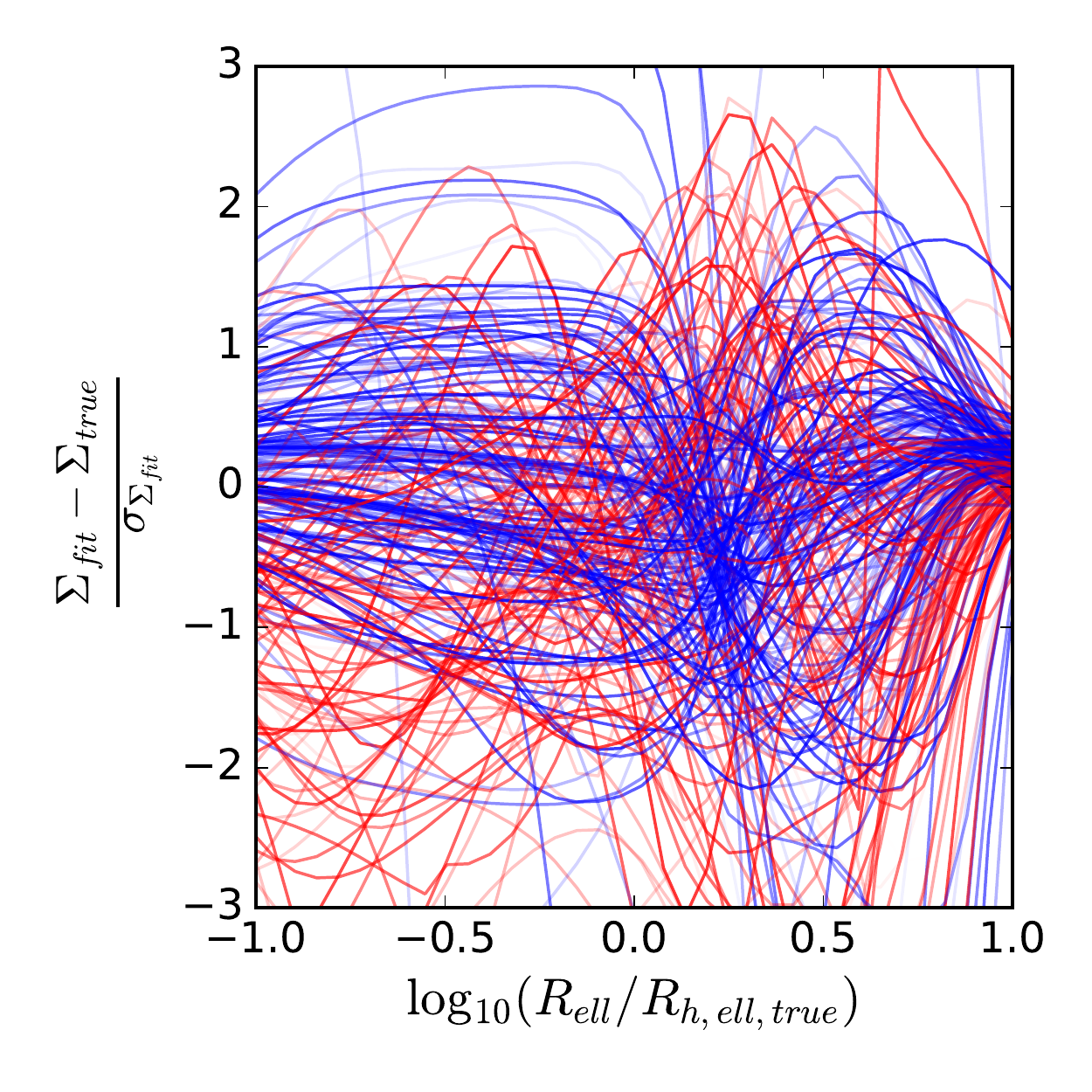}
    \end{tabular}  
    \end{centering}
\caption{ Performance of fits to mock data drawn from $\alpha\beta\gamma$ models \citep{abgmodel} with $\alpha=2$, $\beta=5$ (left panels) or $\beta=9$ (right panels), and $\gamma=0$ (blue lines) or $\gamma=1$ (red lines), for fits of 3-Plummer (top panels) and 3-Steeper (bottom panels) profiles.  As a function of projected radius (normalized by the true 2D elliptical halflight radius), panels show residuals (normalized by range of the posterior's 68\% credible interval) between the median projected stellar density calculated from the posterior probability distribution function, and the true projected density of the input model.} 
\label{resids_ell}
\end{figure*}

For all mock data sets, Figure \ref{resids_ell} shows profiles of normalized residuals $(\Sigma_{\rm fit}(R)-\Sigma_{\rm true}(R))/\sigma_{\Sigma_{\rm fit}(R)}$---i.e., as a function of elliptical radius, the difference between the median stellar density obtained from our posterior probability distribution function and the true stellar density, divided by the 68\% credible interval of the posterior. Left-hand panels portray cases where the input mock profile has $\beta=5$, while right-hand panels correspond to cases where the input has the steeper outer profile of  $\beta=9$.  Top panels show results for fits that assume the 3-Plummer model, whereas bottom panels show results for fits that assume the 3-Steeper model.  Blue/red curves represent fits to input models with central cores/cusps ($\gamma=0/\gamma=1$).  

First, we find that when the fitted model takes the same form as the input model used to generate the mock data, the residuals scatter as expected about the true profiles, with no discernible bias.  This is the case for the 3-Plummer fits to models with Plummer inputs $(\alpha,\beta,\gamma)=(2,5,0)$ (blue curves in upper left of Figure \ref{resids_ell}), and 3-Steeper fits to models with Steeper inputs $(\alpha,\beta,\gamma)=(2,9,0)$ (blue curves in lower-right panels).  Residuals for these fits are generally confined to the region $|(\Sigma_{\rm fit}-\Sigma_{\rm true})|/\sigma_{\Sigma_{\rm fit}}\la 2$, indicating the scatter expected from statistical fluctuations.  

More interesting are the systematic errors that we observe when the input model violates the assumptions of the model adopted in the fit.  For example, fitting a 3-Plummer model to data generated from models with steeper $\beta_{\rm true}=9$ gives
large systematic errors that alternate between over- and under-estimation (upper-right panel of Figure \ref{resids_ell}).  Conversely, fitting a 3-Steeper model to data generated with $\beta_{\rm true}=5$ also gives systematic errors, albeit less dramatically so (lower-left panel).  The reason for the difference in severity of systematic errors between these two cases of mismatch is that the Steeper model allows for the outer-Plummer slope of $d\log\nu/d\log r=-5$ at finite radius, while there is no radius in a Plummer profile that achieves a slope steeper than $d\log\nu/d\log r=-5$.  

All else being equal, the degree of systematic error is larger for cases where the input model has a central cusp as opposed to a core (red versus blue curves in Figure \ref{resids_ell}).  This follows from the fact that each individual Plummer and Steeper component is cored ($\gamma=0$), so no finite sum of such components can exactly reproduce a cusped central profile.  Nevertheless, we shall find that our 3-component models have sufficient flexibility to separate cored from cusped profiles in a meaningful way (Section \ref{subsec:corecusp}).  

\subsection{Separation of cored and cusped stellar density profiles}
\label{subsec:corecusp}

Given the intrinsic `coredness' of the Plummer and Steeper profiles, it is worth investigating how well the superposition of such profiles can distinguish cores from stellar cusps.

\begin{figure}
\plotone{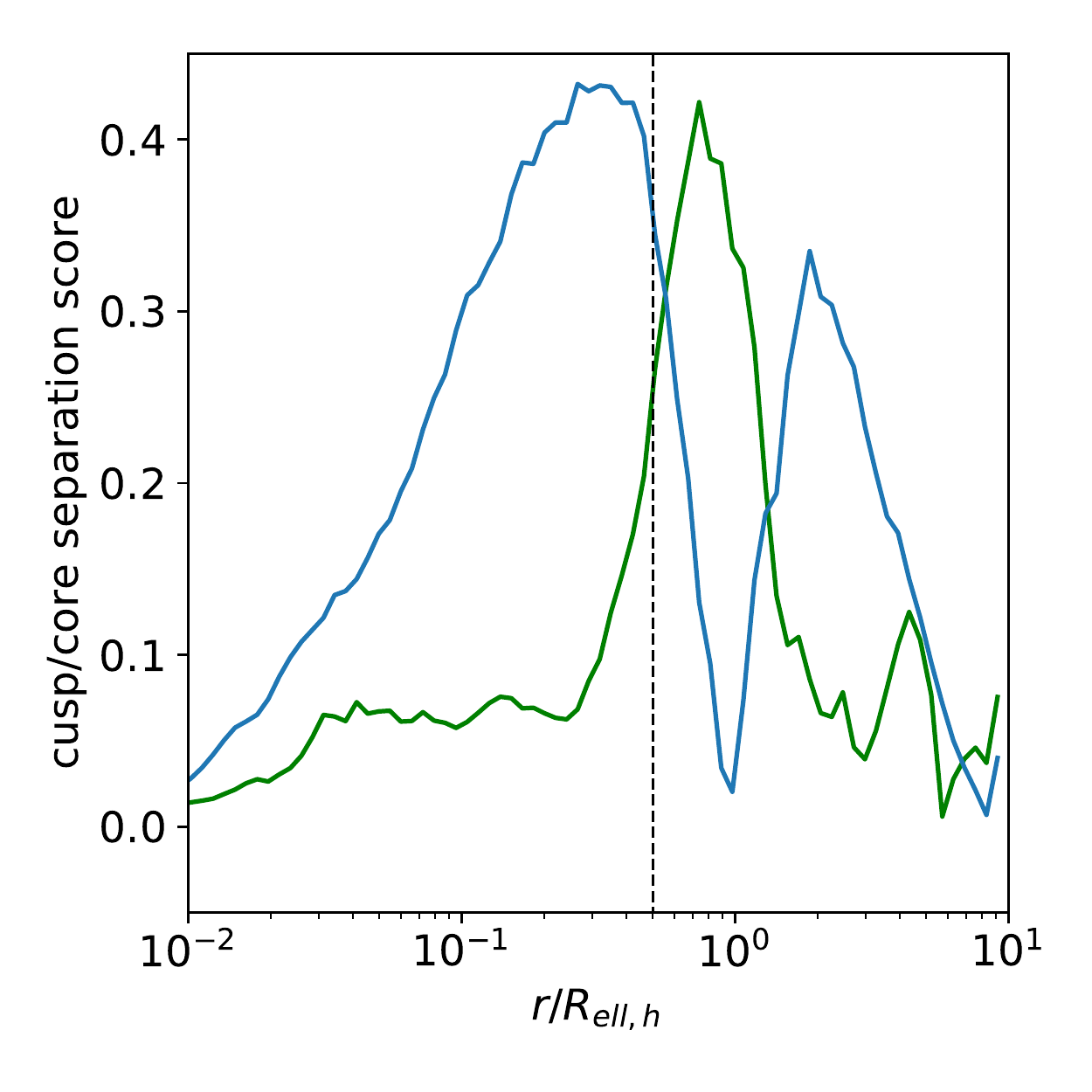}
\caption{\label{fig:ccscore} Cusp-Core separation score, as a function of the 3D radius (normalized by the projected half-light radius). The blue line refers to the score for the 3-Plummer model, and the green for the 3-Steeper. The dashed line represents $r=0.5R_{h}$.}
\end{figure}

We use our mock data sets to identify the optimal radius for making this distinction. For a given fraction of the fitted  projected half-light radius between $0.001R_{h}$ and $10R_{h}$, we assign a core / cusp separation ``score''  in the following manner.  For each mock data set, we record the logarithmic slope of the deprojected 3-Component profile\footnote{We evaluate the logarithmic slope of the 3D stellar density profile obtained by deprojecting the `circularized' version of the elliptical profile that results from replacing elliptical radius $R_e$ with `circularized' radius  $R=R_e\sqrt{1-\epsilon}$.} at that fraction of the halflight radius (taking the median from the posterior probability distribution).  The core/cusp separation score is then the difference between the median logarithmic slope obtained from all cored input models and the median obtained from all cusped input models, divided by the quadrature sum of standard deviations obtained for each of the two classes of model. This score is plotted as a function of 3D radius in Figure \ref{fig:ccscore}, for both the 3-Plummer (blue line) and 3-Steeper (green line) models. While both scores are maximized in slightly different locations, we find that both remain near their peaks at 3D radius $r\sim 0.5R_{h}$, where $R_h$ is the inferred projected halflight radius.

\begin{figure*}
\begin{centering}
    \begin{tabular}{@{}ccc@{}}
        \includegraphics[scale=0.75]{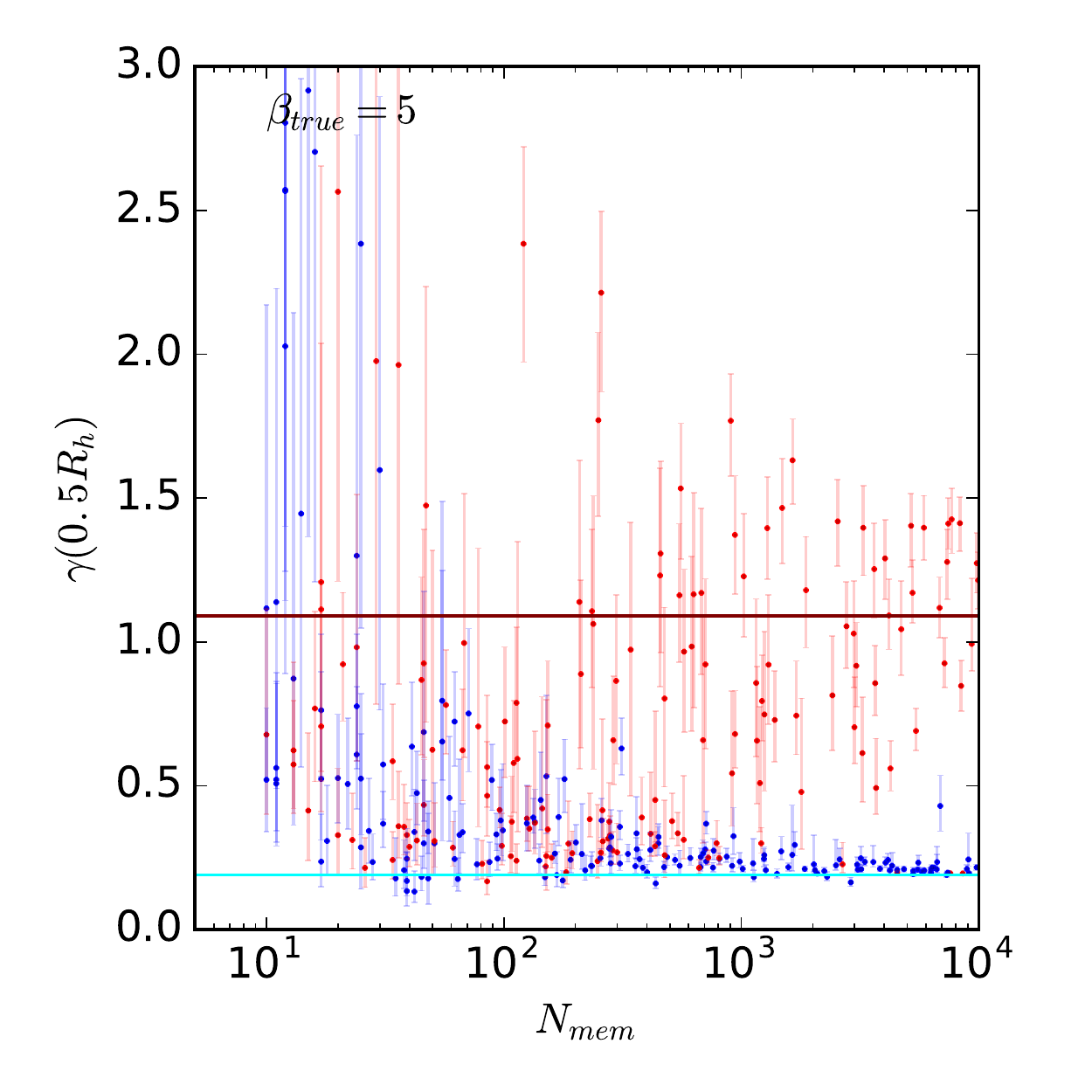}&\hspace{-0in}\includegraphics[scale=0.75]{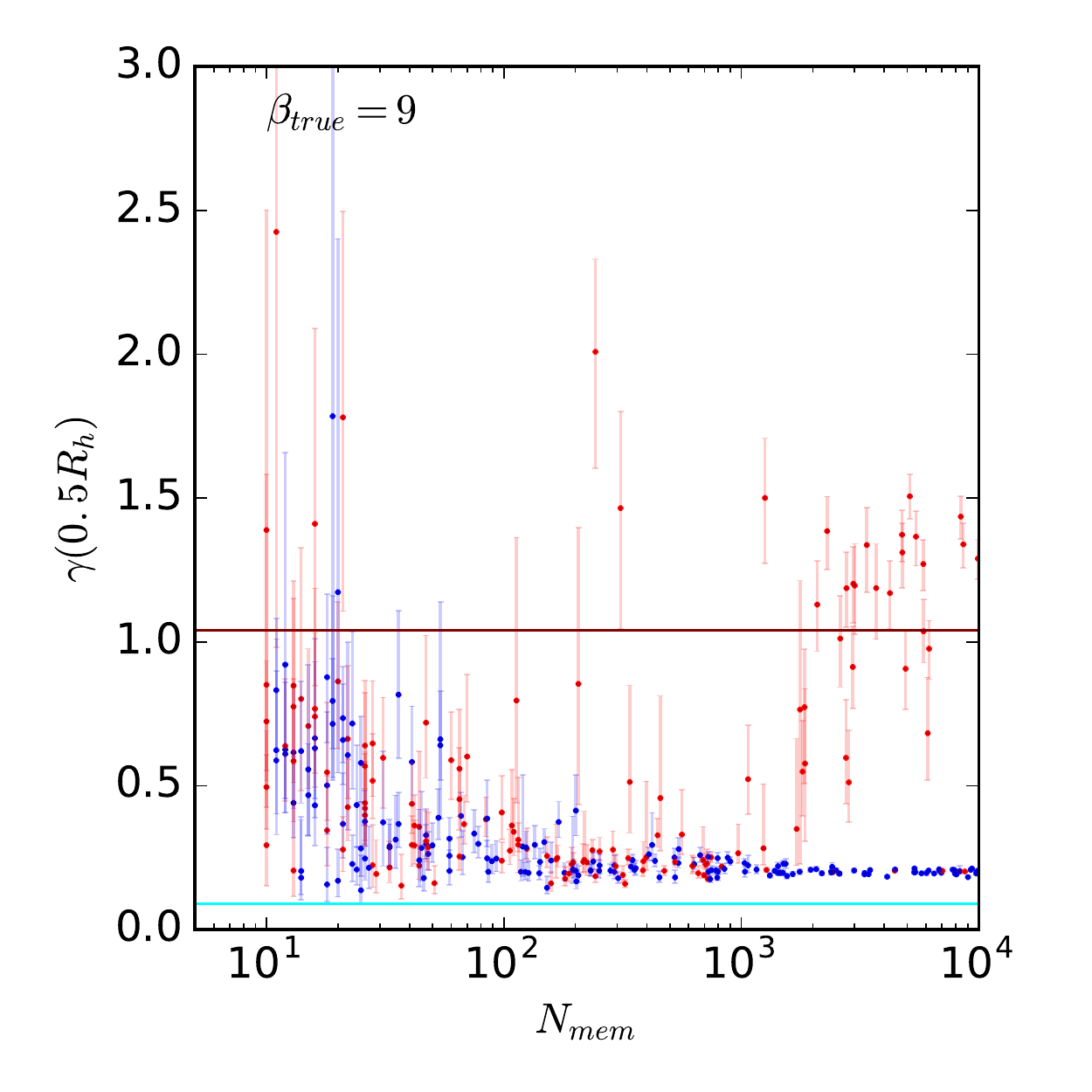}
        \\
        \includegraphics[scale=0.75]{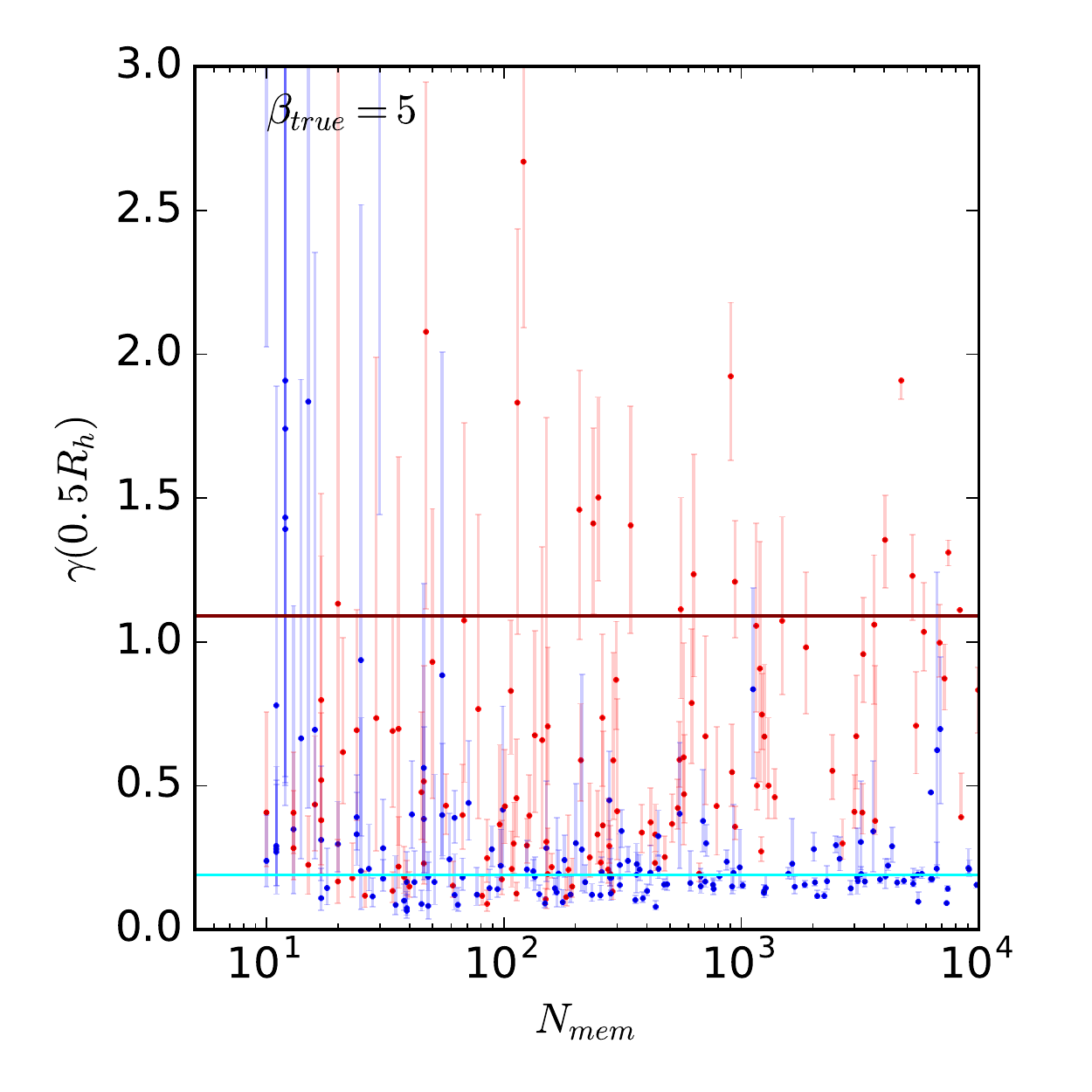}&\hspace{-0in}\includegraphics[scale=0.75]{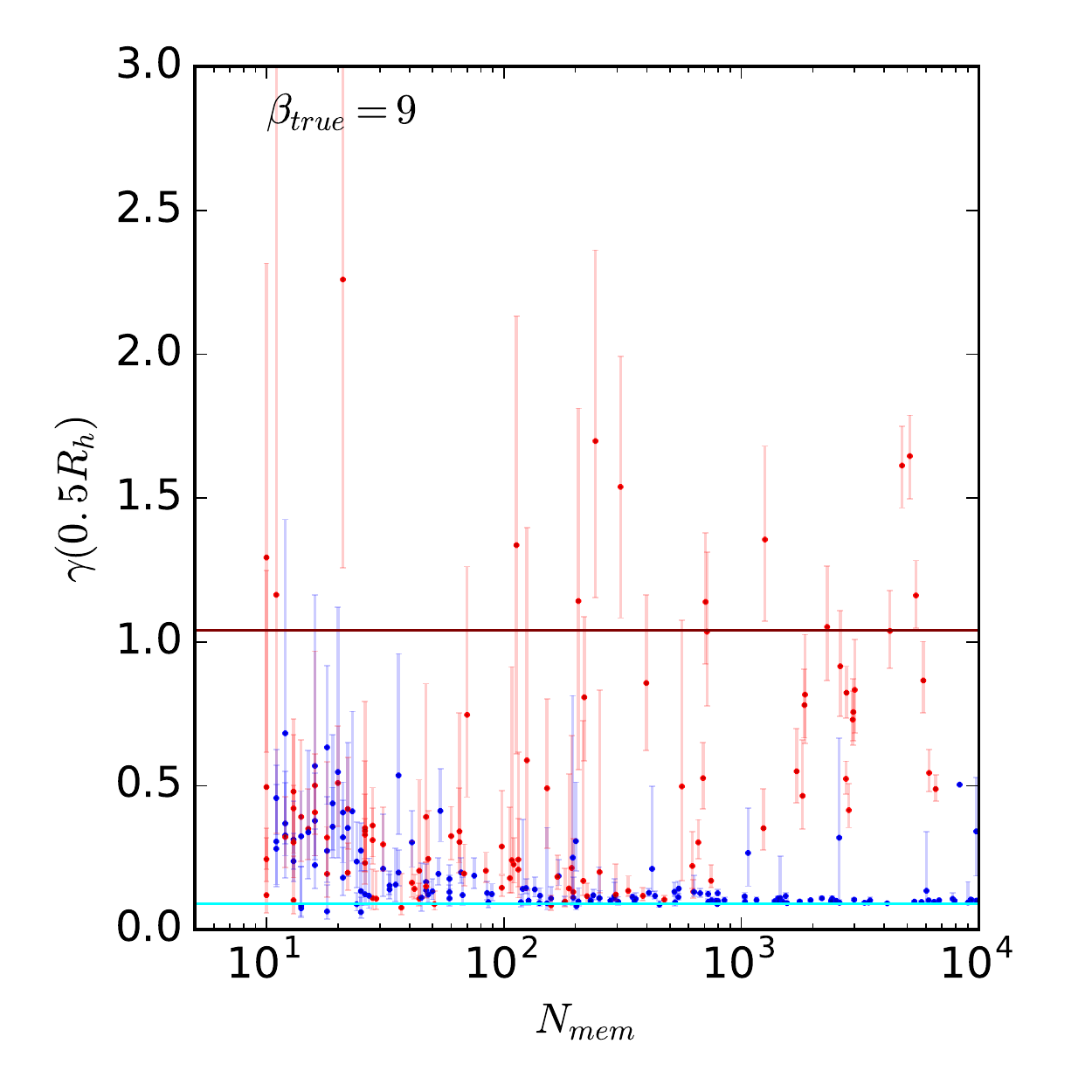}
    \end{tabular}  
    \end{centering}
\caption{Separation of  cored and cusped stellar density profiles using mock data sets.  Plotted as a function of the number of member stars is the negative of the logarithmic slope, $\gamma(0.5R_h)\equiv -d\log\nu/d\log r|_{r=0.5R_h}$, of the stellar density profile evaluated at radius $r=0.5R_h$, where $R_h$ is the median (of the posterior distrubtion) fitted projected halflight radius.  As in Figure \ref{resids_ell}, panels correspond to fits of  3-Plummer (top panels) and 3-Steeper (bottom panels) profiles to mock data sets drawn from $\alpha\beta\gamma$ models with input parameters $\alpha=2$,   $\beta=$5 (left panels) or $\beta=9 $ (right panels), and $\gamma=$0 (blue points) or $\gamma=1$ (red points).  Cyan/red lines represent the true logarithmic slope at $R=0.5R_h$ for the input models with central cores/cusps ($\gamma=0/\gamma=1$).  
} 
\label{fig:mock_gamma_vs_nmem}
\end{figure*}

For each mock data set and as a function of the number of mock members, Figure \ref{fig:mock_gamma_vs_nmem} shows the logarithmic slope that we measure at $r=0.5 R_{h}$ for our 3-Plummer and 3-Steeper fits. For sufficient numbers of galaxy member stars ($N_{mem} \ga 300$), our fits to mock data generated from cusped models (red points) tend to separate from those to data generated from cored models (blue points).  This distinction holds for the 3-component versions of both Plummer and Steeper models. Henceforth, we take $\gamma(0.5R_h)$ as a useful indicator of the central slope of the stellar density profile.

\subsection{Model Selection}
\label{subsec:modelselection}
In order to examine when and if models with more flexibility than the standard single-component Plummer profile are required, we now compare our 3-Plummer and 3-Steeper models to the standard 1-Plummer model and 1-Steeper model in terms of model selection.  For this purpose we use the Bayesian `evidence', or marginalized likelihood---i.e., the integral over the parameter space of the likelihood multiplied by the prior.  Specifically we quantify the `Bayes factor' $B\equiv \log_{10}(E_{M_{1}}/E_{M_{2}})$, where $E_{M_{1}}$ and $E_{M_{2}}$ are the evidences (calculated by MultiNest) for fits to two models $M_{1}$ and $M_{2}$, respectively.  The Bayes factor naturally favors simpler models---owing to their smaller prior volumes---unless additional complexity provides a significantly better fit to the data.  When comparing two models, a Bayes factor of $1/2$ is regarded as ``substantial" support for Model 1 over Model 2 \citep{bayesfactor}.	Regardless of such a subjective criterion, our mock data let us examine what Bayes factors reliably identify data sets that require more modeling complexity than is afforded by the assumption of a  standard 1-Plummer profile.

\begin{figure*}
\begin{centering}
    \begin{tabular}{@{}ccc@{}}
        \includegraphics[scale=0.75]{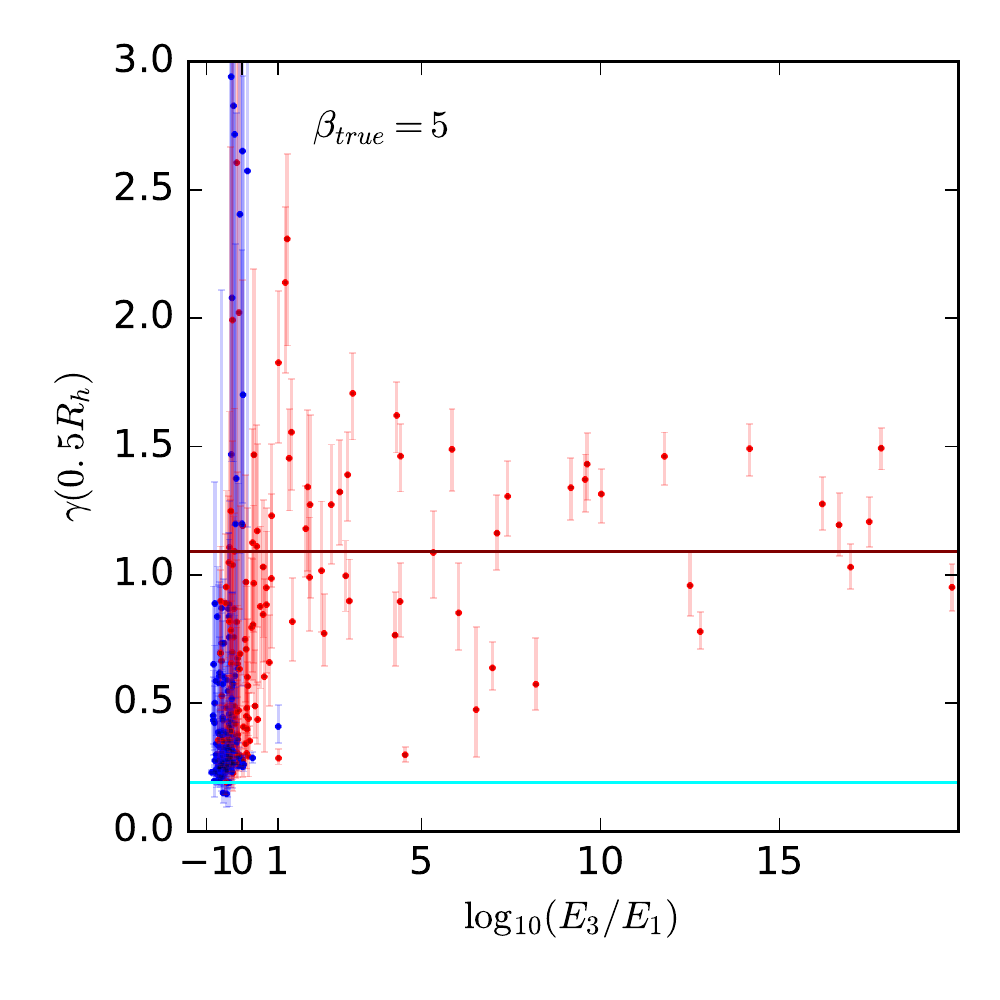}&\hspace{-0in}\includegraphics[scale=0.75]{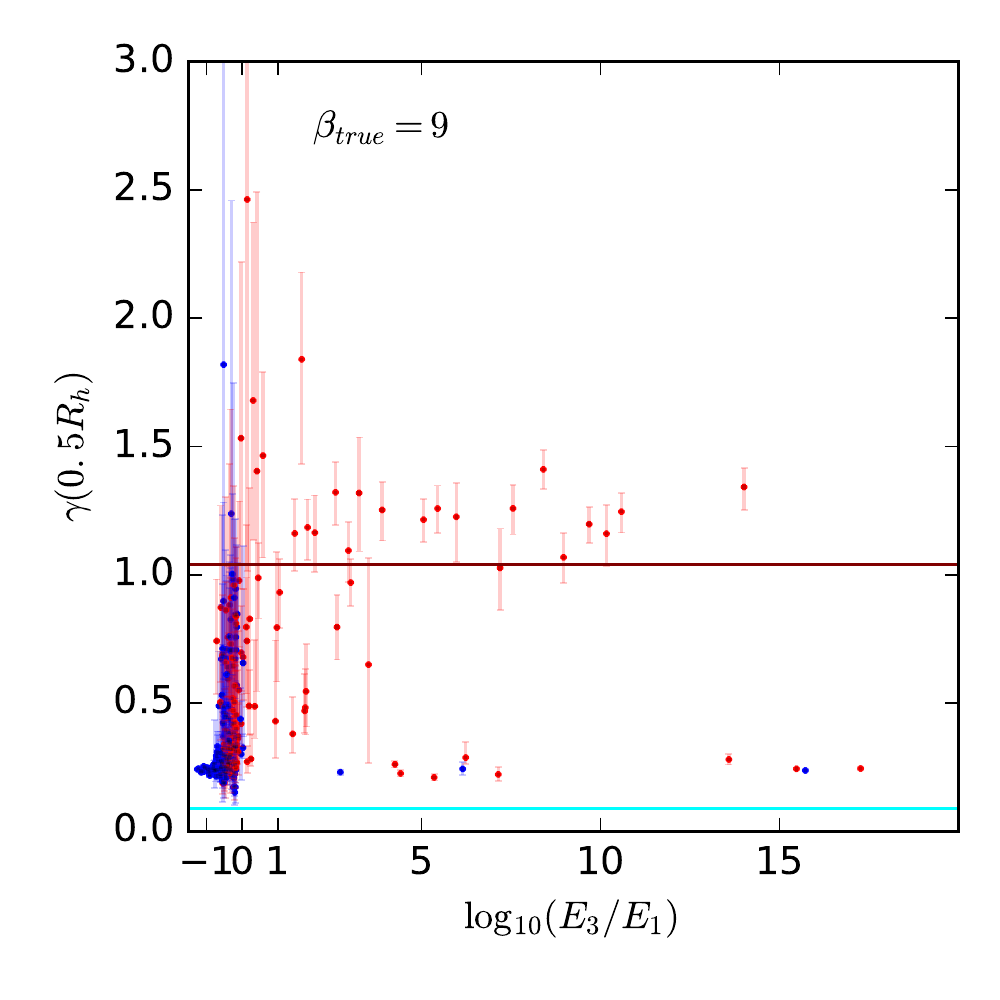}
        \\
        \includegraphics[scale=0.75]{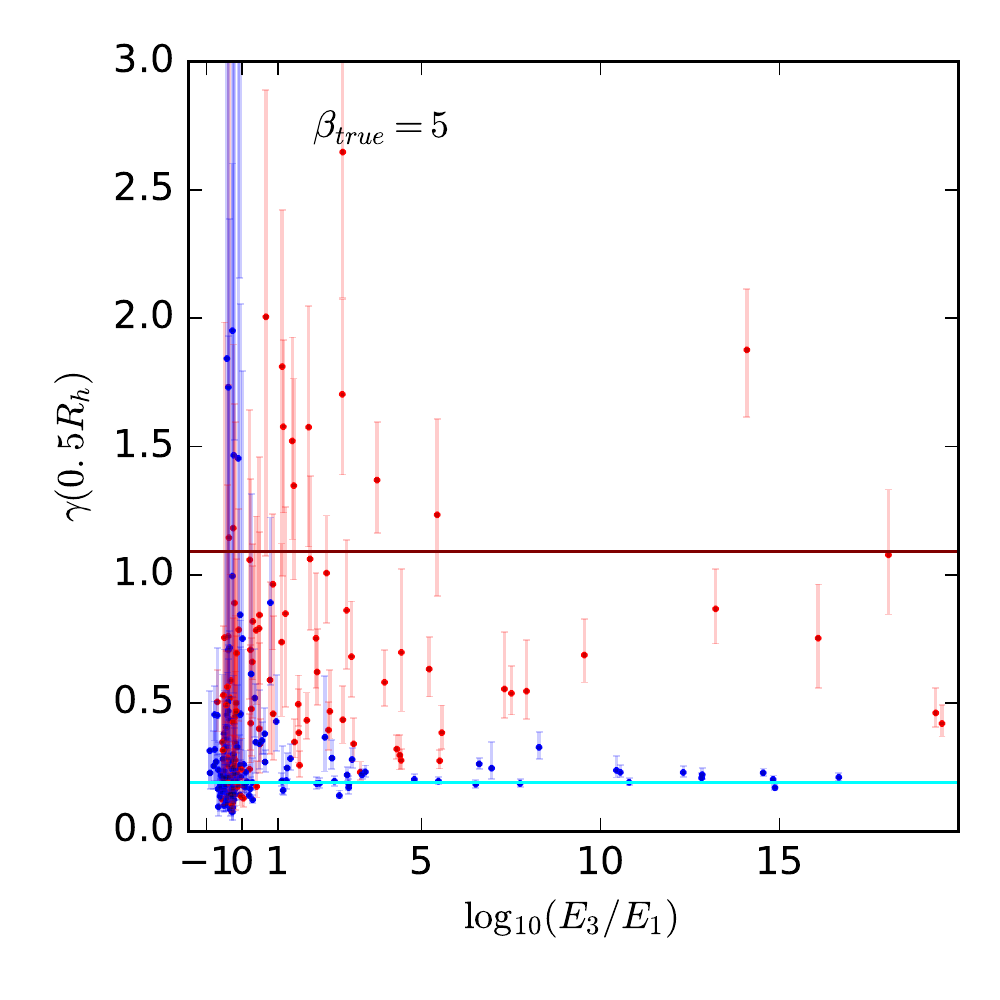}&\hspace{-0in}\includegraphics[scale=0.75]{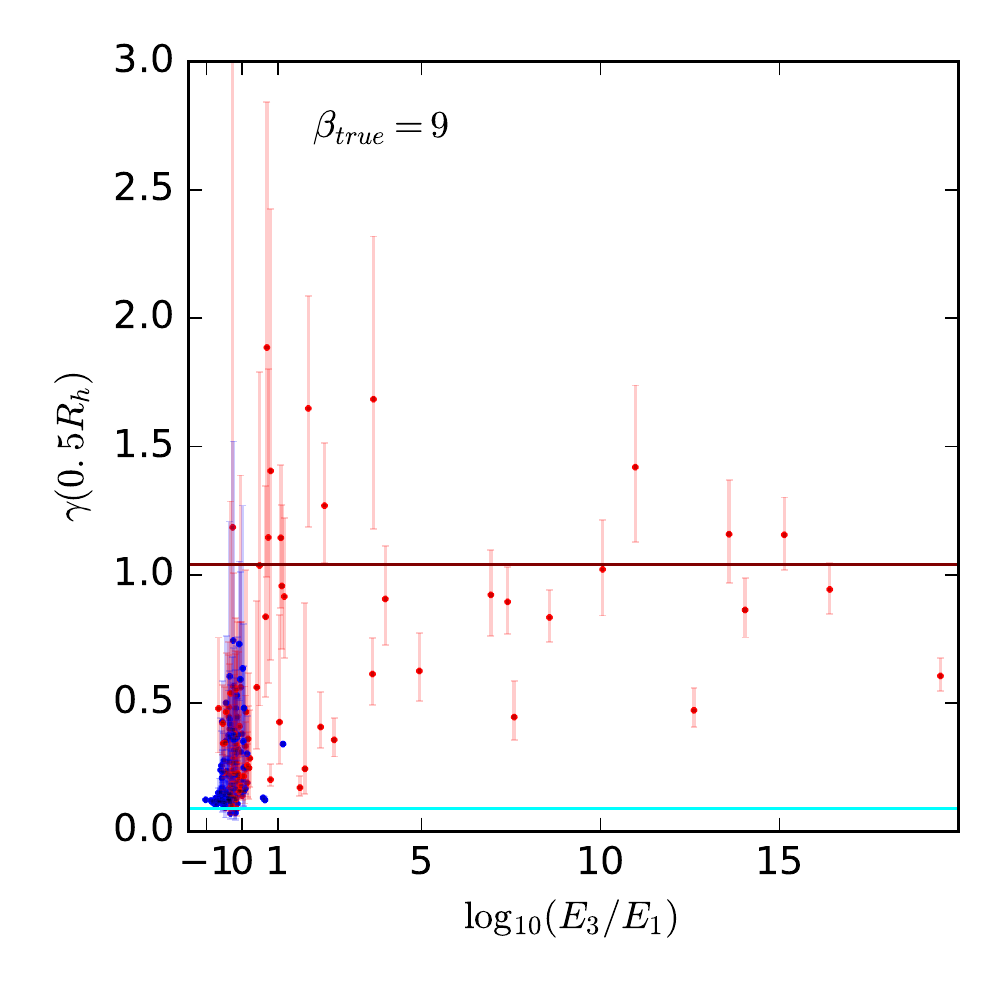}
    \end{tabular}  
    \end{centering}
\caption{Same as Figure \ref{fig:mock_gamma_vs_nmem}, except the slope parameter is plotted as a function of the logarithm of the ratio of evidences comparing 3-Component to 1-Component models.
} 
\label{fig:mock_gamma_vs_bayes}
\end{figure*}

For each mock data set, Figure \ref{fig:mock_gamma_vs_bayes} plots the slope $\gamma(0.5R_h)$ estimated from our 3-component fits versus the Bayes factor that compares 1-component and 3-component fits. Not surprisingly, the Bayes factor indicates strong support for 3-component models over 1-component models when the logarithmic slope is constrained to be cuspy ($\gamma(0.5R_h)\ga 0.5$)--i.e. behavior that our 1-component models cannot capture.  Exceptions can occur when the adopted model assumes the wrong outer slope (upper-right and lower-left panels of Figure \ref{fig:mock_gamma_vs_bayes}); reassuringly, these misleading Bayes factors disappear when we assume the correct outer slopes (upper-left and lower-right panels).  

\begin{figure*}
\plottwo{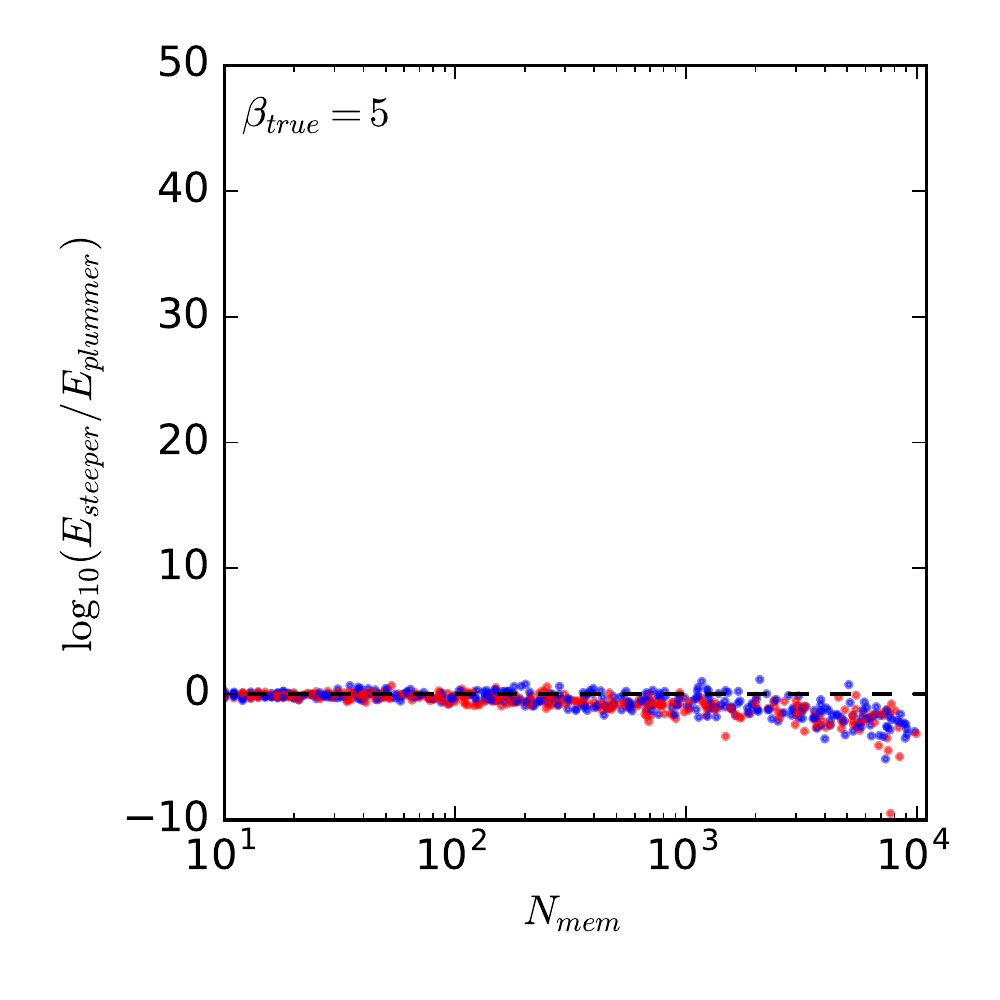}{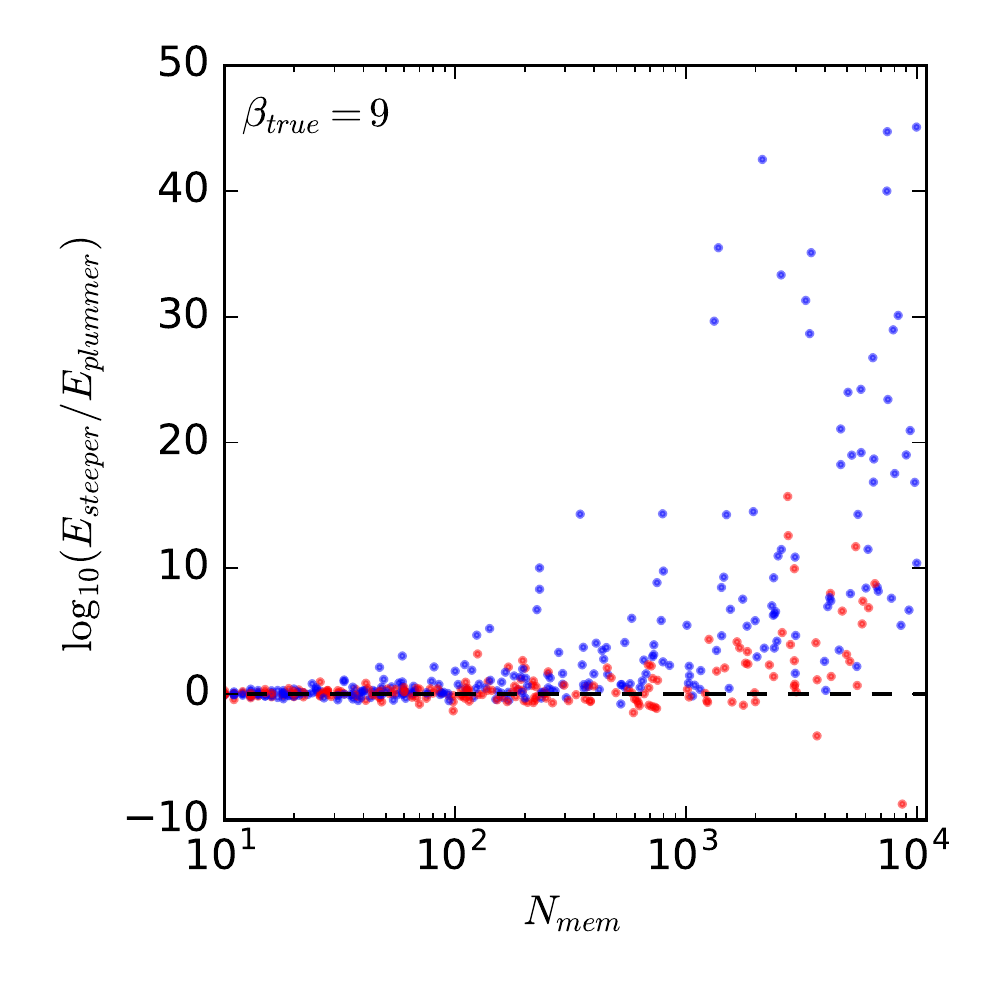}
\caption{\label{fig:steeper_vs_e3_evidence}Evidence ratios of fits to mock data of 3-Steeper profiles as compared to 3-Plummer profiles (larger values favor the 3-Steeper model). Mock data is drawn from an $\alpha\beta\gamma$ \citep{abgmodel} profile with $\alpha=2$, $\beta=$5 (left panel) or 9 (right panel), and $\gamma=$0 (blue points) or 1 (red points). Plotted as a function of the number of member stars in the mock data sets. In all cases, fits allowed for flattened morphology. The dashed line represents a log-evidence ratio of 0.}
\end{figure*}

In order to gauge our ability to discern \textit{outer} slopes, 
Figure \ref{fig:steeper_vs_e3_evidence} displays Bayes factors that compare evidences of 3-Steeper and 3-Plummer models. We find that, as expected, 3-Steeper profiles tend to have negative $\log_{10}\frac{E_{Steeper}}{E_{Plummer}}$ ratios when fitting  mock data generated from models with $\beta=5$ (left hand panel of Figure \ref{fig:steeper_vs_e3_evidence}), and a positive value when fitting to mock data drawn from a model with $\beta=9$ (right panel). This behavior holds for both cored and cusped input models (blue and red points).  In general, we find that evidence ratios reliably indicate significant support for the correct outer slope when samples include $\ga 300$ member stars (not including contaminants).

\subsection{Halflight radius}
\begin{figure*}
\plottwo{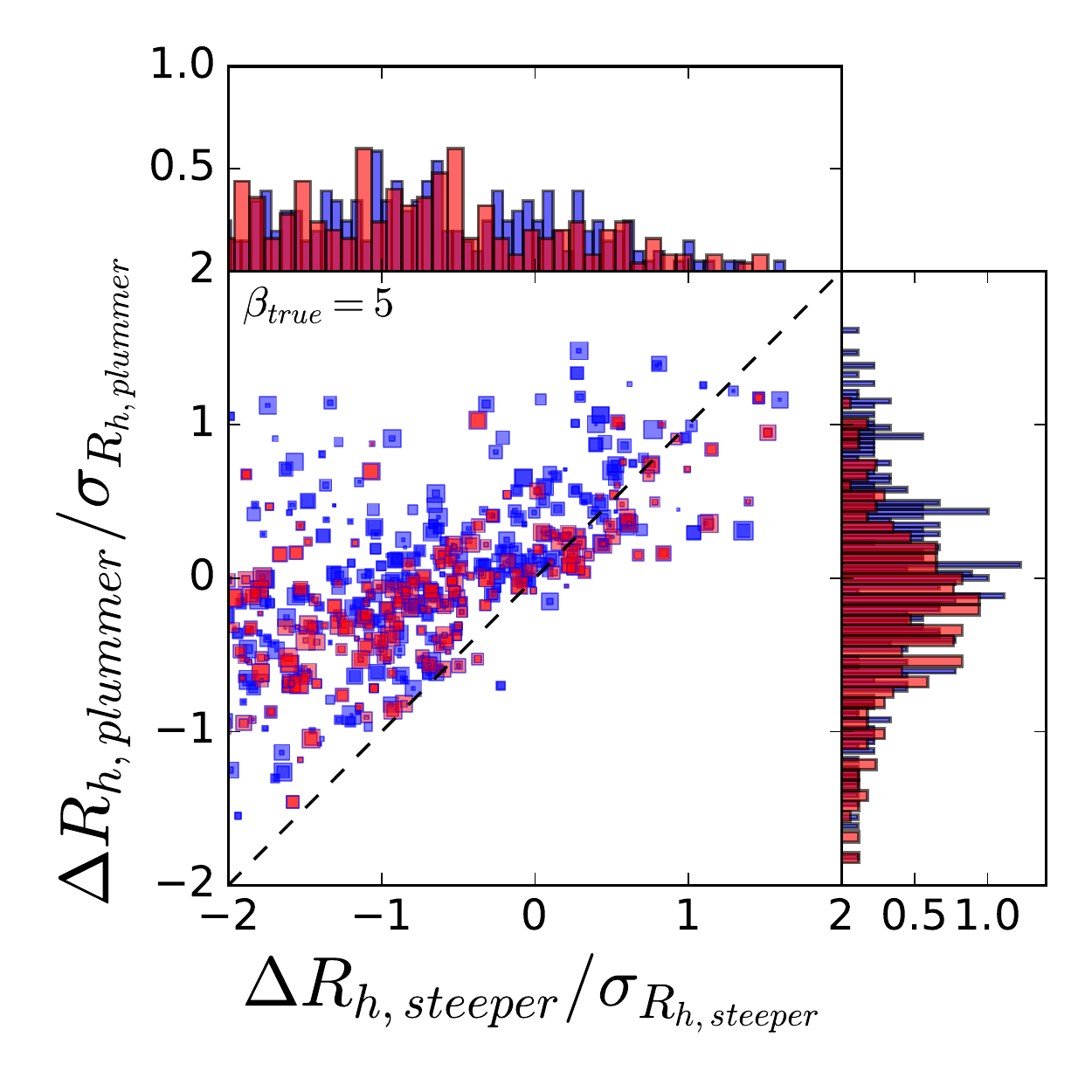}{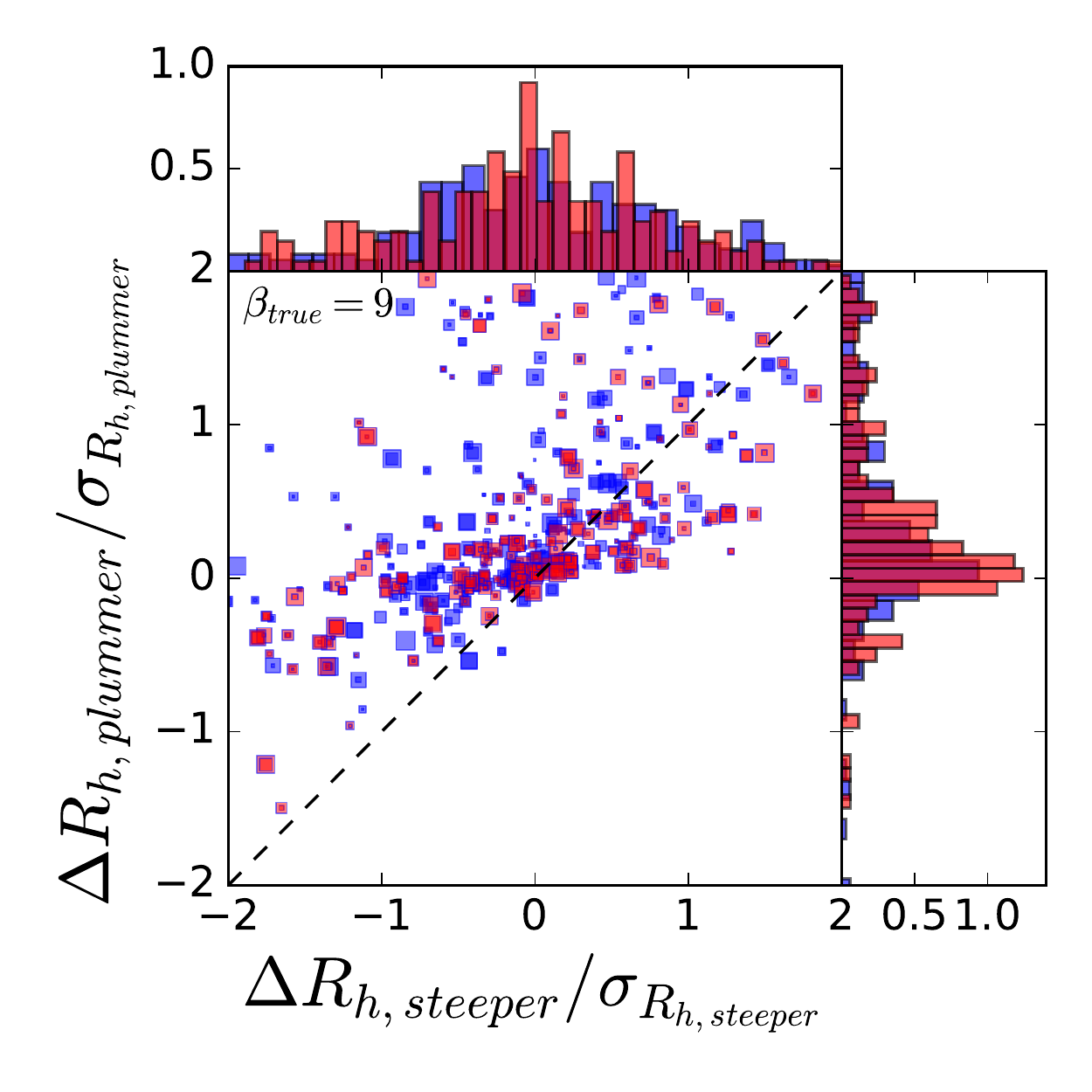}
\caption{\label{rh_plot_steeper} Comparison of estimates of halflight radii obtained from 3-Plummer versus 3-Steeper fits to mock data generated from models with input parameters $\alpha=2$, $\beta=5$ (left) or $\beta=9$ (right) and $\gamma=0$ (blue) or $\gamma=1$ (red).  Plotted points indicate the difference between the estimated halflight radius (median of the posterior probability distribution function) and the true value ($\Delta R_h\equiv R_{h,estimate}-R_{h,true}$), normalized by the 68\% credible interval of the posterior.  Histograms show normalized distributions of residuals for both sets of models. Marker size is proportional to the number of member stars in the mock data set. Black dashed lines indicate the 1:1 relation.} 

\end{figure*}

For many investigations of dwarf galaxy dynamics, the stellar density profile is summarized by the halflight radius---for example, crude dynamical mass estimators typically express the mass as function of the halflight radius and global velocity dispersion \citep[e.g., ][]{walker09,wolf10}. Figure \ref{rh_plot_steeper} compares our fitted estimates of halflight radii to true values.  We find that when the fitted model assumes the correct outer slope, estimates of the halflight radius tend to be unbiased, regardless of whether the input model has a core or a cusp.  The lack of bias when fitting to cusped inputs testifies to the ability of the 3-component models to fit such data sets.  However, when the fitted model assumes a steeper (resp: shallower) slope than the model used to generate the data, the  halflight radius tends to be under- (resp: over-) estimated.  

\section{Application to Known dSphs}
\label{sec:dsph}
Informed by our tests with mock data, we now apply our methodology to known dSph satellites of the Milky Way.  For each dSph we fit  elliptical 1-Plummer, 3-Plummer, 1-Steeper, and 3-Steeper models to publicly available imaging survey data.

\subsection{Data}
We use wide-field survey data from the PanStarrs-1 (PS1) survey \citep{PS1overview}, the Sloan Digital Sky Survey (SDSS) \citep{SDSS14}, the Dark Energy Survey (DES) \citep{DESy1}, and the Dark Energy Camera Legacy Survey (DECaLS or DCLS) \citep{decals-overview}.  We also use some of the public catalogs of \citet[][`M18' hereafter]{M18a}, who provide deeper $g$- and $r$-band followup imaging for many of the dSphs discovered in these surveys, albeit over more limited areas that often provide only partial coverage. We find that these restricted fields in the M18 catalogs can bias our measurements, and therefore we use the M18 data only for dwarfs for which the available data cover a radius more than four times wider than the galaxy's half light radius (as measured in the PS1 or SDSS surveys). The M18 fields which pass this criteria are Bootes II, Coma Berenices, Canes Venatici II, Hercules, Leo V, Pisces II, and Willman I.  In total, the combination of adopted SDSS, PS1, DES and \citet{M18a} data cover 42 known dSph satellites of the Milky Way.  We omit 2 of these galaxies from our sample. First, our fits for Pegasus III did not yield a significant detection in any of our catalogs, as the numbers of member stars above our adopted magnitude limits were consistent with zero.\footnote{Our adopted magnitude limits are based on survey completeness limits (see below), which in some cases are shallower than the data used for the original discoveries of these systems.}  We further exclude Sculptor, for which the available DES Year-1 data provide non-uniform spatial coverage and the M18 data cover only the central regions.  For similar reasons, we exclude the PS1 data for Ursa Major II, for which we use only the SDSS data. In SDSS, we discard Sextans I and Segue II, which are fully covered by PS1 but bisected by the edge of the SDSS footprint. We omit Leo II in SDSS because of obvious crowding in its center. For Bootes III, we only make a significant detection in the deeper DECaLS data. The 40 galaxies for which we present results are listed in Table \ref{galaxiesall}, along with previously-published central coordinates, distances, metallicities derived from isochrone fitting, and the maximum radius of the field centered on the dSph (see below).

\startlongtable
\begin{deluxetable*}{cccccccc}
\tabletypesize{\scriptsize}
\tablecaption{Galaxies Analyzed\label{galaxiesall}}
\tablehead{\colhead{Name} & \colhead{Survey(s)}& \colhead{RA ($^{\circ}$ )} & \colhead{DEC ($^{\circ}$ )} & \colhead{Distance (kpc)} & \colhead{ Isochrone [Fe/H], Age (Gyr)} & \colhead{Field Radius$^{c}$ ($^{\circ}$ )} & \colhead{Reference}         }
\startdata
Bootes I & PS1, DECaLS & 210.025 & 14.5 & 66$^{+2}_{-2}$ & -2.55,13.7 & 2.0&(1,23,24) \\
Bootes II & PS1, SDSS, M18, DECaLS & 209.5 & 12.85 & 42$^{+1}_{-1}$ & -1.79,13.0 & 1.5,0.4& (1,25) \\
Bootes III & DECaLS & 209.3 & 26.8 & 52 $^{+3.6}_{-3.6}$& -2.0 & 1.0 & (2,3,4) \\
Coma Berenices & PS1, SDSS, M18, DECaLS & 186.746 & 23.904 & 44$^{+4}_{-4}$ & -2.60,13.9 &1.5,0.8& (1,26) \\
Crater 2 & PS1 & 177.310 & -18.413 & 117.5$^{+1.1}_{-1.1}$ & -1.98, 10.0 & 2.0 & (20, 21) \\
Canes Venatici I & PS1, SDSS & 202.0146 & 33.556 & 218$^{+10}_{-10}$ & -1.98,12.6 &1.5& (1,24)\\
Canes Venatici II & PS1, SDSS, M18 & 194.292 & 34.321 & 160$^{+4}_{-4}$ & -2.21,13.7 &1.0,0.225& (6,24)\\
Draco & PS1, SDSS & 260.0516 & 57.915 & 76$^{+6}_{-6}$ & -1.93 &2.0& (1) \\
Draco II & PS1 & 238.198 & 64.565 & 20$^{+3}_{-3}$ & -2.2,12.0 & 1.5& (5)\\
Hercules & PS1,SDSS, M18, DECaLS & 247.758 & 12.792 & 132$^{+12}_{-12}$ & -2.41,15.0 &1.5,0.4& (1,27) \\
Leo I & PS1 & 152.117 & 12.3064 & 254$^{+15}_{-15}$ & -1.43, 6.4 &1.5& (1,22) \\
Leo II &PS1, DECaLS &  152.117 & 12.306 & 233$^{+14}_{-14}$ & -1.62, 8.8 &1.5& (1,22)\\
Leo IV & PS1, SDSS, DECaLS & 173.238 & -0.533 & 154$^{+6}_{-6}$ & -2.54,13.7 & 1.5&(1,24)\\
Leo V & SDSS, M18, DECaLS & 172.79 & 2.22 & 178$^{+10}_{-10}$ & -2.00 & 1.5, 0.19 & (1)\\
Pisces II & SDSS, M18& 344.629 & 5.9525 & 183$^{+15}_{-15}$ & -1.9 & 1.0, 0.19 & (6,7)\\
Sagittarius II & PS1 & 298.169 & -22.068 & 65$^{+5}_{-5}$ & -2.2,12.0 & 0.5 & (5) \\
Segue I & SDSS, DECaLS & 151.767 & 16.082 & 23$^{+2}_{-2}$ & -2.72 &1.5& (1) \\
Segue II & PS1, DECaLS & 34.817 & 20.1753 & 35$^{+2}_{-2}$ & -2.0 & 1.5 & (8,9)\\
Sextans I & PS1 & 153.263 & -1.615 & 86$^{+4}_{-4}$ & -1.93, 12.0 &2.0& (1,22)\\
Triangulum II & PS1 & 33.323 & 36.1783 & 30$^{+2}_{-2}$ & -1.93,13.0 & 1.0 & (10,11) \\
Ursa Major I & PS1, SDSS & 158.72 & 51.92 & 97$^{+4}_{-4}$ & -2.18 &1.5& (1)\\
Ursa Major II & SDSS & 132.875 & 63.13 & 32$^{+4}_{-4}$ & -2.47 &1.5& (1)\\
Ursa Minor & PS1 & 227.285 & 67.223 & 76$^{+3}_{-3}$ & -2.13, 12.0 &2.0& (1,22)\\
Willman I & PS1, SDSS, M18 & 162.338 & 51.05 & 38$^{+7}_{-7}$ & -2.1 &1.5, 0.41& (1)\\
Cetus II & DES & 19.47 & -17.42 & 30$^{+3}_{-3}$ & -1.8, 10.9$^{a}$   &1.0& (13)\\
Columba I & DES & 82.86 & -28.03 & 182$^{+18}_{-18}$ & -2.1, 12.0$^{b}$   &1.0& (13)\\ 
Eridanus III & DES &  35.690 & -52.284 &87$^{+4}_{-4}$ & -1.8 &1.0 & (14, 18)\\
Fornax & DES & 39.9970 & -34.449 & 147$^{+12}_{-12}$ & -.99   &7.5& (12)\\
Grus I  & DES & 344.177 & -50.163 & 120$^{+11}_{-11}$ & -1.42 &1.0&  (14, 15)\\
Grus II & DES & -46.44 & -51.94 & 53$^{+5}_{-5}$ & -1.8, 12.5$^{a}$ & 1.0 & (13) \\
Horologium I & DES & 43.882 & -54.119 & 79 & -1.8   &2.0& (14, 16)\\
Horologium II & DES & 49.134 & -50.0181 & 78$^{+5}_{-5}$ & -2.0,13.5 &1.0& (14, 17) \\
Phoenix II & DES & 354.998 & -54.406 & 83 & -1.8 &1.0& (14)\\
Pictor I & DES & 70.948 & -50.283 & 114 & -1.8 &1.0&   (14)\\
Reticulum II & DES & 53.925 & -54.049 & 30 & -2.67   &2.0& (16,19)\\
Reticulum III & DES & 56.36 & -60.45 & 92$^{+13}_{-13}$ & -2.0$^{b}$   &1.5& (14)\\
Tucana II & DES & 342.98 & -58.57 & 57$^{+5}_{-5}$ & -1.8 &1.0& (14)\\
Tucana III & DES & 359.15 & -59.6 & 25$^{+2}_{-2}$ & -2.1, 10.9$^{a}$   &1.0& (13)\\
Tucana IV & DES & 0.73 & -60.85 & 48$^{+4}_{-4}$ & -2.1, 11.6$^{a}$   &0.7& (13)\\
Tucana V & DES & 354.35 & -63.27 & 55$^{+9}_{-0}$ & -1.6, 10.9 & 1.0&(13, 18)\\
\enddata
\tablecomments{Objects without a listed stellar population age represent entries for which no published value could be located. We therefore assumed an age of 12.0 Gyr for those systems. For the four objects without stated distance uncertainties, \citet{beasts} report errors of 0.1-0.2 magnitudes in distance modulus.  \\$^{a}$ Metallicities were converted from z-values given in \citet{desy2-dsphs} via http://astro.wsu.edu/cgi-bin/XYZ.pl, which uses \citet{Asplund2006} for the conversion.\\$^{b}$ Metallicity not determined, but given that all other ultrafaint DSGs are metal-poor, we believe that this is a fair estimate. This isochrone also corresponds to a red giant branch feature.\\$^{c}$ Where two field radii are listed, the first corresponds to the SDSS/PS1/DES/DECaLS surveys, and the second to the catalogs provided by \citet{M18a}.
\\References: \\ 1: \citet{mcc}; 2: \citet{booiii-discovery}; 3: \citet{boo-iii-more}; 4: \citet{boo-iii-red-clump}; 5: \citet{sagii-dracoii}; 6: \citet{pisc-ii-cvn-ii}; 7: \citet{pisc-ii-discovery}; 8: \citet{segii}; 9: \citet{seg-ii-least-massive}; 10: \citet{tria-ii-discovery}; 11: \citet{more-tria-ii}; 12: \citet{mcc}; 13: \citet{beasts}; 14: \citet{desy2-dsphs}; 15: \citet{tucii-grusi}; 16: \citet{retii-hori}; 17: \citet{horii-more}; 18: \citet{eridiii-tucv}; 19: \citet{retii-mw}; 20: \citet{crater2_discovery}; 21: \citet{crater2_spec}; 22: \citet{ages1}; 23: \citet{booi-age}; 24: \citet{booi-ages-2}; 25: \citet{booii-age}; 26: \citet{comab-age}; 27: \citet{herc-age}}

\end{deluxetable*}

After querying the survey databases, we applied the following cuts to separate stars from background galaxies. For PS1, we required that the difference between the aperture magnitude and the PSF magnitude be less than 0.2 magnitudes. For DES and DECaLS, we required that $spread$\_$I < 0.003$, which follows the method described by \citet{beasts}. For SDSS, we selected objects classified as stars from Data Release 9 (flags $mode=1$ and $type=6$).  For the M18 data, following those authors, we retained only the sources having $-0.4<\mathrm{sharp}<0.4$ and $\chi<3$.  For all catalogs, reddening corrections were applied according to \citet{SF2011}, based on the dust maps of \citet{SFD1998}.

A few fields had small gaps in their coverage, typically due to bright Milky Way stars.  To account for this, we estimated the number of missing stars and subtracted that from $N_{stars,model}$ in the likelihood function (Equation \ref{eq:llf}).  The typical size of the gaps was about 1\% of the field, with the largest hole,  2\% of the total field area, in Segue I (SDSS). In most cases, the gaps were far away ($\approx 1^{o} $) from the galaxy in question, where the background contamination dominates, so we only applied a correction to the parameter describing the density of Milky Way contaminant stars. In cases where the gaps were closer to the galaxy, but not near the center (between 4 and 15 times the half light radius of the galaxy), we approximated the number of missing stars at that point as $N_{missing}=\Sigma(\vec{R})\times Area_{missing}$, treating $\Sigma(\vec{R})$ as constant over the entire gap and equal to the value at its center. $Area_{missing}$ was determined by manually drawing a polygon around the gap. The only galaxy with a hole less than 4 $R_{h}$ from its center was Leo I in SDSS and DECaLS, caused by the bright star Regulus. In the SDSS data, we numerically integrated the projected stellar density over the polygon drawn around Regulus; for DECaLS, the hole was wider than Leo I istelf, so we omitted Leo I in this survey. Each data set in DECaLS had many gaps due to bright sources, but they were so numerous that it became infeasible to manually inscribe all of them in polygons.

For each galaxy, we used the isochrones from the Dartmouth Stellar Evolution Database \citep{isochrones} to generate isochrones with ages and metallicities corresponding to those listed in Table \ref{galaxiesall}. We eliminated all objects $\geq \sqrt{.35^{2}+\sigma_{g}^{2}+\sigma_{i}^{2}}$ magnitudes away from any point on the isochrone in $g-i$ color and $\geq \sqrt{.1^{2}+\sigma_{r}^{2}}$ in $r$-band magnitude, where $\sigma_{g},\sigma_{i},$ and $\sigma_{r}$ are the $1\sigma$ magnitude errors in those respective bands.  The arbitrary width of this cut is not expected to significantly affect the fit results, as it primarily sets the level of background contamination. We then used the surveys' published 95\% completeness limits to discard stars with $r$-band magnitudes fainter than than 23.5 in DES, 23.4 in DECaLS, and 22.0 in SDSS and PS1.  For the M18 data, we used a limiting magnitude of $r=24.7$.

After performing initial fits to the data using a wide field of view (typically 2 degrees, as high as 7.5 degrees for larger galaxies such as Sextans and Fornax) we re-ran the fits with smaller fields of view, with the field radius chosen to be $0.5^{\circ}$ past the point where the dSph's stellar density fell below the contaminant density in the initial fit.  Given the more limited fields of view available in the M18 data, we simply chose the radius of the field of view to be the largest that provides spatially complete coverage. Because we were interested in the outer parts of galaxies, we did not analyze any M18 data sets where the M18 field of view was smaller than 4 times the galaxy's half light radius when measured in PS1 or SDSS. Table \ref{galaxiesall} reports the radius of the field used in the final fits for each galaxy.

\vspace{10 mm}

\subsection{Results}
\label{sec:results}

 Table \ref{ellresults} lists the structural parameters obtained from our fits of the elliptical 1-Plummer and 3-Plummer models.  Columns 3 and 5 include the numbers of member stars brighter than the imposed magnitude limits, for the 3-Plummer and 1-Plummer models, respectively.  Columns 6 and 8 list the inferred ellipticities, while columns 7 and 9 list the corresponding position angles.  Columns 2 and 4 list `circularized' projected halflight radii, obtained by calculating the semi-major axis of the ellipse that encloses half of the stars, and then multiplying by $\sqrt{1-\epsilon}$.  Column 10 lists the Bayes factors that compare evidences calculated for the two models (larger values favor the 3-Plummer model).  Column 11 lists the logarithmic slope parameter $\gamma(0.5R_h)$.
 
 Table \ref{superplummertable} lists the structural parameters obtained from our fits of the flattened 1-Steeper and 3-Steeper models. Columns 3 and 5 include the numbers of member stars brighter than the imposed magnitude limits. Columns 6 and 8 list the inferred ellipticities, while columns 7 and 9 list the corresponding position angles.  Columns 2 and 4 list `circularized' projected halflight radii.  Column 10 lists the Bayes factors that compare evidences calculated for the 3-Steeper and 3-Plummer models (larger values favor the 3-Steeper model), and column 11 lists the Bayes factor comparing 1-Steeper and 1-Plummer models. Column 12 lists the Bayes factor comparing 3-Steeper models to 1-Steeper models. Column 12 lists the logarithmic slope parameter $\gamma(0.5R_h)$.  For all fits, random samples from the full posterior probability  distributions are available online at the Zenodo database\footnote{https://doi.org/10.5281/zenodo.3625137}.

 \startlongtable
\begin{deluxetable*}{lllllllllll}
\setlength{\tabcolsep}{0pt} 

\tabletypesize{\tiny}

\tablecaption{Structural parameters inferred from flattened 1-Plummer and 3-Plummer models\label{ellresults}}\vspace{2 mm}
\tablehead{\\ \colhead{Galaxy}   & \colhead{$R_{h,3c}$(')} & \colhead{$N_{mem,3c}$}& \colhead{$R_{h,1c}$(')} & \colhead{$N_{mem,1c}$} & \colhead{$\epsilon_{3c}$}& \colhead{$\theta_{0,3c}$}& \colhead{$\epsilon_{1c}$}& \colhead{$\theta_{0,1c}$} &\colhead{ $\log_{10}[\frac{E_3}{E_1}]$} &\colhead{$\gamma(0.5R_h)$}}
\startdata
\input{table_3.txt}
\enddata
\tablecomments{Values quoted represent the median$^{+1\sigma}_{-1\sigma}$ values of the posterior distribution.  $\theta_{0}$ is listed in degrees East of North. $N_{members}$ refers to the number of member stars brighter than our magnitude limits.\vspace{10 mm}}
\end{deluxetable*}

\startlongtable
\begin{deluxetable*}{lllllllllcccc}
\setlength{\tabcolsep}{0pt} 

\tabletypesize{\tiny}
\tablecaption{Structural parameters inferred from flattened 1-Component and 3-Component `Steeper' models\label{superplummertable}}
\tablehead{\\ \colhead{Galaxy}   & \colhead{$R_{h,3c}$(')} & \colhead{$N_{mem,3c}$}& \colhead{$R_{h,1c}$(')} & \colhead{$N_{mem,1c}$} & \colhead{$\epsilon_{3c}$}& \colhead{$\theta_{0,3c}$}& \colhead{$\epsilon_{1c}$}& \colhead{$\theta_{0,1c}$} &\colhead{$\log_{10}[\frac{E_3,stp}{E_3,plm}]$} &\colhead{$\log_{10}[\frac{E_1,stp}{E_1,plm}]$}&\colhead{$\log_{10}[\frac{E_3,stp}{E_1,stp}]$}&\colhead{$\gamma(0.5R_h)$}}
\startdata
\input{table_4.txt}
\enddata
\tablecomments{Values quoted represent the median$^{+1\sigma}_{-1\sigma}$ values of the posterior distribution. $\theta_{0}$ is listed in degrees East of North. $N_{members}$ refers to the number of member stars brighter than our magnitude limits. }
\end{deluxetable*}

\begin{figure*}
\plotone{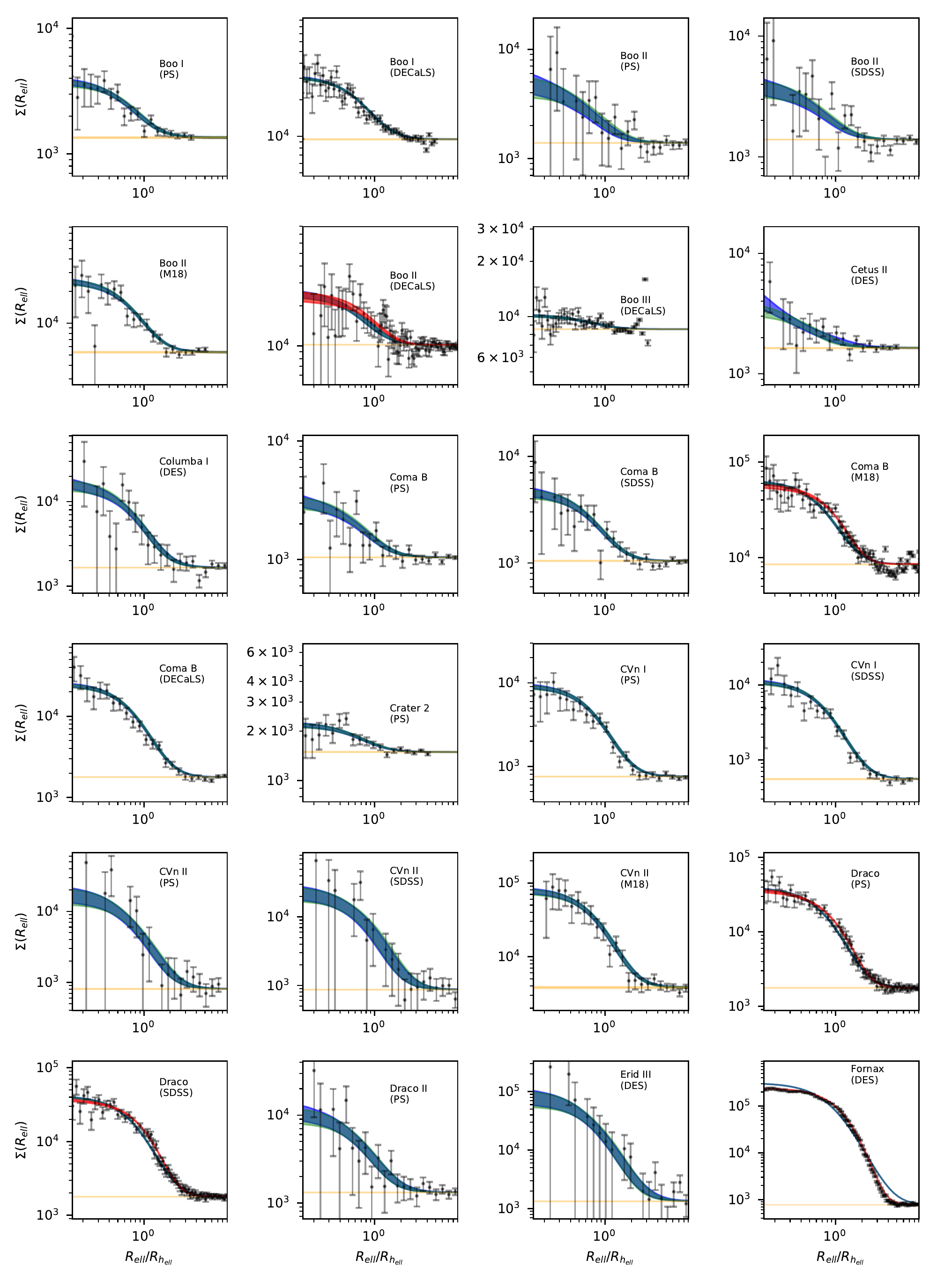}
\caption{\label{fig:all_fits_1} Projected stellar density profiles of dSphs, as a function of the elliptical radii (normalized by $R_{h,ell}$, the semi-major axis of the ellipse enclosing half the stars in our 3-Plummer model). Data points represent empirical profiles estimated by counting stars within elliptical annuli. The blue band represents the 68\% posterior credible interval for the 3-Plummer fit to unbinned data, and the green represents the same for the 1-Plummer fit. The orange band is the 68\% interval for the projected density of contaminating background stars (from the 3-Plummer model). In cases where the 3-Steeper model is favored with $B\geq 0.5$ over the 3-Plummer model, the red band represents the 68\% posterior credible interval for the 3-Steeper fit. }
\end{figure*}

\begin{figure*}
\plotone{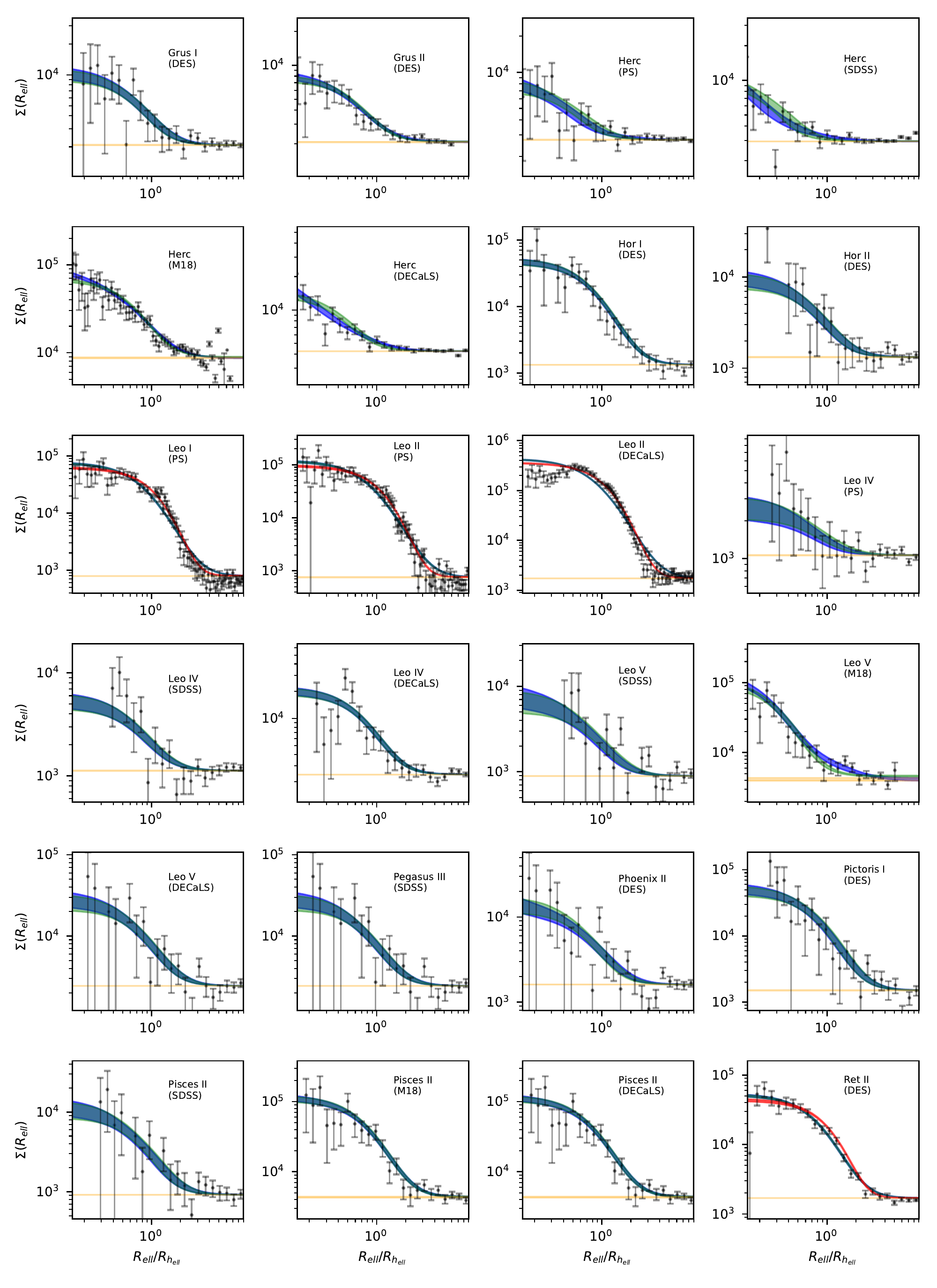}
\caption{\label{fig:all_fits_2} Projected stellar density profiles of dSphs, continued. Colors and points represent the same quantities as in Figure \ref{fig:all_fits_1}.}
\end{figure*}

\begin{figure*}
\plotone{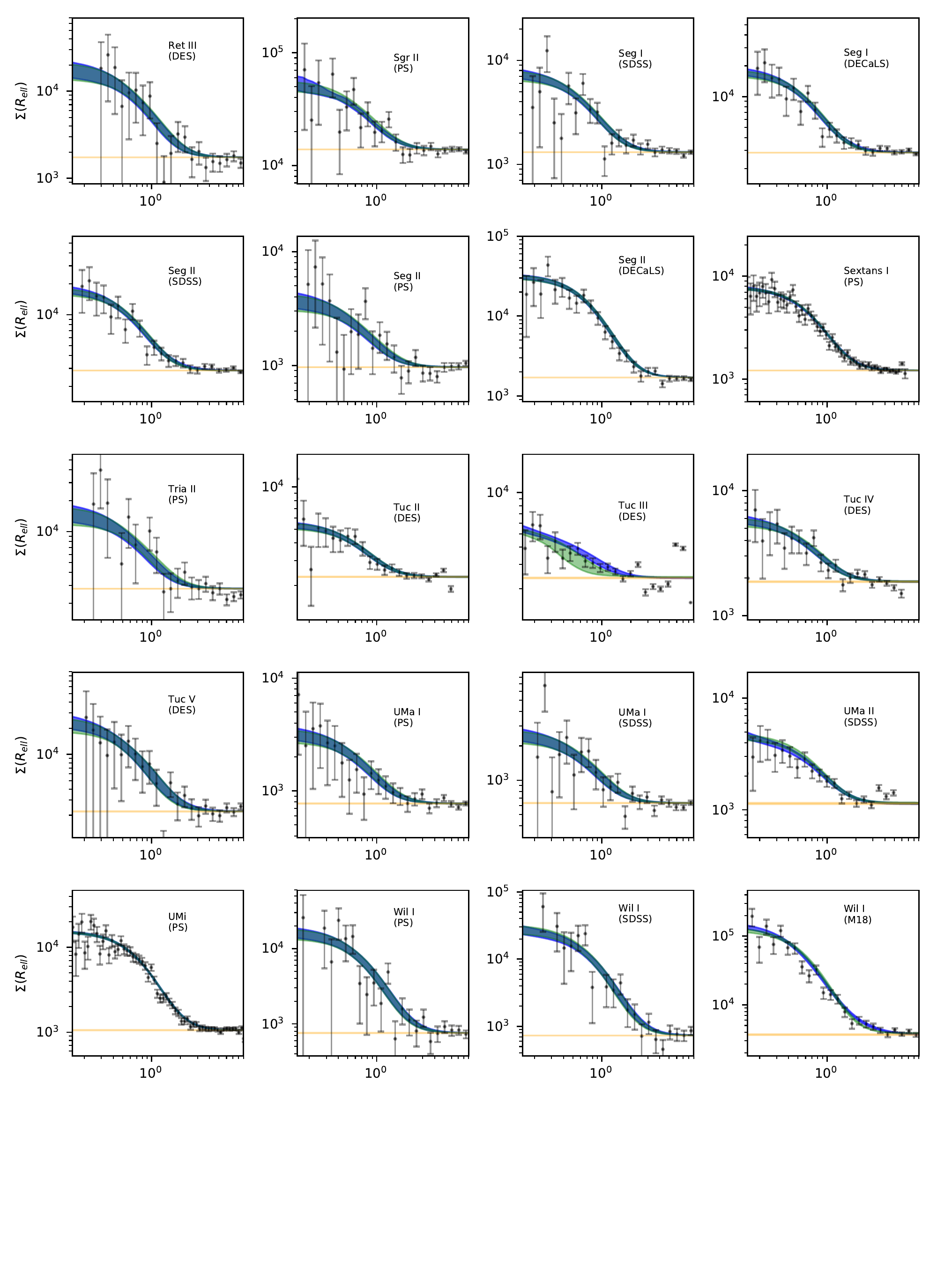}
\caption{\label{fig:all_fits_3} Projected stellar density profiles of dSphs, continued. Colors and points represent the same quantities as in Figure \ref{fig:all_fits_1}.}
\end{figure*}

Figures \ref{fig:all_fits_1}-\ref{fig:all_fits_3} illustrate projected density profiles that we infer for each dSph. Data points indicate empirical profiles derived by counting stars in elliptical annular bins. Colored bands represent 68\% credible intervals derived from posterior probability distributions for our 1-Plummer (green) and 3-Plummer (blue) fits. In cases where the evidence ratio indicates that the 3-Steeper model is preferable to the 3-Plummer model ($B>0.5$), the 68\% credible interval for the 3-Steeper model is shown in red. We must emphasize that the binned profiles are only an estimation of the stellar density in the nearby region, and the binned values can be sensitive to bin location, bin width, and artifacts such as gaps due to bright stars (which are accounted for in the fits). This is one of the reasons that we fit to discrete data consisting of positions of individual stars, and not directly to binned profiles. Nevertheless, the panels show that our inferences generally agree well with the binned profiles.

\begin{figure}
\plotone{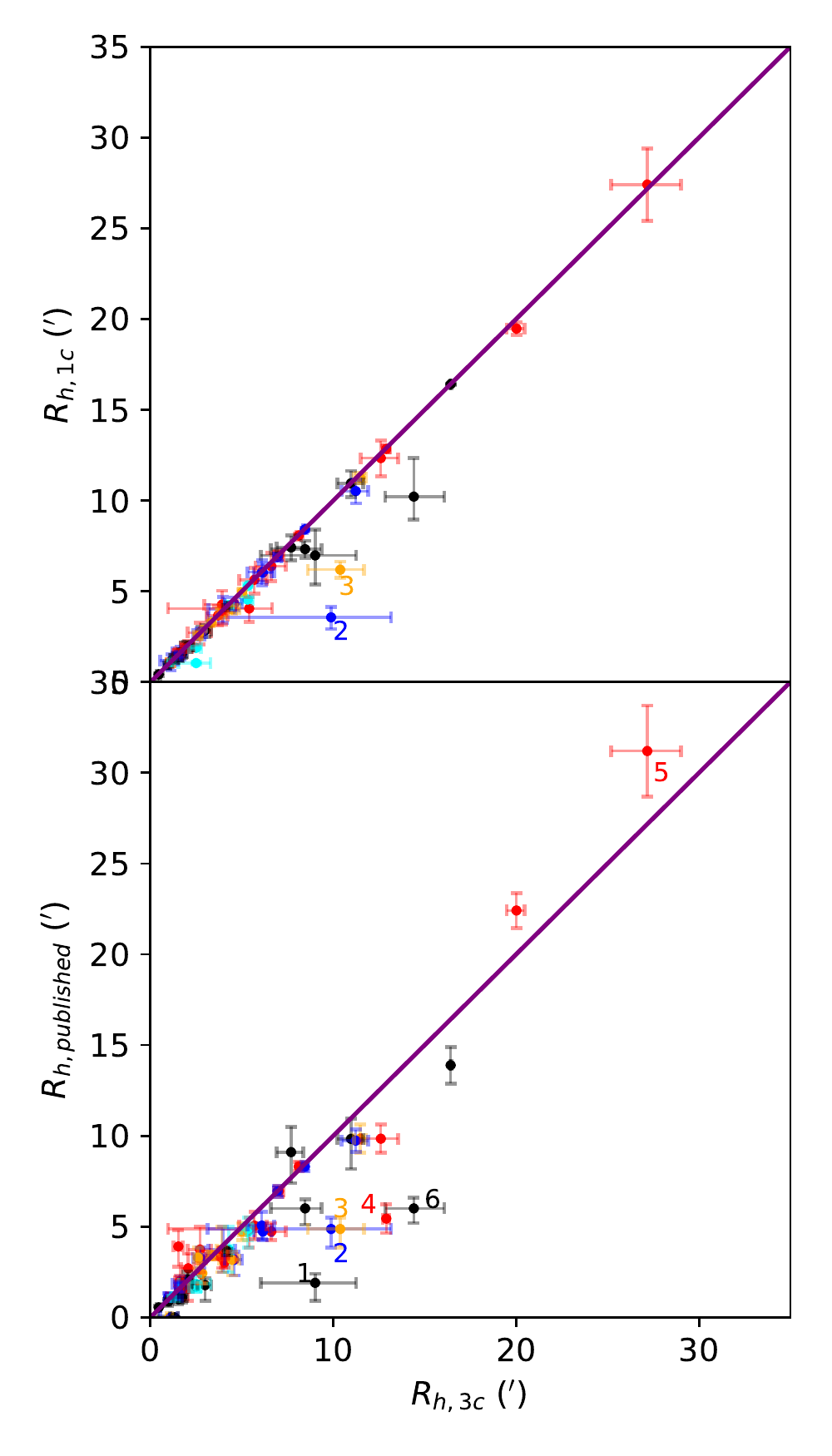}
\caption{\label{fig:rhalf} Comparison of circularized projected halflight radii amongst our 1-component, 3-component, and previously published measurements (from the references listed in Table 2).  Blue/red/black/cyan/orange markers represent measurements using data from the SDSS/PS1/DES/M18/DECaLS surveys. In cases where the 3-Steeper model is favored with $B\geq 0.5$ over the 3-Plummer model, the plotted halflight radii are those obtained from the 3-Steeper model. 1: Cetus II (DES); 2: Herc (SDSS); 3: Herc (DECaLS); 4: UMi (PS); 5: Crater 2 (PS); 6: Tuc III (DES).}
\end{figure}

Figure \ref{fig:rhalf} compares our estimates of circularized projected halflight radii, $R_{h,3c}\equiv R_{h,ell}\sqrt{1-\epsilon}$, obtained from our 3-component fits to those obtained from our 1-component fits (top panel), and to the circularized halflight radii published in the references listed in Table \ref{galaxiesall} (bottom panel). In cases where the 3-Steeper model is favored with $B> 0.5$ over the 3-Plummer model, the plotted halflight radii are those obtained from the Steeper model; otherwise the Plummer values are plotted.  For the most part, agreement is good to within statistical errors.  However, we do notice that in some cases our 3-component fits yield systematically larger halflight radii (albeit within 1-2 statistical errorbars) than our 1-component fits.  For Cetus II, Ursa Minor and Tucana III,  our fits yield halflight radii significantly larger than the  previously-published values.  For the case of Ursa Minor, the plotted previously-published value of $8.2\pm 1.2$ arcmin is the one listed by \citet{mcc}, which originally comes from the exponential profile fit by \citet{irwin_1995}.  Our estimate derived from PS1 data is $\sim 50\%$ larger than the \citet{irwin_1995} value, but agrees well with M18's estimate of $\sim 12.3$ arcmin for a 1-Plummer model.

\begin{figure}
\plotone{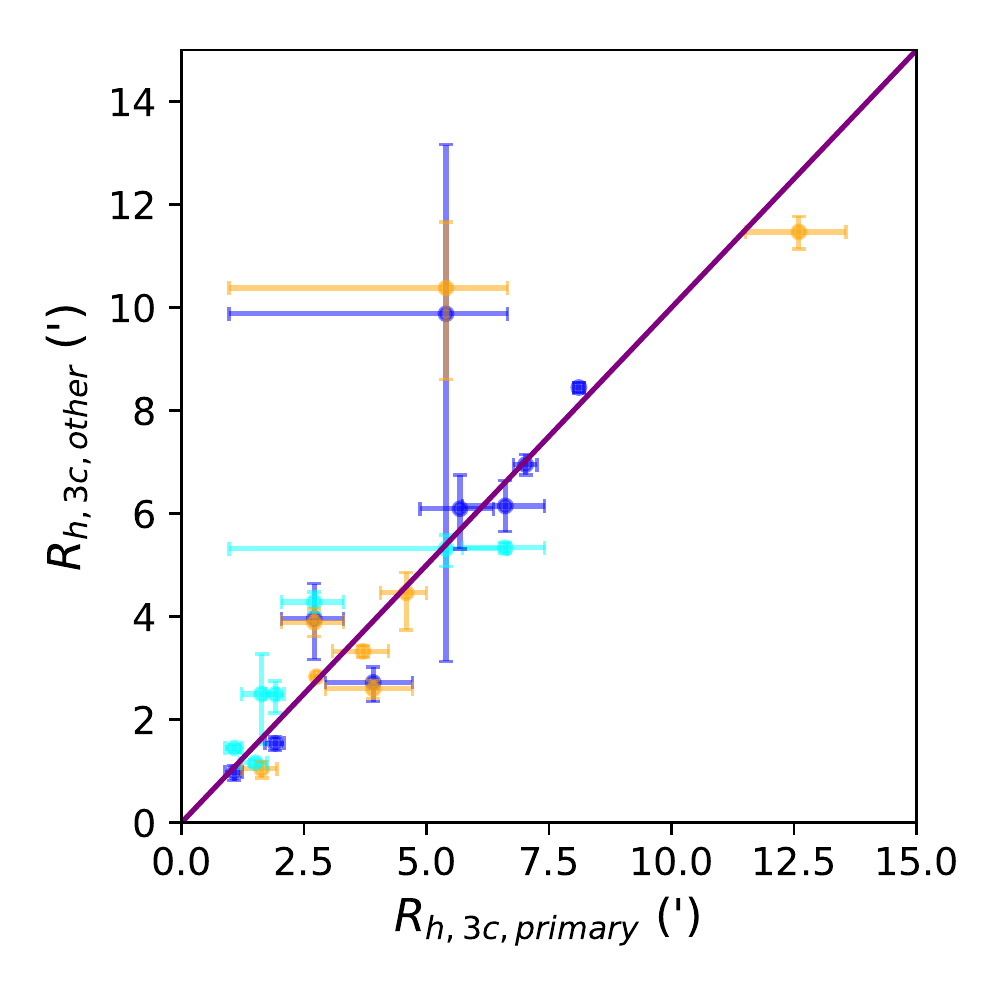}
\caption{\label{fig:rhmult} Comparison of half-light radii for galaxies found in multiple surveys. Half-light radii for the `primary' survey (PS1, or SDSS if the galaxy does not fall into the PS1 footprint) are plotted on the horizontal axis. The half-light radii for the same galaxy in other surveys is plotted on the vertical axis. Blue/cyan/orange points represent `other' surveys of SDSS/M18/DECaLS, respectively. The purple line represents a 1:1 ratio. Error bars indicate the $1\sigma$ width of the posterior.}
\end{figure}

Because some galaxies have data in multiple surveys, we can compare measurements taken in one survey to others. This is done for inferred half-light radii in Figure \ref{fig:rhmult}. We find that almost every galaxy falls along the 1:1 relation, indicating that measurements are generally stable between surveys. The two conspicuous points that lie far away from the 1:1 line both belong to Hercules; their values for the half-light radius contain large uncertainties, so the deviation between surveys is not significant.

\begin{figure*}
\plottwo{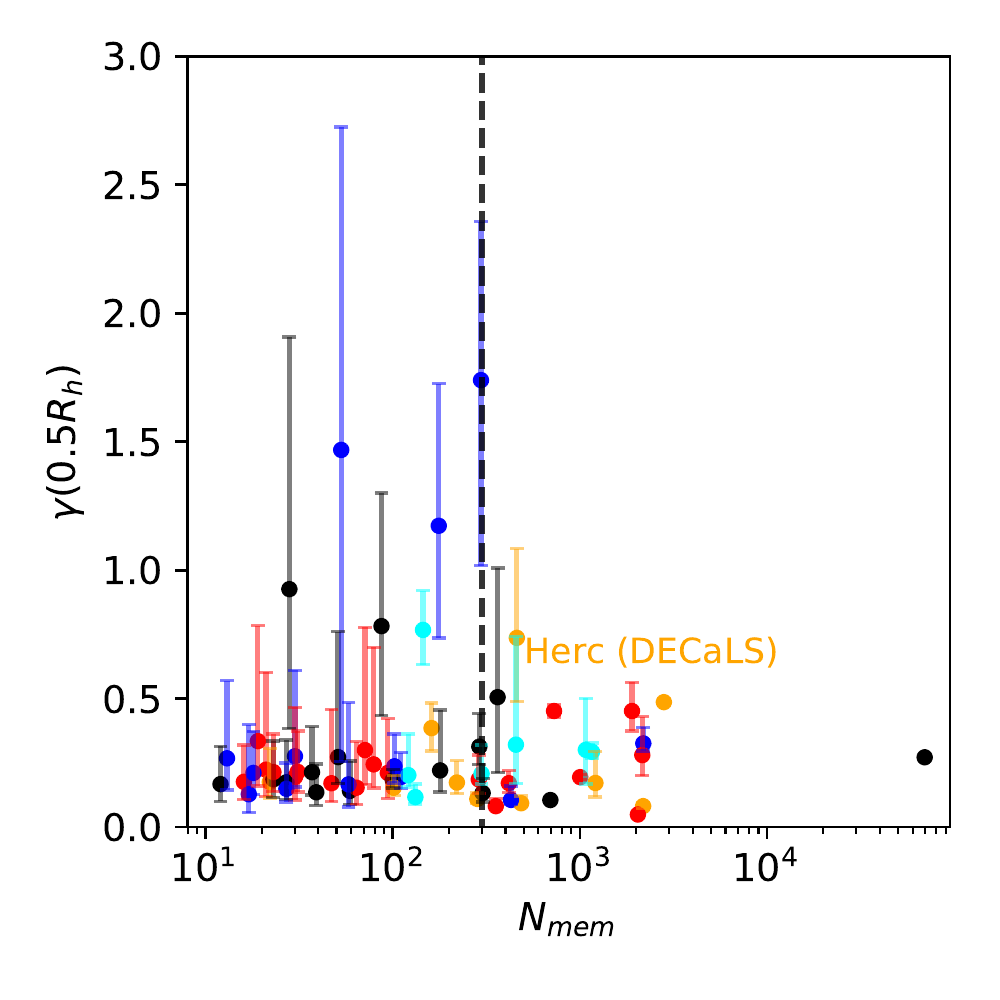}{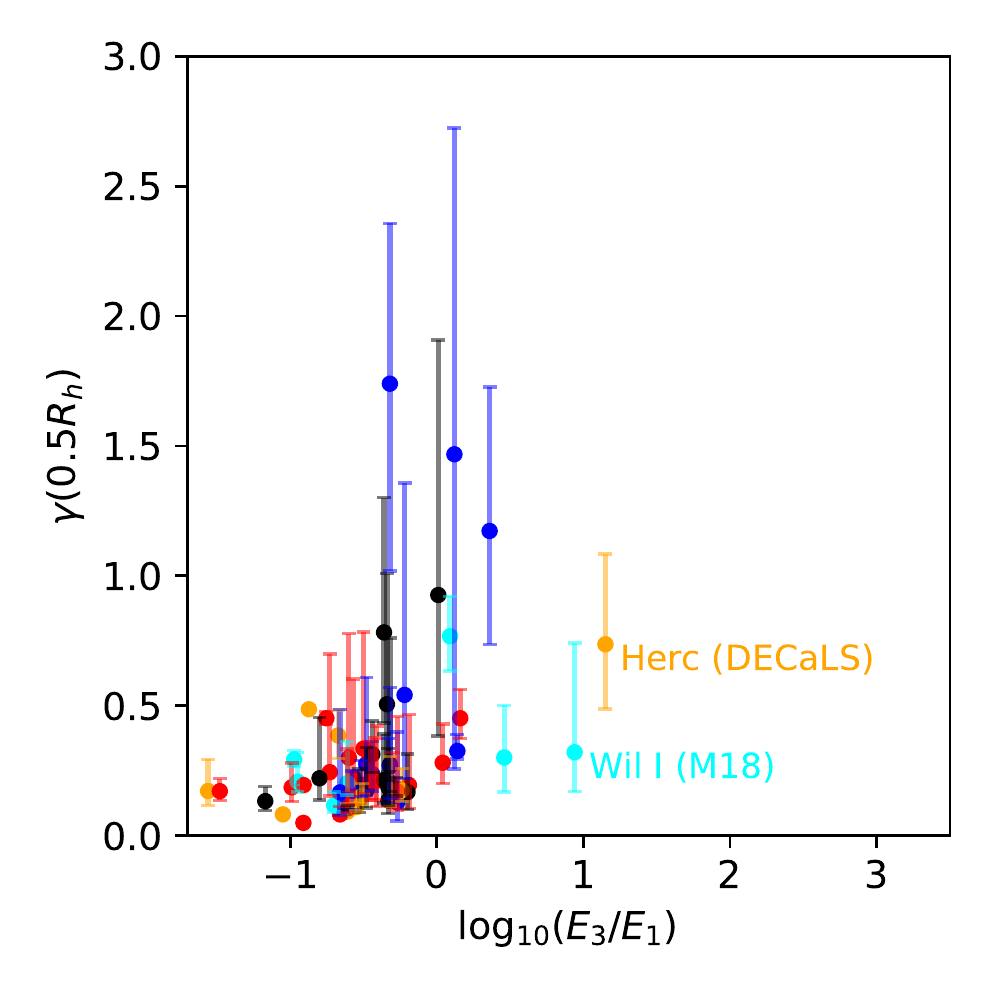}
\caption{\label{fig:ev_vs_gamma} Logarithmic slope parameter, $\gamma(0.5R_h)$ obtained from fits to real dSphs, versus inferred number of member stars (left) and Bayes factor $B\equiv E_3/E_1$ that informs comparison between 3-component and 1-component models (right). Blue/red/black/cyan/orange points represent fits to data sets from SDSS/PS1/DES/M18/DECaLS, respectively. We show results for the Plummer model unless the 3-Steeper model is favored over the 3-Plummer model with $B> 0.5$, in which cases we show results for the Steeper model. The vertical dashed line in the left hand panel marks our cutoff value $N_{mem}=300$.  }
\end{figure*}

The left-hand panel of Figure \ref{fig:ev_vs_gamma} plots estimates of the logarithmic slope parameter, $\gamma(0.5 R_h)$, for 3-component fits versus the inferred number of member stars.  The plotted results are those obtained from fits that assume Plummer (resp: Steeper)  profiles if the evidence ratio that compares Steeper to Plummer profiles is $\leq 10^{1/2})$ (resp: $ >10^{1/2}$).   Comparing to the corresponding results obtained for mock data sets (lower-right panels of Figures \ref{fig:mock_gamma_vs_nmem} and \ref{fig:mock_gamma_vs_bayes}, we find that the regions where we successfully detected cuspy stellar density profiles in mock systems are sparsely populated by the real dSphs ($\gamma(0.5 R_h)\ga 0.5$, $N_{\rm mem}\ga 300$, $B_{3c,1c}\ga 0.5$). Perhaps the main reason is that there are relatively few real dSphs for which we detect $N_{\rm mem}\ga 300$ members brighter than the surveys' 95\%-completeness  magnitude limits. Thus it is only for these dSphs that our modeling is sensitive to the slope of the inner stellar density profile. Our estimates of $\gamma(0.5R_h)$ and $B_{3c,1c}$ remain compatible with cored profiles and provide no significant evidence in favor of steep central stellar cusps. We emphasize that, for the vast majority of systems, where we detect fewer than $\sim 300$ member stars, our fits do not rule out the presence of cusps; we simply need deeper data in order to detect any stellar cusps that may exist in these systems.

 The right-hand panel of Figure \ref{fig:ev_vs_gamma} plots the evidence ratios, $B_{3c,1c}=\log_{10}[E_3/E_1]$, comparing 3-component and 1-component models (again, the plotted results are for Plummer models unless the 3-Steeper model is favored over the 3-Plummer model with an evidence ratio exceeding $10^{1/2}$). For the most part, the additional flexibility of the 3-component models does not overcome the larger prior space compared to the 1-component models, resulting in negative Bayes factors. 

\begin{figure*}
\plotone{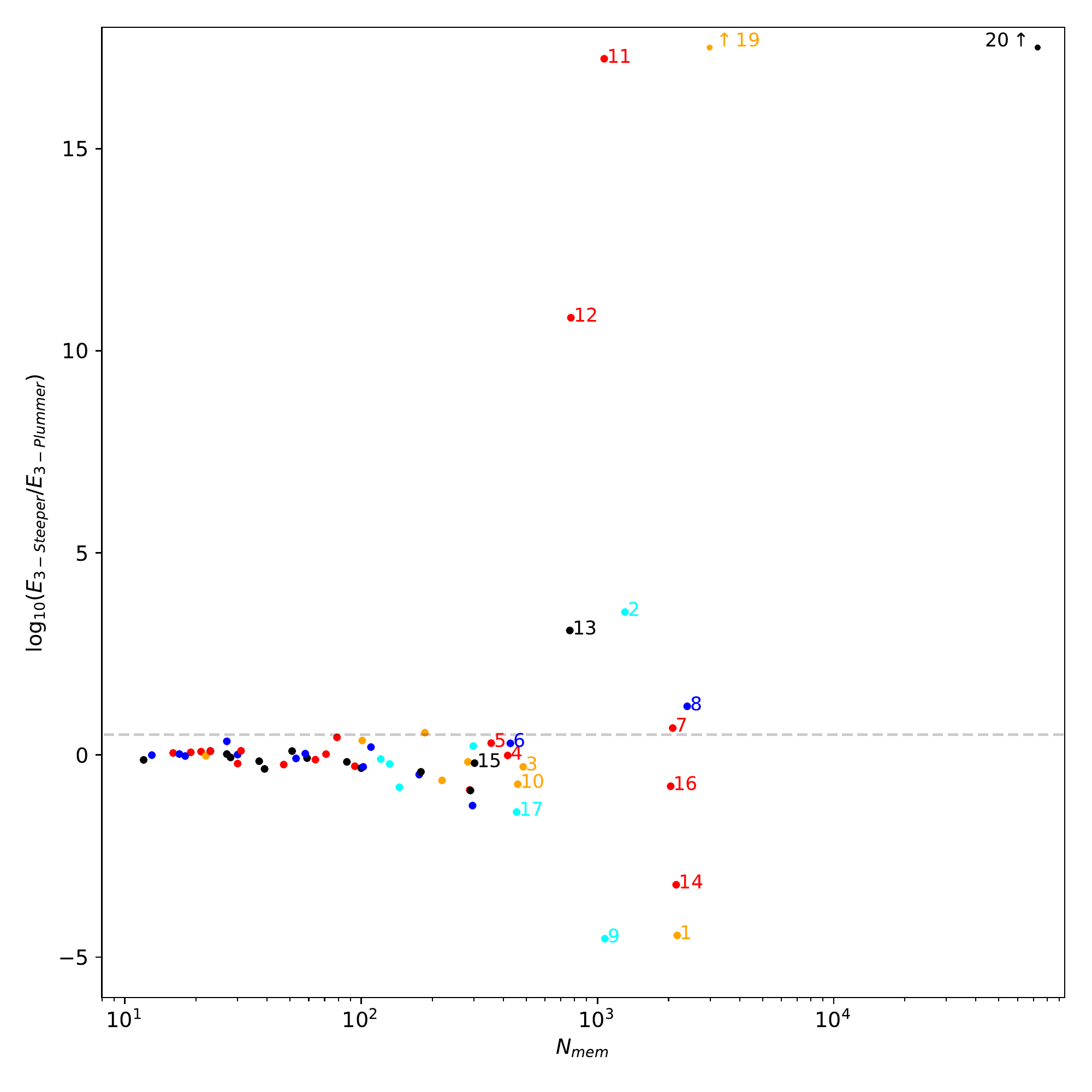}
\caption{\label{fig:BF_e3_s3} Bayes factor $B=\log_{10}(\frac{E_{3-Steeper}}{E_{3-Plummer}})$ that informs comparison between 3-Steeper and 3-Plummer models plotted against inferred number of member stars. Blue/red/black/cyan/orange points represent fits to data sets from SDSS/PS1/DES/M18/DECaLS, respectively. Fits with $N_{mem}>300$ are numbered as follows: 1: Boo I (DECaLS); 2: Coma B (M18); 3: Coma B (DECaLS); 4: Crater 2 (PS); 5: CVn I (PS); 6: CVn I (SDSS); 7: Draco (PS); 8: Draco (SDSS); 9: Herc (M18); 10: Herc (DECaLS); 11: Leo I (PS); 12: Leo II (PS); 13: Ret II (DES); 14: Sextans I (PS); 15: Tuc II (DES); 16: UMi (PS); 17: Wil I (M18); 19: Leo II (DECaLS); 20: Fornax (DES). The arrows next to Leo II in DECaLS ($B=50.84$) and Fornax in DES ($B=941.47$) indicate that those galaxies lie too high off the y-axis to show without distorting the scale of other points.}
\end{figure*}

    Figure \ref{fig:BF_e3_s3} plots the Bayes factor comparing the evidence of 3-Steeper fits to that of 3-Plummer fits, as a function of the number of member stars detected in the galaxy. Similarly to Figure \ref{fig:steeper_vs_e3_evidence}, the corresponding plot for mock data, there is no strong model selection for galaxies with less than 300 member stars. For dSphs above this cut, there are eight data sets with $B>0.5$: Draco (PS), Draco (SDSS), Coma Berenices (M18), Leo I (PS1), Leo II (PS1 \& DECaLS), Reticulum II (DES), and Fornax (DES). For Fornax, visual inspection of Figure \ref{fig:all_fits_1} demonstrates how the binned data dramatically undershoots the 3-Plummer profile, but matches the 3-Steeper (red) profile. Although less dramatic, it is still clear from Figure \ref{fig:all_fits_2} that the 3-Plummer profile exceeds the data in the cases of Leo I and Leo II. The difference between the 3-Plummer and 3-Steeper profile for Draco and Reticulum II is less pronounced, and they have accordingly lower Bayes factors. For Coma Berenices, the steeper profile is only favored in the M18 data; this may be due to the increased number of member stars detected in the M18 data, or to inhomogeneity across the M18 field for this galaxy. 

\subsection{Discussion}
\label{subsec:discussion}
We now discuss some specific results of our fits to the real dSph data sets.  We begin by considering a few anomalous and/or unexpected results, and then discuss detections of non-standard stellar density profiles.

First, the halflight radii that we estimate for two galaxies---Cetus II and Tucana III---differ significantly from those of previous studies.  For Cetus II, the halflight radius of $R_{h}=9.01^{+2.98}_{-2.24}$ arcminutes that we measure (from the 3-Plummer model) is several times larger than the one originally reported by \citet{desy2-dsphs}. Our measurement possibly sheds light on the recent result by \citet{cetus_ii_paper}, who do not detect any overdensity around the location of Cetus II. This might be because they look for an overdensity within the original, smaller half light radius. Indeed, their entire field of view ($5.5'$ across) is smaller than the size of the half-light radius we find for Cetus II. As opposed to the value of $[Fe/H]=-1.9$ that we adopt from the previous literature, \citet{cetus_ii_paper} also find that the stars in Cetus II appear to follow an isochrone of $[Fe/H]=-1.28$. In order to make sure that this is not the cause of the discrepancy between their results and ours, we repeated the isochrone cut and 3-Plummer fit using an isochrone with $[Fe/H]=-1.28$. We found that the half-light radius for this fit was nearly identical to the fit using the original isochrone. We tentatively conclude that despite the non-detection in \citet{cetus_ii_paper}, Cetus II may still exist as a bound halo object with a larger size than reported by \citet{desy2-dsphs}; deeper wide-field (or wider deep-field) photometry and/or follow-up spectroscopy is required in order to settle this issue regarding the nature of Cetus II. 

We also find a much larger half-light radius for Tucana III than originally reported by \citet{desy2-dsphs}. Unlike for the other dSphs, our estimate is far larger for the Plummer profiles ($R_{h}=14.40^{+1.57}_{-1.67}$ and $10.22^{+1.27}_{-2.12}$ arcminutes for the 3-Plummer and 1-Plummer models, respectively) than for the Steeper models ($R_{h}=5.08^{+0.56}_{-0.41}$ arcminutes for the 3-Steeper model and $10.11^{+0.98}_{-1.77}$ arcminutes for the 1-Steeper model). This larger, model-dependent half light radius that we obtain for Tuc III is almost certainly due to contamination by stars from a tidal stream emanating from Tucana III \citep{li2019}, which was noticed by \citet{desy2-dsphs} and deliberately excluded from their analysis that provided the previously-published halflight radius. We measure a high ellipticity ($0.81^{+.02}_{-.03}$) for the 3-Plummer model and our 3-Plummer position angle ($83.87^{+1.20}_{-1.21}$ degrees East of North) matches the orientation of the stream \citep{tuciiistream}, further suggesting that our fit is contaminated by stream stars.  Such contamination is further reflected in the stellar density profiles shown in Figure \ref{fig:all_fits_3}, where the density around Tuc III rises again after falling to the fitted background level.  As such, none of the models we have fitted here can be considered to provide an adequate description of Tuc III.  

\subsubsection{Beyond the Standard Plummer profile}
The slope parameter $\gamma(0.5 R_h)$ and the Bayes factors comparing 1-component versus 3-component and Plummer versus Steeper models provide quantitative criteria with which to identify galaxies for which the standard 1-Plummer model is disfavored.  Our mock data sets indicate that we can reliably identify centrally cusped and/or steepened outer profiles when the number of detected member stars exceeds a few hundred.  There are fifteen galaxies for which the available survey data meet or exceed this requirement: Boo I, Boo III, Coma B, Cra 2, CVn I, Dra, Herc, Leo I, Leo II, Sex I, UMi, For, Ret II, Tuc II, and Wil I.  

None of these galaxies present compelling evidence for cuspy central stellar density profiles. While Hercules has a high value of $\gamma(0.5R_h)=0.74^{+0.35}_{-0.25}$ in the DECaLS data with the 3-Plummer model and a Bayes factor $B_{3c,1c}>0.5$, the deeper M18 data fails to replicate this, with $\gamma(0.5R_h)=0.30^{+0.20}_{-0.13}$.  

Of the galaxies with more than 300 detected member stars, four clearly favor the Steeper model over the Plummer model.  For each of Leo I, Leo II, and Fornax, the Bayes factor comparing Steeper to Plummer is $B>10$ for both 1-component and 3-component models.  For Fornax, the comparison is dramatic, favoring the Steeper model with $B>900$.  This result is compatible with the relatively poor fits of Plummer and exponential models to Fornax star-count data by \citet{irwin_1995} and \citet{MYfornax}.  Indeed, \citet{MYfornax} considered a variety of different analytic models---Plummer, exponential, Sersic and King---for Fornax and found that none could fit Fornax (DES) data with $\chi^2<1.4$ per degree of freedom.  Additionally, Reticulum II favors the steeper model with $B>2.9$ for both 3-component and 1-component models.  
Another two galaxies, Draco and Coma Berenices, present ambiguous results. For Draco, the SDSS data favor the Steeper model, with $B>1.2$ for the 3-component models. However, the PS1 data for Draco favor the 3-steeper model at only $B=0.67$ over the 3-Plummer model, and the 1-Steeper model is  disfavored ($B=-0.34$) compared to the 1-Plummer model. Neither 3-component model is strongly preferred ($B\leq 0.3$) over its corresponding 1-component model.  For Coma Berenices, the Steeper model is favored at $B\ga 3.5$ only for the M18 data; for SDSS and PS1 data, the Steeper model is disfavored, with Bayes factor no higher than $B\sim 0.2$. Five galaxies---Crater 2, Hercules, Ursa Minor, Willman I, and Sextans I---all favor the Plummer model with $B<-0.5$ in the 3-component case.

With data from the references in Table \ref{galaxiesall}, we examined these galaxies' luminosities, metallicities, surface brightnesses, sizes, and mass-to-light ratios for correlations with the outer profile steepness that we infer. Since the results of \citet{tidessim} indicate that tidal effects could alter the shape of a dSph's stellar profile, we also compared the previously mentioned properties to perihelion distance and current distance from the Milky Way center \citep{peri}. We found no significant correlation of any of these properties with the outer stellar density profile.  We also considered whether outer profile steepness might be related to membership within the ``Vast Polar Structure'' of dwarf galaxies surrounding the Milky Way \citep[][]{vpos2009,vpos2012,vposnew}. While the galaxies with $B>0.5$ lie close to this structure (3,6,17,26 and 45 kpc away for Coma Berenices, Reticulum II, Fornax, Leo II, and Leo I, respectively), there is no difference in the distance for those galaxies with $B<-0.5$ (2, 33, and 71 kpc for Sextans, Ursa Minor, and Hercules; we could not find a published distance for Willman I). If outer steepness is correlated with some property of dSphs, discerning this will require deeper imaging to bring more dSphs above the $N_{mem}>300$ minimum.

\section{SUMMARY}
\label{sec:conclusion}
In this paper, we have presented a flexible model consisting of a combination of three generalized Plummer profiles to the stellar distribution of dSphs. We demonstrate with mock data sets that our model is capable of fitting and also differentiating both cored and cusped profiles. We fit 1-Component and 3-Component Plummer and `Steeper' profiles to 40 dSphs using catalogs from the DES, SDSS, PS1, DECaLS, and M18 surveys. Summary statistics are listed in Tables \ref{ellresults} and \ref{superplummertable}. We also make available the maximum-likelihood posterior distributions themselves to allow other researchers to accurately quantify the profile uncertainties, rather than relying on summary statistics.\\
Given the importance of the cusp / core problem to the study of dSphs, we investigated whether any of the dSphs showed evidence for cusps in their stellar profiles. This is significant, as the stellar profile is an input into the kinematic equations that estimate mass density profiles. Using the value of the logarithmic slope at $R=0.5 R_{h}$, we found that we our 3-Plummer model can discriminate cusps from cores in dSphs that have roughly 300 or more member stars. We find no definitive evidence for stellar cusps in any of the fifteen dSphs that exceed this threshold, although most of the ultrafaint galaxies in our sample fall below this threshold 300. We also apply a Bayes factor comparison between the 3-Steeper and 3-Plummer fits, and find that the 3-Steeper model is favored with $B\geq0.5$ in four galaxies (Leo I, Leo II, Reticulum II, and Fornax).

\section{ACKNOWLEDGEMENTS}
We would like to thank Evan Tucker, Sergey Koposov, Andrew Pace, and the anonymous referee for providing helpful comments. M.G.W. acknowledges funding from National Science Foundation grants AST-1813881 and AST-1909584.  This paper makes use of the Whole Sky Database (wsdb) created by Sergey Koposov and maintained at Carnegie Mellon University by Sergey Koposov and the Institute of Astronomy, Cambridge by Sergey Koposov, Vasily Belokurov and Wyn Evans with financial support from the Science \& Technology Facilities Council (STFC), the European Research Council (ERC). This paper also made use of the Python package $dustmaps$ \citep{dustmaps}.

\begin{small}
\begin{footnotesize}

\bibliography{draft}

\end{footnotesize}
\end{small}

\end{document}

%% file: table_3.txt
BooI$_{PS}$ & $ 12.60^{+1.09}_{-0.96} $ & $ 288^{+26}_{-24} $ & $ 12.34^{+1.00}_{-0.96} $ & $ 285^{+26}_{-25} $ & $ 0.28^{+0.07}_{-0.07} $ & $ 25.53^{+9.61}_{-7.17} $ & $ 0.28^{+0.07}_{-0.08} $ & $ 26.03^{+10.31}_{-7.15} $ & -0.99 & $ 0.19^{+0.09}_{-0.06} $\\[2 mm]
BooI$_{DCLS}$ & $ 11.47^{+0.33}_{-0.30} $ & $ 2176^{+62}_{-58} $ & $ 11.26^{+0.29}_{-0.26} $ & $ 2150^{+59}_{-59} $ & $ 0.42^{+0.02}_{-0.02} $ & $ 6.36^{+1.70}_{-1.66} $ & $ 0.42^{+0.02}_{-0.02} $ & $ 6.30^{+1.81}_{-1.72} $ & -1.05 & $ 0.08^{+0.01}_{-.01} $\\[2 mm]
BooII$_{PS}$ & $ 2.71^{+0.67}_{-0.60} $ & $ 19^{+6}_{-5} $ & $ 2.73^{+0.66}_{-0.55} $ & $ 20^{+5}_{-5} $ & $ 0.32^{+0.12}_{-0.11} $ & $ 279.54^{+22.02}_{-209.05} $ & $ 0.32^{+0.14}_{-0.12} $ & $ 275.83^{+20.49}_{-204.36} $ & -0.50 & $ 0.33^{+0.45}_{-0.17} $\\[2 mm]
BooII$_{SDSS}$ & $ 3.96^{+0.79}_{-0.68} $ & $ 30^{+7}_{-7} $ & $ 4.02^{+0.76}_{-0.67} $ & $ 31^{+8}_{-6} $ & $ 0.26^{+0.11}_{-0.09} $ & $ 277.32^{+27.02}_{-219.09} $ & $ 0.27^{+0.11}_{-0.10} $ & $ 87.05^{+213.11}_{-30.86} $ & -0.48 & $ 0.28^{+0.33}_{-0.12} $\\[2 mm]
BooII$_{M18}$ & $ 4.28^{+0.21}_{-0.19} $ & $ 298^{+15}_{-15} $ & $ 4.23^{+0.21}_{-0.20} $ & $ 297^{+16}_{-16} $ & $ 0.11^{+0.05}_{-0.04} $ & $ 89.53^{+197.04}_{-19.95} $ & $ 0.11^{+0.05}_{-0.04} $ & $ 87.98^{+197.04}_{-18.04} $ & -0.95 & $ 0.21^{+0.11}_{-0.04} $\\[2 mm]
BooII$_{DCLS}$ & $ 3.89^{+0.27}_{-0.25} $ & $ 186^{+16}_{-14} $ & $ 3.82^{+0.28}_{-0.25} $ & $ 184^{+17}_{-16} $ & $ 0.25^{+0.11}_{-0.09} $ & $ 279.21^{+8.22}_{-189.54} $ & $ 0.25^{+0.10}_{-0.10} $ & $ 281.02^{+8.76}_{-8.43} $ & -0.63 & $ 0.17^{+0.07}_{-0.05} $\\[2 mm]
BooIII$_{DCLS}$ & $ 32.45^{+2.15}_{-2.05} $ & $ 1210^{+108}_{-108} $ & $ 33.03^{+2.59}_{-2.40} $ & $ 1261^{+136}_{-122} $ & $ 0.34^{+0.07}_{-0.08} $ & $ 282.39^{+5.88}_{-5.83} $ & $ 0.33^{+0.08}_{-0.09} $ & $ 278.91^{+6.56}_{-8.51} $ & -1.56 & $ 0.17^{+0.12}_{-0.06} $\\[2 mm]
CetII$_{DES}$ & $ 9.01^{+2.98}_{-2.24} $ & $ 87^{+26}_{-22} $ & $ 6.98^{+1.61}_{-1.43} $ & $ 73^{+19}_{-17} $ & $ 0.50^{+0.13}_{-0.15} $ & $ 62.51^{+7.67}_{-7.18} $ & $ 0.40^{+0.14}_{-0.14} $ & $ 63.63^{+14.31}_{-10.70} $ & -0.36 & $ 0.78^{+0.52}_{-0.35} $\\[2 mm]
ColI$_{DES}$ & $ 2.09^{+0.21}_{-0.19} $ & $ 51^{+6}_{-5} $ & $ 2.01^{+0.19}_{-0.17} $ & $ 50^{+5}_{-5} $ & $ 0.22^{+0.09}_{-0.08} $ & $ 67.62^{+15.50}_{-12.57} $ & $ 0.21^{+0.09}_{-0.07} $ & $ 69.12^{+15.70}_{-13.05} $ & -0.32 & $ 0.27^{+0.49}_{-0.10} $\\[2 mm]
ComaB$_{PS}$ & $ 6.61^{+0.87}_{-0.80} $ & $ 71^{+10}_{-10} $ & $ 6.39^{+0.82}_{-0.74} $ & $ 70^{+10}_{-9} $ & $ 0.27^{+0.11}_{-0.10} $ & $ 281.41^{+13.31}_{-194.98} $ & $ 0.28^{+0.12}_{-0.10} $ & $ 280.33^{+12.48}_{-192.29} $ & -0.60 & $ 0.30^{+0.48}_{-0.13} $\\[2 mm]
ComaB$_{SDSS}$ & $ 6.15^{+0.50}_{-0.49} $ & $ 110^{+10}_{-10} $ & $ 6.03^{+0.53}_{-0.51} $ & $ 109^{+10}_{-10} $ & $ 0.17^{+0.07}_{-0.06} $ & $ 296.87^{+16.49}_{-24.84} $ & $ 0.17^{+0.08}_{-0.07} $ & $ 296.02^{+15.50}_{-22.84} $ & -0.52 & $ 0.19^{+0.10}_{-0.04} $\\[2 mm]
ComaB$_{M18}$ & $ 5.34^{+0.10}_{-0.10} $ & $ 1307^{+27}_{-27} $ & $ 5.32^{+0.10}_{-0.10} $ & $ 1305^{+29}_{-27} $ & $ 0.30^{+0.02}_{-0.02} $ & $ 281.95^{+2.26}_{-2.06} $ & $ 0.30^{+0.02}_{-0.02} $ & $ 281.99^{+2.28}_{-2.13} $ & -0.99 & $ 0.11^{+0.02}_{-0.01} $\\[2 mm]
ComaB$_{DCLS}$ & $ 5.01^{+0.16}_{-0.15} $ & $ 485^{+15}_{-15} $ & $ 4.94^{+0.16}_{-0.15} $ & $ 480^{+15}_{-15} $ & $ 0.39^{+0.03}_{-0.03} $ & $ 294.93^{+2.25}_{-2.25} $ & $ 0.39^{+0.03}_{-0.03} $ & $ 294.86^{+2.34}_{-2.32} $ & -0.61 & $ 0.09^{+0.03}_{-0.01} $\\[2 mm]
Cra2$_{PS}$ & $ 27.17^{+1.97}_{-1.83} $ & $ 417^{+38}_{-38} $ & $ 27.41^{+2.00}_{-2.00} $ & $ 426^{+44}_{-41} $ & $ 0.18^{+0.07}_{-0.06} $ & $ 70.46^{+204.98}_{-16.02} $ & $ 0.18^{+0.08}_{-0.07} $ & $ 70.25^{+200.74}_{-15.99} $ & -1.48 & $ 0.17^{+0.05}_{-0.04} $\\[2 mm]
CVnI$_{PS}$ & $ 7.02^{+0.25}_{-0.23} $ & $ 355^{+12}_{-13} $ & $ 6.97^{+0.24}_{-0.24} $ & $ 353^{+13}_{-13} $ & $ 0.44^{+0.03}_{-0.03} $ & $ 79.10^{+2.26}_{-2.21} $ & $ 0.44^{+0.03}_{-0.03} $ & $ 79.06^{+2.27}_{-2.25} $ & -0.66 & $ 0.08^{+0.03}_{-0.01} $\\[2 mm]
CVnI$_{SDSS}$ & $ 6.95^{+0.20}_{-0.20} $ & $ 428^{+12}_{-13} $ & $ 6.88^{+0.20}_{-0.20} $ & $ 426^{+12}_{-13} $ & $ 0.36^{+0.03}_{-0.03} $ & $ 72.92^{+2.50}_{-2.56} $ & $ 0.36^{+0.03}_{-0.03} $ & $ 73.16^{+2.66}_{-2.52} $ & -0.64 & $ 0.11^{+0.05}_{-0.02} $\\[2 mm]
CVnII$_{PS}$ & $ 1.08^{+0.20}_{-0.15} $ & $ 16^{+2}_{-3} $ & $ 1.05^{+0.19}_{-0.15} $ & $ 15^{+3}_{-2} $ & $ 0.35^{+0.13}_{-0.12} $ & $ 303.70^{+30.05}_{-277.44} $ & $ 0.36^{+0.13}_{-0.13} $ & $ 299.53^{+31.02}_{-271.87} $ & -0.30 & $ 0.18^{+0.14}_{-0.07} $\\[2 mm]
CVnII$_{SDSS}$ & $ 0.97^{+0.16}_{-0.13} $ & $ 17^{+3}_{-2} $ & $ 0.95^{+0.14}_{-0.12} $ & $ 17^{+2}_{-3} $ & $ 0.66^{+0.09}_{-0.11} $ & $ 9.23^{+10.05}_{-3.65} $ & $ 0.67^{+0.09}_{-0.12} $ & $ 9.12^{+10.47}_{-3.72} $ & -0.27 & $ 0.13^{+0.27}_{-0.07} $\\[2 mm]
CVnII$_{M18}$ & $ 1.45^{+0.08}_{-0.07} $ & $ 132^{+7}_{-6} $ & $ 1.44^{+0.08}_{-0.07} $ & $ 132^{+7}_{-7} $ & $ 0.36^{+0.05}_{-0.06} $ & $ 14.46^{+4.66}_{-3.99} $ & $ 0.37^{+0.05}_{-0.06} $ & $ 14.56^{+5.12}_{-4.11} $ & -0.70 & $ 0.12^{+0.05}_{-0.03} $\\[2 mm]
Dra$_{PS}$ & $ 8.11^{+0.11}_{-0.11} $ & $ 2083^{+28}_{-26} $ & $ 8.07^{+0.10}_{-0.10} $ & $ 2077^{+27}_{-27} $ & $ 0.30^{+0.01}_{-0.01} $ & $ 85.37^{+1.39}_{-1.38} $ & $ 0.30^{+0.01}_{-0.01} $ & $ 85.53^{+1.40}_{-1.44} $ & -0.84 & $ 0.10^{+.01}_{-.01} $\\[2 mm]
Dra$_{SDSS}$ & $ 8.45^{+0.10}_{-0.10} $ & $ 2398^{+28}_{-29} $ & $ 8.41^{+0.10}_{-0.10} $ & $ 2391^{+30}_{-30} $ & $ 0.34^{+0.01}_{-0.01} $ & $ 272.38^{+1.17}_{-1.18} $ & $ 0.34^{+0.01}_{-0.01} $ & $ 272.37^{+1.16}_{-1.15} $ & -0.80 & $ 0.09^{+.01}_{-.01} $\\[2 mm]
DraII$_{PS}$ & $ 2.06^{+0.39}_{-0.31} $ & $ 30^{+6}_{-4} $ & $ 1.97^{+0.32}_{-0.28} $ & $ 29^{+5}_{-4} $ & $ 0.50^{+0.10}_{-0.12} $ & $ 300.61^{+7.04}_{-8.08} $ & $ 0.49^{+0.10}_{-0.13} $ & $ 300.50^{+7.32}_{-8.69} $ & -0.19 & $ 0.19^{+0.27}_{-0.09} $\\[2 mm]
EriIII$_{DES}$ & $ 0.45^{+0.11}_{-0.08} $ & $ 12^{+2}_{-2} $ & $ 0.42^{+0.08}_{-0.07} $ & $ 11^{+3}_{-1} $ & $ 0.43^{+0.11}_{-0.13} $ & $ 284.45^{+11.22}_{-195.26} $ & $ 0.42^{+0.12}_{-0.13} $ & $ 284.55^{+11.82}_{-14.19} $ & -0.20 & $ 0.17^{+0.15}_{-0.07} $\\[2 mm]
For$_{DES}$ & $ 16.42^{+0.03}_{-0.03} $ & $ 72995^{+108}_{-106} $ & $ 16.41^{+0.03}_{-0.03} $ & $ 72990^{+110}_{-109} $ & $ 0.31^{+.010}_{-.010} $ & $ 40.73^{+0.17}_{-0.17} $ & $ 0.31^{+.010}_{-.010} $ & $ 40.74^{+0.18}_{-0.18} $ & -2.25 & $ 0.09^{+.010}_{-.010} $\\[2 mm]
GrusI$_{DES}$ & $ 2.99^{+0.41}_{-0.33} $ & $ 59^{+8}_{-7} $ & $ 2.84^{+0.35}_{-0.28} $ & $ 57^{+7}_{-7} $ & $ 0.46^{+0.08}_{-0.10} $ & $ 339.90^{+8.97}_{-23.05} $ & $ 0.45^{+0.09}_{-0.10} $ & $ 337.32^{+10.47}_{-33.69} $ & -0.53 & $ 0.14^{+0.12}_{-0.05} $\\[2 mm]
GrusII$_{DES}$ & $ 8.46^{+1.88}_{-0.89} $ & $ 290^{+41}_{-28} $ & $ 7.32^{+0.50}_{-0.47} $ & $ 258^{+20}_{-18} $ & $ 0.12^{+0.05}_{-0.04} $ & $ 59.58^{+225.13}_{-19.55} $ & $ 0.12^{+0.06}_{-0.05} $ & $ 67.05^{+228.74}_{-26.25} $ & -0.44 & $ 0.31^{+0.13}_{-0.07} $\\[2 mm]
Herc$_{PS}$ & $ 5.40^{+4.43}_{-1.26} $ & $ 94^{+41}_{-21} $ & $ 4.05^{+0.73}_{-0.61} $ & $ 73^{+13}_{-12} $ & $ 0.60^{+0.09}_{-0.10} $ & $ 276.37^{+4.66}_{-5.61} $ & $ 0.67^{+0.07}_{-0.08} $ & $ 275.49^{+4.44}_{-5.30} $ & -0.41 & $ 0.21^{+0.21}_{-0.10} $\\[2 mm]
Herc$_{SDSS}$ & $ 9.88^{+6.75}_{-3.28} $ & $ 176^{+65}_{-41} $ & $ 3.56^{+0.65}_{-0.58} $ & $ 88^{+15}_{-14} $ & $ 0.56^{+0.08}_{-0.10} $ & $ 279.53^{+4.30}_{-3.86} $ & $ 0.64^{+0.07}_{-0.09} $ & $ 281.22^{+3.97}_{-4.21} $ & 0.36 & $ 1.17^{+0.55}_{-0.44} $\\[2 mm]
Herc$_{M18}$ & $ 5.32^{+0.35}_{-0.26} $ & $ 1075^{+42}_{-39} $ & $ 4.55^{+0.13}_{-0.12} $ & $ 977^{+24}_{-26} $ & $ 0.62^{+0.01}_{-0.01} $ & $ 286.47^{+0.80}_{-0.84} $ & $ 0.62^{+0.01}_{-0.01} $ & $ 286.28^{+0.87}_{-0.85} $ & 0.46 & $ 0.30^{+0.20}_{-0.13} $\\[2 mm]
Herc$_{DCLS}$ & $ 10.38^{+1.78}_{-1.29} $ & $ 460^{+53}_{-44} $ & $ 6.19^{+0.44}_{-0.44} $ & $ 320^{+25}_{-23} $ & $ 0.55^{+0.05}_{-0.05} $ & $ 284.06^{+2.79}_{-2.60} $ & $ 0.58^{+0.05}_{-0.06} $ & $ 283.67^{+2.38}_{-2.21} $ & 1.15 & $ 0.74^{+0.35}_{-0.25} $\\[2 mm]
HorI$_{DES}$ & $ 1.60^{+0.11}_{-0.09} $ & $ 100^{+6}_{-6} $ & $ 1.55^{+0.08}_{-0.08} $ & $ 97^{+6}_{-6} $ & $ 0.13^{+0.05}_{-0.05} $ & $ 68.29^{+16.21}_{-15.18} $ & $ 0.14^{+0.06}_{-0.05} $ & $ 66.59^{+13.41}_{-13.65} $ & -0.28 & $ 0.18^{+0.04}_{-0.03} $\\[2 mm]
HorII$_{DES}$ & $ 2.02^{+0.34}_{-0.28} $ & $ 27^{+5}_{-4} $ & $ 1.97^{+0.30}_{-0.26} $ & $ 26^{+5}_{-4} $ & $ 0.41^{+0.13}_{-0.13} $ & $ 281.60^{+7.66}_{-8.10} $ & $ 0.40^{+0.14}_{-0.14} $ & $ 279.71^{+8.25}_{-189.75} $ & -0.48 & $ 0.17^{+0.16}_{-0.07} $\\[2 mm]
LeoI$_{PS}$ & $ 4.05^{+0.05}_{-0.05} $ & $ 1068^{+16}_{-15} $ & $ 4.04^{+0.05}_{-0.06} $ & $ 1067^{+16}_{-16} $ & $ 0.43^{+0.01}_{-0.01} $ & $ 84.31^{+0.95}_{-0.98} $ & $ 0.43^{+0.01}_{-0.01} $ & $ 84.27^{+1.01}_{-0.97} $ & -0.92 & $ 0.07^{+.01}_{-.010} $\\[2 mm]
LeoII$_{PS}$ & $ 2.75^{+0.04}_{-0.04} $ & $ 772^{+13}_{-14} $ & $ 2.74^{+0.04}_{-0.04} $ & $ 768^{+14}_{-13} $ & $ 0.11^{+0.02}_{-0.02} $ & $ 47.79^{+5.60}_{-5.89} $ & $ 0.11^{+0.02}_{-0.02} $ & $ 47.87^{+5.90}_{-5.98} $ & -0.79 & $ 0.16^{+0.01}_{-0.01} $\\[2 mm]
LeoII$_{DCLS}$ & $ 2.83^{+0.02}_{-0.02} $ & $ 2985^{+26}_{-25} $ & $ 2.82^{+0.02}_{-0.02} $ & $ 2979^{+24}_{-25} $ & $ 0.07^{+0.01}_{-0.01} $ & $ 41.18^{+4.61}_{-4.55} $ & $ 0.07^{+0.01}_{-0.01} $ & $ 41.23^{+4.60}_{-4.81} $ & -1.06 & $ 0.17^{+.01}_{-.010} $\\[2 mm]
LeoIV$_{PS}$ & $ 3.91^{+0.97}_{-0.80} $ & $ 21^{+6}_{-7} $ & $ 4.27^{+1.08}_{-0.76} $ & $ 24^{+6}_{-6} $ & $ 0.40^{+0.15}_{-0.13} $ & $ 81.80^{+235.82}_{-56.91} $ & $ 0.42^{+0.15}_{-0.16} $ & $ 80.67^{+228.51}_{-56.16} $ & -0.57 & $ 0.22^{+0.38}_{-0.10} $\\[2 mm]
LeoIV$_{SDSS}$ & $ 2.72^{+0.37}_{-0.30} $ & $ 27^{+4}_{-4} $ & $ 2.67^{+0.34}_{-0.28} $ & $ 26^{+4}_{-4} $ & $ 0.38^{+0.11}_{-0.11} $ & $ 326.49^{+16.72}_{-278.44} $ & $ 0.38^{+0.12}_{-0.12} $ & $ 323.33^{+18.58}_{-274.67} $ & -0.58 & $ 0.15^{+0.10}_{-0.05} $\\[2 mm]
LeoIV$_{DCLS}$ & $ 2.60^{+0.19}_{-0.16} $ & $ 101^{+8}_{-7} $ & $ 2.54^{+0.16}_{-0.15} $ & $ 99^{+8}_{-7} $ & $ 0.23^{+0.07}_{-0.07} $ & $ 328.20^{+14.70}_{-265.00} $ & $ 0.23^{+0.07}_{-0.08} $ & $ 324.46^{+17.51}_{-260.61} $ & -0.51 & $ 0.15^{+0.05}_{-0.03} $\\[2 mm]
LeoV$_{SDSS}$ & $ 1.63^{+0.41}_{-0.32} $ & $ 13^{+3}_{-2} $ & $ 1.64^{+0.38}_{-0.33} $ & $ 13^{+4}_{-2} $ & $ 0.32^{+0.14}_{-0.11} $ & $ 271.53^{+17.15}_{-203.28} $ & $ 0.33^{+0.13}_{-0.12} $ & $ 274.53^{+15.64}_{-203.28} $ & -0.32 & $ 0.27^{+0.30}_{-0.12} $\\[2 mm]
LeoV$_{M18}$ & $ 2.50^{+0.97}_{-0.78} $ & $ 145^{+22}_{-22} $ & $ 1.04^{+0.08}_{-0.07} $ & $ 92^{+6}_{-6} $ & $ 0.32^{+0.07}_{-0.07} $ & $ 89.64^{+186.52}_{-6.59} $ & $ 0.23^{+0.08}_{-0.08} $ & $ 277.15^{+9.16}_{-189.60} $ & 0.09 & $ 0.77^{+0.15}_{-0.13} $\\[2 mm]
LeoV$_{DCLS}$ & $ 1.04^{+0.17}_{-0.14} $ & $ 22^{+4}_{-3} $ & $ 1.02^{+0.16}_{-0.14} $ & $ 22^{+3}_{-4} $ & $ 0.33^{+0.11}_{-0.12} $ & $ 80.45^{+194.17}_{-11.69} $ & $ 0.32^{+0.12}_{-0.11} $ & $ 80.75^{+194.26}_{-12.49} $ & -0.32 & $ 0.18^{+0.13}_{-0.06} $\\[2 mm]
PegII$_{SDSS}$ & $ 1.12^{+0.60}_{-0.37} $ & $ 7^{+4}_{-3} $ & $ 1.17^{+0.54}_{-0.34} $ & $ 8^{+3}_{-3} $ & $ 0.42^{+0.15}_{-0.14} $ & $ 274.25^{+33.05}_{-236.14} $ & $ 0.43^{+0.16}_{-0.16} $ & $ 279.37^{+26.29}_{-236.15} $ & -0.22 & $ 0.54^{+0.82}_{-0.32} $\\[2 mm]
PhoeII$_{DES}$ & $ 1.74^{+0.28}_{-0.24} $ & $ 28^{+5}_{-4} $ & $ 1.58^{+0.26}_{-0.22} $ & $ 27^{+4}_{-5} $ & $ 0.31^{+0.12}_{-0.11} $ & $ 282.90^{+13.88}_{-11.15} $ & $ 0.27^{+0.11}_{-0.10} $ & $ 285.28^{+15.31}_{-14.54} $ & .010 & $ 0.93^{+0.98}_{-0.54} $\\[2 mm]
PictI$_{DES}$ & $ 0.98^{+0.11}_{-0.10} $ & $ 39^{+4}_{-4} $ & $ 0.96^{+0.10}_{-0.09} $ & $ 38^{+4}_{-4} $ & $ 0.43^{+0.07}_{-0.09} $ & $ 54.32^{+6.24}_{-6.16} $ & $ 0.44^{+0.07}_{-0.09} $ & $ 55.23^{+6.06}_{-6.33} $ & -0.33 & $ 0.14^{+0.11}_{-0.05} $\\[2 mm]
PiscII$_{SDSS}$ & $ 1.51^{+0.31}_{-0.24} $ & $ 18^{+4}_{-3} $ & $ 1.42^{+0.25}_{-0.20} $ & $ 18^{+3}_{-3} $ & $ 0.33^{+0.11}_{-0.11} $ & $ 79.13^{+199.52}_{-18.57} $ & $ 0.34^{+0.12}_{-0.12} $ & $ 81.14^{+197.36}_{-18.23} $ & -0.33 & $ 0.21^{+0.16}_{-0.08} $\\[2 mm]
PiscII$_{M18}$ & $ 1.16^{+0.07}_{-0.06} $ & $ 121^{+6}_{-6} $ & $ 1.13^{+0.06}_{-0.06} $ & $ 118^{+7}_{-6} $ & $ 0.18^{+0.06}_{-0.06} $ & $ 277.77^{+9.57}_{-191.77} $ & $ 0.18^{+0.06}_{-0.07} $ & $ 277.28^{+9.27}_{-191.25} $ & -0.61 & $ 0.20^{+0.16}_{-0.05} $\\[2 mm]
PiscII$_{DCLS}$ & $ 57.47^{+6.08}_{-6.59} $ & $ 742^{+137}_{-125} $ & $ 1.34^{+0.14}_{-0.13} $ & $ 37^{+4}_{-4} $ & $ 0.25^{+0.07}_{-0.07} $ & $ 55.45^{+10.07}_{-9.43} $ & $ 0.31^{+0.10}_{-0.10} $ & $ 78.08^{+195.17}_{-13.25} $ & 1.66 & $ 1.08^{+0.41}_{-0.30} $\\[2 mm]
RetII$_{DES}$ & $ 4.18^{+0.08}_{-0.08} $ & $ 764^{+15}_{-15} $ & $ 4.15^{+0.08}_{-0.08} $ & $ 759^{+16}_{-15} $ & $ 0.60^{+0.01}_{-0.01} $ & $ 69.51^{+0.80}_{-0.77} $ & $ 0.60^{+0.01}_{-0.01} $ & $ 69.51^{+0.80}_{-0.80} $ & -0.77 & $ 0.04^{+0.01}_{-.010} $\\[2 mm]
RetIII$_{DES}$ & $ 1.32^{+0.23}_{-0.20} $ & $ 23^{+3}_{-4} $ & $ 1.29^{+0.23}_{-0.18} $ & $ 23^{+3}_{-4} $ & $ 0.33^{+0.11}_{-0.11} $ & $ 44.26^{+233.02}_{-15.89} $ & $ 0.33^{+0.12}_{-0.11} $ & $ 49.96^{+242.22}_{-18.49} $ & -0.33 & $ 0.19^{+0.15}_{-0.07} $\\[2 mm]
SgrII$_{PS}$ & $ 1.60^{+0.26}_{-0.18} $ & $ 79^{+7}_{-4} $ & $ 1.63^{+0.15}_{-0.14} $ & $ 85^{+10}_{-8} $ & $ 0.27^{+0.05}_{-0.08} $ & $ 72.79^{+214.38}_{-17.46} $ & $ 0.24^{+0.10}_{-0.09} $ & $ 74.90^{+14.69}_{-13.02} $ & -0.73 & $ 0.24^{+0.46}_{-0.09} $\\[2 mm]
SegI$_{SDSS}$ & $ 4.58^{+0.53}_{-0.41} $ & $ 102^{+11}_{-9} $ & $ 4.32^{+0.39}_{-0.35} $ & $ 97^{+9}_{-8} $ & $ 0.19^{+0.07}_{-0.06} $ & $ 286.74^{+14.26}_{-16.39} $ & $ 0.20^{+0.08}_{-0.08} $ & $ 286.52^{+15.73}_{-14.75} $ & -0.44 & $ 0.24^{+0.13}_{-0.07} $\\[2 mm]
SegI$_{DCLS}$ & $ 4.46^{+0.72}_{-0.40} $ & $ 220^{+27}_{-19} $ & $ 4.00^{+0.25}_{-0.23} $ & $ 200^{+12}_{-13} $ & $ 0.33^{+0.05}_{-0.05} $ & $ 63.82^{+6.55}_{-6.45} $ & $ 0.33^{+0.05}_{-0.06} $ & $ 62.36^{+6.25}_{-6.52} $ & -0.24 & $ 0.17^{+0.09}_{-0.04} $\\[2 mm]
SegII$_{SDSS}$ & $ 64.78^{+1.44}_{-1.49} $ & $ 1194^{+79}_{-73} $ & $ 3.91^{+0.30}_{-0.28} $ & $ 88^{+8}_{-7} $ & $ 0.42^{+0.02}_{-0.02} $ & $ 7.52^{+6.56}_{-2.97} $ & $ 0.20^{+0.08}_{-0.07} $ & $ 316.53^{+18.11}_{-37.24} $ & 9.84 & $ 3.10^{+0.11}_{-0.13} $\\[2 mm]
SegII$_{PS}$ & $ 3.70^{+0.62}_{-0.52} $ & $ 31^{+6}_{-6} $ & $ 3.67^{+0.57}_{-0.51} $ & $ 31^{+6}_{-6} $ & $ 0.25^{+0.10}_{-0.09} $ & $ 285.46^{+24.51}_{-218.97} $ & $ 0.26^{+0.11}_{-0.10} $ & $ 283.94^{+24.08}_{-221.31} $ & -0.42 & $ 0.21^{+0.16}_{-0.08} $\\[2 mm]
SegII$_{DCLS}$ & $ 3.32^{+0.12}_{-0.12} $ & $ 283^{+11}_{-11} $ & $ 3.28^{+0.11}_{-0.11} $ & $ 279^{+11}_{-10} $ & $ 0.32^{+0.03}_{-0.04} $ & $ 342.69^{+3.62}_{-3.96} $ & $ 0.32^{+0.03}_{-0.04} $ & $ 342.59^{+3.84}_{-4.01} $ & -0.55 & $ 0.11^{+0.02}_{-0.02} $\\[2 mm]
SexI$_{PS}$ & $ 20.02^{+0.53}_{-0.43} $ & $ 2154^{+53}_{-45} $ & $ 19.48^{+0.35}_{-0.35} $ & $ 2111^{+40}_{-41} $ & $ 0.18^{+0.02}_{-0.02} $ & $ 69.66^{+3.49}_{-3.47} $ & $ 0.18^{+0.02}_{-0.02} $ & $ 69.38^{+3.61}_{-3.51} $ & 0.04 & $ 0.28^{+0.15}_{-0.08} $\\[2 mm]
TriaII$_{PS}$ & $ 1.53^{+0.26}_{-0.23} $ & $ 23^{+5}_{-4} $ & $ 1.55^{+0.24}_{-0.24} $ & $ 24^{+4}_{-5} $ & $ 0.25^{+0.11}_{-0.09} $ & $ 298.55^{+25.84}_{-219.82} $ & $ 0.26^{+0.13}_{-0.10} $ & $ 300.72^{+27.55}_{-222.65} $ & -0.45 & $ 0.21^{+0.15}_{-0.08} $\\[2 mm]
TucII$_{DES}$ & $ 10.97^{+0.74}_{-0.67} $ & $ 302^{+24}_{-22} $ & $ 10.95^{+0.75}_{-0.69} $ & $ 303^{+25}_{-23} $ & $ 0.34^{+0.06}_{-0.07} $ & $ 274.76^{+6.45}_{-187.91} $ & $ 0.34^{+0.07}_{-0.08} $ & $ 274.02^{+6.81}_{-187.37} $ & -1.17 & $ 0.13^{+0.06}_{-0.04} $\\[2 mm]
TucIII$_{DES}$ & $ 14.40^{+1.57}_{-1.67} $ & $ 363^{+34}_{-37} $ & $ 10.22^{+1.27}_{-2.12} $ & $ 272^{+31}_{-47} $ & $ 0.81^{+0.02}_{-0.03} $ & $ 83.87^{+1.20}_{-1.21} $ & $ 0.73^{+0.05}_{-0.15} $ & $ 83.17^{+2.07}_{-1.66} $ & -0.34 & $ 0.51^{+0.50}_{-0.29} $\\[2 mm]
TucIV$_{DES}$ & $ 7.69^{+0.81}_{-0.65} $ & $ 179^{+18}_{-17} $ & $ 7.42^{+0.71}_{-0.65} $ & $ 174^{+18}_{-16} $ & $ 0.34^{+0.07}_{-0.07} $ & $ 336.19^{+8.46}_{-23.77} $ & $ 0.35^{+0.07}_{-0.09} $ & $ 333.32^{+9.26}_{-54.63} $ & -0.80 & $ 0.22^{+0.23}_{-0.08} $\\[2 mm]
TucV$_{DES}$ & $ 1.53^{+0.29}_{-0.22} $ & $ 37^{+6}_{-6} $ & $ 1.39^{+0.20}_{-0.18} $ & $ 33^{+5}_{-4} $ & $ 0.38^{+0.10}_{-0.12} $ & $ 40.85^{+10.03}_{-7.65} $ & $ 0.37^{+0.12}_{-0.13} $ & $ 41.60^{+11.90}_{-8.36} $ & -0.36 & $ 0.21^{+0.18}_{-0.09} $\\[2 mm]
UMaI$_{PS}$ & $ 5.68^{+0.82}_{-0.68} $ & $ 64^{+9}_{-8} $ & $ 5.64^{+0.75}_{-0.65} $ & $ 64^{+9}_{-8} $ & $ 0.49^{+0.08}_{-0.10} $ & $ 270.33^{+6.93}_{-188.46} $ & $ 0.50^{+0.09}_{-0.11} $ & $ 88.05^{+187.63}_{-7.84} $ & -0.61 & $ 0.15^{+0.18}_{-0.07} $\\[2 mm]
UMaI$_{SDSS}$ & $ 6.09^{+0.78}_{-0.66} $ & $ 58^{+8}_{-8} $ & $ 6.06^{+0.72}_{-0.63} $ & $ 57^{+8}_{-7} $ & $ 0.58^{+0.07}_{-0.08} $ & $ 78.57^{+4.98}_{-4.79} $ & $ 0.59^{+0.07}_{-0.09} $ & $ 77.97^{+5.43}_{-4.83} $ & -0.66 & $ 0.17^{+0.32}_{-0.09} $\\[2 mm]
UMaII$_{SDSS}$ & $ 11.23^{+0.77}_{-0.69} $ & $ 296^{+20}_{-19} $ & $ 10.52^{+0.68}_{-0.63} $ & $ 284^{+19}_{-19} $ & $ 0.59^{+0.04}_{-0.04} $ & $ 284.23^{+2.09}_{-2.22} $ & $ 0.62^{+0.03}_{-0.03} $ & $ 284.24^{+2.11}_{-2.09} $ & -0.32 & $ 1.74^{+0.62}_{-0.72} $\\[2 mm]
UMi$_{PS}$ & $ 12.90^{+0.19}_{-0.17} $ & $ 2041^{+30}_{-28} $ & $ 12.85^{+0.19}_{-0.18} $ & $ 2036^{+30}_{-30} $ & $ 0.53^{+.01}_{-.01} $ & $ 52.05^{+0.69}_{-0.68} $ & $ 0.53^{+.01}_{-.01} $ & $ 52.00^{+0.70}_{-0.68} $ & -0.91 & $ 0.05^{+0.01}_{-.010} $\\[2 mm]
WilI$_{PS}$ & $ 1.91^{+0.20}_{-0.18} $ & $ 47^{+5}_{-4} $ & $ 1.85^{+0.17}_{-0.16} $ & $ 45^{+5}_{-4} $ & $ 0.44^{+0.08}_{-0.10} $ & $ 78.69^{+4.96}_{-4.75} $ & $ 0.43^{+0.08}_{-0.10} $ & $ 78.65^{+5.27}_{-5.13} $ & -0.27 & $ 0.17^{+0.29}_{-0.07} $\\[2 mm]
WilI$_{SDSS}$ & $ 1.54^{+0.14}_{-0.12} $ & $ 53^{+5}_{-4} $ & $ 1.48^{+0.12}_{-0.11} $ & $ 52^{+4}_{-4} $ & $ 0.54^{+0.05}_{-0.06} $ & $ 77.14^{+3.09}_{-3.32} $ & $ 0.53^{+0.06}_{-0.07} $ & $ 76.08^{+3.32}_{-3.33} $ & 0.12 & $ 1.47^{+1.26}_{-1.21} $\\[2 mm]
WilI$_{M18}$ & $ 2.49^{+0.36}_{-0.25} $ & $ 455^{+32}_{-28} $ & $ 1.88^{+0.06}_{-0.06} $ & $ 380^{+11}_{-12} $ & $ 0.50^{+0.02}_{-0.02} $ & $ 77.86^{+1.63}_{-1.60} $ & $ 0.51^{+0.02}_{-0.02} $ & $ 77.03^{+1.54}_{-1.55} $ & 0.94 & $ 0.32^{+0.42}_{-0.15} $\\[2 mm]

%% file: table_4.txt
BooI$_{PS}$ & $ 9.18^{+0.52}_{-0.49} $ & $ 216^{+17(60)}_{-16(53)} $ & $ 12.69^{+0.94}_{-0.88} $ & $ 240^{+22(85)}_{-20(67)} $ & $ 0.23^{+0.07}_{-0.07} $ & $ 27.86^{+14.02}_{-9.03} $ & $ 0.28^{+0.07}_{-0.08} $ & $ 25.99^{+11.20}_{-7.75} $ & -0.86 & -0.42 & -1.45 & $ 0.40^{+0.10}_{-0.08} $\\[2 mm]
BooI$_{DCLS}$ & $ 8.85^{+0.17}_{-0.16} $ & $ 1737^{+43(157)}_{-41(146)} $ & $ 11.79^{+0.27}_{-0.26} $ & $ 1848^{+52(186)}_{-48(172)} $ & $ 0.37^{+0.02}_{-0.02} $ & $ 6.10^{+2.09}_{-1.92} $ & $ 0.41^{+0.02}_{-0.02} $ & $ 5.78^{+1.98}_{-1.76} $ & -4.47 & -1.89 & -3.62 & $ 0.24^{+0.03}_{-0.02} $\\[2 mm]
BooII$_{PS}$ & $ 2.49^{+0.51}_{-0.47} $ & $ 17^{+5(20)}_{-4(14)} $ & $ 1.72^{+0.36}_{-0.30} $ & $ 19^{+5(20)}_{-5(15)} $ & $ 0.32^{+0.13}_{-0.11} $ & $ 278.84^{+18.95}_{-207.44} $ & $ 0.31^{+0.14}_{-0.12} $ & $ 276.45^{+19.73}_{-205.28} $ & 0.07 & -0.06 & -0.38 & $ 0.49^{+0.40}_{-0.19} $\\[2 mm]
BooII$_{SDSS}$ & $ 3.70^{+0.57}_{-0.58} $ & $ 29^{+6(23)}_{-7(22)} $ & $ 2.52^{+0.41}_{-0.39} $ & $ 30^{+7(26)}_{-6(20)} $ & $ 0.26^{+0.11}_{-0.09} $ & $ 87.35^{+216.81}_{-37.20} $ & $ 0.25^{+0.11}_{-0.10} $ & $ 88.11^{+212.59}_{-34.81} $ & .01 & 0.06 & -0.54 & $ 0.45^{+0.24}_{-0.14} $\\[2 mm]
BooII$_{M18}$ & $ 3.61^{+0.14}_{-0.14} $ & $ 257^{+13(48)}_{-12(43)} $ & $ 9.08^{+0.38}_{-0.35} $ & $ 260^{+13(49)}_{-14(47)} $ & $ 0.11^{+0.05}_{-0.04} $ & $ 271.37^{+17.07}_{-200.76} $ & $ 0.11^{+0.05}_{-0.04} $ & $ 274.99^{+14.65}_{-198.84} $ & 0.22 & 0.30 & -1.04 & $ 0.48^{+0.06}_{-0.05} $\\[2 mm]
BooII$_{DCLS}$ & $ 3.29^{+0.20}_{-0.18} $ & $ 161^{+13(49)}_{-13(43)} $ & $ 2.18^{+0.13}_{-0.12} $ & $ 160^{+14(50)}_{-13(44)} $ & $ 0.24^{+0.10}_{-0.08} $ & $ 281.10^{+8.54}_{-8.72} $ & $ 0.22^{+0.10}_{-0.08} $ & $ 279.74^{+9.45}_{-190.05} $ & 0.55 & 0.59 & -0.68 & $ 0.39^{+0.10}_{-0.09} $\\[2 mm]
BooIII$_{DCLS}$ & $ 18.79^{+0.75}_{-0.77} $ & $ 654^{+53(196)}_{-54(193)} $ & $ 17.19^{+1.01}_{-0.97} $ & $ 1089^{+104(407)}_{-99(328)} $ & $ 0.29^{+0.07}_{-0.08} $ & $ 287.98^{+6.49}_{-6.43} $ & $ 0.38^{+0.07}_{-0.08} $ & $ 282.89^{+5.43}_{-5.74} $ & -2.11 & 0.64 & -4.31 & $ 0.32^{+0.07}_{-0.06} $\\[2 mm]
CetII$_{DES}$ & $ 6.57^{+1.09}_{-1.10} $ & $ 67^{+13(51)}_{-13(47)} $ & $ 6.36^{+1.21}_{-1.11} $ & $ 66^{+16(70)}_{-14(47)} $ & $ 0.37^{+0.12}_{-0.12} $ & $ 61.61^{+11.05}_{-9.76} $ & $ 0.36^{+0.14}_{-0.14} $ & $ 62.60^{+15.33}_{-11.47} $ & -0.18 & -0.15 & -0.38 & $ 1.24^{+0.85}_{-0.56} $\\[2 mm]
ColI$_{DES}$ & $ 1.82^{+0.16}_{-0.14} $ & $ 45^{+4(17)}_{-4(14)} $ & $ 1.78^{+0.15}_{-0.14} $ & $ 44^{+4(17)}_{-4(15)} $ & $ 0.20^{+0.08}_{-0.07} $ & $ 67.23^{+17.30}_{-14.18} $ & $ 0.20^{+0.09}_{-0.08} $ & $ 69.14^{+17.01}_{-13.47} $ & 0.10 & -0.02 & -0.22 & $ 0.46^{+0.12}_{-0.10} $\\[2 mm]
ComaB$_{PS}$ & $ 5.80^{+0.65}_{-0.61} $ & $ 63^{+8(33)}_{-8(27)} $ & $ 3.84^{+0.39}_{-0.41} $ & $ 63^{+8(33)}_{-9(27)} $ & $ 0.26^{+0.11}_{-0.09} $ & $ 279.30^{+14.71}_{-199.71} $ & $ 0.27^{+0.11}_{-0.10} $ & $ 279.20^{+13.57}_{-195.81} $ & 0.02 & -0.12 & -0.46 & $ 0.47^{+0.20}_{-0.13} $\\[2 mm]
ComaB$_{SDSS}$ & $ 5.61^{+0.39}_{-0.37} $ & $ 100^{+9(33)}_{-8(28)} $ & $ 3.70^{+0.27}_{-0.25} $ & $ 100^{+9(34)}_{-9(29)} $ & $ 0.15^{+0.06}_{-0.05} $ & $ 287.07^{+21.15}_{-219.84} $ & $ 0.15^{+0.06}_{-0.06} $ & $ 282.50^{+22.20}_{-216.01} $ & 0.20 & 0.16 & -0.48 & $ 0.50^{+0.12}_{-0.09} $\\[2 mm]
ComaB$_{M18}$ & $ 4.71^{+0.08}_{-0.08} $ & $ 1160^{+25(88)}_{-23(82)} $ & $ 5.87^{+0.10}_{-0.10} $ & $ 1157^{+25(89)}_{-25(86)} $ & $ 0.32^{+0.02}_{-0.02} $ & $ 283.21^{+1.96}_{-1.87} $ & $ 0.32^{+0.02}_{-0.02} $ & $ 283.13^{+2.04}_{-1.90} $ & 3.54 & 3.51 & -0.98 & $ 0.29^{+0.03}_{-0.03} $\\[2 mm]
ComaB$_{DCLS}$ & $ 4.83^{+0.15}_{-0.15} $ & $ 459^{+15(54)}_{-15(51)} $ & $ 3.08^{+0.08}_{-0.08} $ & $ 434^{+13(50)}_{-13(46)} $ & $ 0.38^{+0.03}_{-0.03} $ & $ 294.43^{+2.25}_{-2.35} $ & $ 0.37^{+0.03}_{-0.03} $ & $ 293.95^{+2.45}_{-2.41} $ & -0.30 & -1.37 & 0.48 & $ 0.65^{+0.16}_{-0.13} $\\[2 mm]
Cra2$_{PS}$ & $ 20.00^{+1.06}_{-0.98} $ & $ 315^{+27(100)}_{-26(93)} $ & $ 10.66^{+0.75}_{-0.64} $ & $ 345^{+32(132)}_{-32(107)} $ & $ 0.17^{+0.07}_{-0.06} $ & $ 64.86^{+16.76}_{-12.97} $ & $ 0.19^{+0.08}_{-0.08} $ & $ 64.24^{+17.44}_{-14.19} $ & -0.02 & 0.13 & -1.61 & $ 0.41^{+0.07}_{-0.06} $\\[2 mm]
CVnI$_{PS}$ & $ 6.27^{+0.19}_{-0.18} $ & $ 315^{+12(42)}_{-10(37)} $ & $ 4.14^{+0.13}_{-0.12} $ & $ 312^{+11(40)}_{-11(38)} $ & $ 0.44^{+0.03}_{-0.03} $ & $ 79.43^{+2.27}_{-2.23} $ & $ 0.44^{+0.03}_{-0.03} $ & $ 79.59^{+2.34}_{-2.34} $ & 0.29 & 0.25 & -0.61 & $ 0.23^{+0.07}_{-0.03} $\\[2 mm]
CVnI$_{SDSS}$ & $ 6.49^{+0.17}_{-0.17} $ & $ 392^{+12(42)}_{-11(39)} $ & $ 4.23^{+0.10}_{-0.10} $ & $ 384^{+11(39)}_{-11(39)} $ & $ 0.37^{+0.03}_{-0.03} $ & $ 73.42^{+2.31}_{-2.29} $ & $ 0.37^{+0.03}_{-0.03} $ & $ 74.33^{+2.55}_{-2.47} $ & 0.29 & -0.31 & -0.06 & $ 0.88^{+0.34}_{-0.39} $\\[2 mm]
CVnII$_{PS}$ & $ 0.90^{+0.15}_{-0.12} $ & $ 13^{+3(10)}_{-2(6)} $ & $ 0.60^{+0.10}_{-0.08} $ & $ 13^{+2(9)}_{-2(6)} $ & $ 0.39^{+0.12}_{-0.12} $ & $ 297.88^{+38.29}_{-278.17} $ & $ 0.38^{+0.13}_{-0.14} $ & $ 293.52^{+38.10}_{-269.98} $ & 0.05 & -0.02 & -0.23 & $ 0.29^{+0.14}_{-0.10} $\\[2 mm]
CVnII$_{SDSS}$ & $ 0.85^{+0.11}_{-0.10} $ & $ 15^{+2(10)}_{-2(7)} $ & $ 0.57^{+0.07}_{-0.06} $ & $ 15^{+2(8)}_{-2(7)} $ & $ 0.63^{+0.08}_{-0.11} $ & $ 10.04^{+12.25}_{-4.07} $ & $ 0.65^{+0.08}_{-0.11} $ & $ 9.96^{+11.65}_{-4.14} $ & 0.02 & -0.06 & -0.19 & $ 0.16^{+0.13}_{-0.07} $\\[2 mm]
CVnII$_{M18}$ & $ 1.28^{+0.06}_{-0.05} $ & $ 117^{+6(22)}_{-6(20)} $ & $ 5.80^{+0.26}_{-0.26} $ & $ 116^{+6(22)}_{-6(21)} $ & $ 0.38^{+0.05}_{-0.05} $ & $ 13.98^{+4.19}_{-3.58} $ & $ 0.38^{+0.05}_{-0.05} $ & $ 13.89^{+4.25}_{-3.62} $ & -0.23 & -0.17 & -0.74 & $ 0.27^{+0.07}_{-0.05} $\\[2 mm]
Dra$_{PS}$ & $ 7.45^{+0.09}_{-0.09} $ & $ 1898^{+27(96)}_{-26(94)} $ & $ 3.67^{+0.04}_{-0.04} $ & $ 1859^{+25(88)}_{-23(85)} $ & $ 0.30^{+0.01}_{-0.01} $ & $ 85.57^{+1.31}_{-1.32} $ & $ 0.30^{+0.01}_{-0.01} $ & $ 85.85^{+1.34}_{-1.42} $ & 0.67 & -0.34 & 0.16 & $ 0.45^{+0.11}_{-0.08} $\\[2 mm]
Dra$_{SDSS}$ & $ 7.73^{+0.09}_{-0.09} $ & $ 2181^{+35(124)}_{-32(107)} $ & $ 3.81^{+0.04}_{-0.04} $ & $ 2134^{+26(93)}_{-26(90)} $ & $ 0.33^{+0.01}_{-0.01} $ & $ 272.54^{+1.07}_{-1.09} $ & $ 0.33^{+0.01}_{-0.01} $ & $ 272.34^{+1.19}_{-1.18} $ & 1.20 & 0.27 & 0.14 & $ 0.33^{+0.06}_{-0.03} $\\[2 mm]
DraII$_{PS}$ & $ 1.82^{+0.31}_{-0.26} $ & $ 26^{+5(21)}_{-4(12)} $ & $ 1.14^{+0.18}_{-0.14} $ & $ 25^{+3(16)}_{-4(12)} $ & $ 0.48^{+0.10}_{-0.12} $ & $ 300.16^{+7.29}_{-8.48} $ & $ 0.46^{+0.11}_{-0.12} $ & $ 297.74^{+8.45}_{-9.34} $ & -0.22 & -0.18 & -0.24 & $ 0.33^{+0.20}_{-0.13} $\\[2 mm]
EriIII$_{DES}$ & $ 0.39^{+0.09}_{-0.07} $ & $ 11^{+2(11)}_{-2(6)} $ & $ 0.36^{+0.07}_{-0.05} $ & $ 10^{+1(7)}_{-2(5)} $ & $ 0.44^{+0.10}_{-0.12} $ & $ 281.14^{+12.35}_{-194.73} $ & $ 0.42^{+0.11}_{-0.13} $ & $ 279.42^{+14.71}_{-195.20} $ & -0.13 & -0.18 & -0.14 & $ 0.33^{+0.16}_{-0.12} $\\[2 mm]
For$_{DES}$ & $ 15.97^{+0.02}_{-0.02} $ & $ 69573^{+106(374)}_{-104(369)} $ & $ 2.13^{+.010}_{-.010} $ & $ 69559^{+107(381)}_{-104(377)} $ & $ 0.31^{+.010}_{-.010} $ & $ 40.60^{+0.17}_{-0.16} $ & $ 0.31^{+.010}_{-.010} $ & $ 40.59^{+0.16}_{-0.16} $ & 941.47 & 941.07 & -1.85 & $ 0.27^{+.010}_{-.010} $\\[2 mm]
GrusI$_{DES}$ & $ 2.42^{+0.27}_{-0.22} $ & $ 48^{+6(28)}_{-5(17)} $ & $ 2.39^{+0.25}_{-0.22} $ & $ 47^{+6(24)}_{-5(17)} $ & $ 0.41^{+0.09}_{-0.10} $ & $ 335.18^{+11.61}_{-45.87} $ & $ 0.41^{+0.09}_{-0.11} $ & $ 332.84^{+12.75}_{-260.38} $ & -0.07 & -0.10 & -0.50 & $ 0.29^{+0.12}_{-0.09} $\\[2 mm]
GrusII$_{DES}$ & $ 6.67^{+0.56}_{-0.48} $ & $ 232^{+21(78)}_{-18(57)} $ & $ 6.21^{+0.40}_{-0.37} $ & $ 215^{+16(59)}_{-14(49)} $ & $ 0.12^{+0.05}_{-0.04} $ & $ 69.89^{+237.19}_{-30.03} $ & $ 0.13^{+0.06}_{-0.05} $ & $ 66.42^{+231.37}_{-27.66} $ & -0.87 & -0.91 & -0.41 & $ 0.70^{+0.21}_{-0.14} $\\[2 mm]
Herc$_{PS}$ & $ 3.61^{+0.75}_{-0.58} $ & $ 63^{+13(58)}_{-10(31)} $ & $ 2.29^{+0.42}_{-0.33} $ & $ 61^{+11(49)}_{-10(30)} $ & $ 0.65^{+0.07}_{-0.08} $ & $ 274.65^{+4.58}_{-186.52} $ & $ 0.67^{+0.06}_{-0.08} $ & $ 274.79^{+5.04}_{-186.59} $ & -0.27 & -0.31 & -0.39 & $ 0.22^{+0.18}_{-0.10} $\\[2 mm]
Herc$_{SDSS}$ & $ 5.78^{+1.26}_{-1.01} $ & $ 121^{+21(85)}_{-19(59)} $ & $ 2.23^{+0.40}_{-0.35} $ & $ 78^{+14(53)}_{-12(39)} $ & $ 0.58^{+0.08}_{-0.10} $ & $ 281.43^{+4.09}_{-3.79} $ & $ 0.66^{+0.07}_{-0.09} $ & $ 282.07^{+3.75}_{-3.98} $ & -0.49 & -0.70 & 0.57 & $ 1.85^{+0.69}_{-0.62} $\\[2 mm]
Herc$_{M18}$ & $ 4.00^{+0.10}_{-0.10} $ & $ 875^{+23(84)}_{-22(81)} $ & $ 9.83^{+0.24}_{-0.24} $ & $ 856^{+24(83)}_{-24(81)} $ & $ 0.59^{+0.01}_{-0.01} $ & $ 286.51^{+0.93}_{-0.94} $ & $ 0.61^{+0.01}_{-0.02} $ & $ 286.28^{+0.90}_{-0.92} $ & -4.55 & -3.76 & -0.31 & $ 1.01^{+0.25}_{-0.23} $\\[2 mm]
Herc$_{DCLS}$ & $ 8.12^{+0.75}_{-0.67} $ & $ 379^{+32(117)}_{-28(96)} $ & $ 3.69^{+0.26}_{-0.24} $ & $ 278^{+20(78)}_{-20(68)} $ & $ 0.55^{+0.05}_{-0.05} $ & $ 285.32^{+2.90}_{-2.63} $ & $ 0.56^{+0.06}_{-0.06} $ & $ 284.04^{+2.55}_{-2.31} $ & -0.73 & -1.61 & 2.05 & $ 1.53^{+0.42}_{-0.38} $\\[2 mm]
HorI$_{DES}$ & $ 1.46^{+0.08}_{-0.07} $ & $ 90^{+5(23)}_{-6(18)} $ & $ 0.72^{+0.04}_{-0.03} $ & $ 87^{+5(20)}_{-5(17)} $ & $ 0.16^{+0.06}_{-0.06} $ & $ 69.34^{+11.37}_{-11.55} $ & $ 0.17^{+0.06}_{-0.06} $ & $ 69.96^{+10.98}_{-10.91} $ & -0.33 & -0.37 & -0.23 & $ 0.45^{+0.08}_{-0.07} $\\[2 mm]
HorII$_{DES}$ & $ 1.74^{+0.26}_{-0.22} $ & $ 23^{+4(16)}_{-3(11)} $ & $ 1.72^{+0.23}_{-0.21} $ & $ 23^{+3(14)}_{-4(11)} $ & $ 0.35^{+0.13}_{-0.12} $ & $ 281.69^{+8.83}_{-11.45} $ & $ 0.33^{+0.14}_{-0.13} $ & $ 281.51^{+9.50}_{-10.72} $ & 0.02 & -0.06 & -0.40 & $ 0.32^{+0.14}_{-0.10} $\\[2 mm]
LeoI$_{PS}$ & $ 3.82^{+0.04}_{-0.04} $ & $ 1003^{+15(53)}_{-14(50)} $ & $ 2.55^{+0.03}_{-0.03} $ & $ 1003^{+15(54)}_{-15(53)} $ & $ 0.43^{+0.01}_{-0.01} $ & $ 84.67^{+0.94}_{-0.93} $ & $ 0.43^{+0.01}_{-0.01} $ & $ 84.69^{+0.93}_{-0.98} $ & 17.22 & 17.21 & -0.90 & $ 0.19^{+0.01}_{-.01} $\\[2 mm]
LeoII$_{PS}$ & $ 2.62^{+0.03}_{-0.03} $ & $ 727^{+13(45)}_{-12(44)} $ & $ 1.75^{+0.02}_{-0.02} $ & $ 725^{+12(46)}_{-13(45)} $ & $ 0.11^{+0.02}_{-0.02} $ & $ 47.82^{+5.43}_{-5.40} $ & $ 0.11^{+0.02}_{-0.02} $ & $ 47.82^{+5.48}_{-5.80} $ & 10.82 & 10.78 & -0.76 & $ 0.45^{+0.03}_{-0.02} $\\[2 mm]
LeoII$_{DCLS}$ & $ 2.65^{+0.02}_{-0.02} $ & $ 2809^{+25(89)}_{-23(84)} $ & $ 1.76^{+0.01}_{-0.01} $ & $ 2801^{+24(82)}_{-23(83)} $ & $ 0.07^{+0.01}_{-0.01} $ & $ 43.98^{+4.43}_{-4.42} $ & $ 0.07^{+0.01}_{-.01} $ & $ 43.94^{+4.47}_{-4.43} $ & 50.84 & 50.65 & -0.88 & $ 0.49^{+0.01}_{-0.01} $\\[2 mm]
LeoIV$_{PS}$ & $ 3.22^{+0.66}_{-0.53} $ & $ 18^{+5(22)}_{-4(16)} $ & $ 2.29^{+0.50}_{-0.38} $ & $ 20^{+5(25)}_{-5(15)} $ & $ 0.37^{+0.15}_{-0.13} $ & $ 87.28^{+233.05}_{-63.34} $ & $ 0.41^{+0.15}_{-0.16} $ & $ 271.73^{+39.94}_{-244.54} $ & 0.08 & .01 & -0.50 & $ 0.31^{+0.17}_{-0.12} $\\[2 mm]
LeoIV$_{SDSS}$ & $ 2.24^{+0.23}_{-0.21} $ & $ 23^{+4(14)}_{-3(11)} $ & $ 1.49^{+0.15}_{-0.14} $ & $ 23^{+4(14)}_{-3(11)} $ & $ 0.40^{+0.09}_{-0.12} $ & $ 332.75^{+12.20}_{-270.91} $ & $ 0.40^{+0.11}_{-0.12} $ & $ 329.98^{+14.01}_{-269.15} $ & 0.34 & 0.25 & -0.48 & $ 0.26^{+0.11}_{-0.08} $\\[2 mm]
LeoIV$_{DCLS}$ & $ 2.22^{+0.13}_{-0.12} $ & $ 88^{+7(26)}_{-6(21)} $ & $ 1.46^{+0.09}_{-0.08} $ & $ 86^{+7(24)}_{-6(21)} $ & $ 0.26^{+0.07}_{-0.08} $ & $ 329.60^{+15.03}_{-301.99} $ & $ 0.26^{+0.07}_{-0.08} $ & $ 325.35^{+18.72}_{-297.74} $ & 0.36 & 0.22 & -0.38 & $ 0.35^{+0.08}_{-0.07} $\\[2 mm]
LeoV$_{SDSS}$ & $ 1.45^{+0.35}_{-0.28} $ & $ 12^{+2(12)}_{-3(8)} $ & $ 0.98^{+0.23}_{-0.18} $ & $ 12^{+3(12)}_{-3(8)} $ & $ 0.35^{+0.13}_{-0.13} $ & $ 275.94^{+15.49}_{-203.95} $ & $ 0.35^{+0.14}_{-0.13} $ & $ 277.41^{+14.20}_{-202.04} $ & .01 & -0.08 & -0.24 & $ 0.38^{+0.26}_{-0.15} $\\[2 mm]
LeoV$_{M18}$ & $ 1.39^{+0.19}_{-0.17} $ & $ 114^{+9(33)}_{-9(35)} $ & $ 4.74^{+0.32}_{-0.30} $ & $ 77^{+5(20)}_{-5(17)} $ & $ 0.30^{+0.07}_{-0.07} $ & $ 272.72^{+6.20}_{-187.12} $ & $ 0.17^{+0.07}_{-0.07} $ & $ 279.46^{+13.06}_{-193.62} $ & -0.79 & -1.42 & 0.72 & $ 1.24^{+0.23}_{-0.20} $\\[2 mm]
LeoV$_{DCLS}$ & $ 0.97^{+0.14}_{-0.12} $ & $ 20^{+3(13)}_{-3(9)} $ & $ 0.63^{+0.09}_{-0.08} $ & $ 19^{+3(13)}_{-3(9)} $ & $ 0.28^{+0.12}_{-0.10} $ & $ 78.64^{+197.81}_{-13.62} $ & $ 0.28^{+0.12}_{-0.10} $ & $ 77.67^{+197.98}_{-15.39} $ & -0.01 & -0.10 & -0.24 & $ 0.38^{+0.14}_{-0.11} $\\[2 mm]
PegII$_{SDSS}$ & $ 1.05^{+0.50}_{-0.30} $ & $ 7^{+3(16)}_{-2(6)} $ & $ 0.68^{+0.30}_{-0.18} $ & $ 7^{+3(16)}_{-2(6)} $ & $ 0.41^{+0.15}_{-0.15} $ & $ 280.15^{+27.02}_{-240.57} $ & $ 0.43^{+0.15}_{-0.16} $ & $ 282.82^{+23.83}_{-235.61} $ & .01 & -0.01 & -0.21 & $ 0.62^{+0.61}_{-0.31} $\\[2 mm]
PhoeII$_{DES}$ & $ 1.64^{+0.20}_{-0.19} $ & $ 26^{+4(15)}_{-3(12)} $ & $ 1.59^{+0.20}_{-0.18} $ & $ 26^{+3(14)}_{-4(13)} $ & $ 0.31^{+0.12}_{-0.11} $ & $ 284.71^{+13.61}_{-10.75} $ & $ 0.27^{+0.11}_{-0.10} $ & $ 289.29^{+15.43}_{-13.29} $ & -0.07 & -0.18 & 0.11 & $ 0.42^{+0.25}_{-0.14} $\\[2 mm]
PictI$_{DES}$ & $ 0.94^{+0.10}_{-0.09} $ & $ 36^{+4(16)}_{-4(12)} $ & $ 0.90^{+0.09}_{-0.08} $ & $ 34^{+3(13)}_{-4(11)} $ & $ 0.45^{+0.07}_{-0.09} $ & $ 55.67^{+5.76}_{-6.14} $ & $ 0.46^{+0.07}_{-0.09} $ & $ 57.80^{+5.83}_{-5.61} $ & -0.35 & -0.57 & -0.12 & $ 0.31^{+0.17}_{-0.10} $\\[2 mm]
PiscII$_{SDSS}$ & $ 1.26^{+0.22}_{-0.18} $ & $ 15^{+3(13)}_{-2(8)} $ & $ 0.82^{+0.14}_{-0.11} $ & $ 15^{+3(11)}_{-3(8)} $ & $ 0.35^{+0.12}_{-0.12} $ & $ 85.48^{+194.89}_{-19.89} $ & $ 0.37^{+0.13}_{-0.13} $ & $ 88.04^{+192.67}_{-19.90} $ & -0.02 & -0.10 & -0.25 & $ 0.35^{+0.19}_{-0.12} $\\[2 mm]
PiscII$_{M18}$ & $ 1.01^{+0.05}_{-0.04} $ & $ 105^{+5(20)}_{-5(18)} $ & $ 5.27^{+0.24}_{-0.22} $ & $ 105^{+5(19)}_{-6(19)} $ & $ 0.18^{+0.06}_{-0.06} $ & $ 276.82^{+9.01}_{-191.77} $ & $ 0.18^{+0.06}_{-0.06} $ & $ 277.37^{+9.38}_{-190.44} $ & -0.11 & -0.11 & -0.60 & $ 0.44^{+0.08}_{-0.06} $\\[2 mm]
PiscII$_{DCLS}$ & $ 1.38^{+3.19}_{-0.22} $ & $ 39^{+27(195)}_{-7(17)} $ & $ 0.75^{+0.07}_{-0.07} $ & $ 31^{+3(14)}_{-4(11)} $ & $ 0.28^{+0.09}_{-0.09} $ & $ 70.80^{+13.83}_{-11.66} $ & $ 0.30^{+0.09}_{-0.10} $ & $ 74.01^{+14.56}_{-12.71} $ & -1.89 & -0.14 & -0.09 & $ 0.51^{+0.15}_{-0.13} $\\[2 mm]
RetII$_{DES}$ & $ 3.84^{+0.07}_{-0.06} $ & $ 695^{+14(51)}_{-15(51)} $ & $ 1.92^{+0.03}_{-0.03} $ & $ 691^{+15(52)}_{-14(50)} $ & $ 0.59^{+0.01}_{-0.01} $ & $ 69.71^{+0.78}_{-0.79} $ & $ 0.59^{+0.01}_{-0.01} $ & $ 69.64^{+0.86}_{-0.78} $ & 3.08 & 2.92 & -0.61 & $ 0.11^{+0.01}_{-.01} $\\[2 mm]
RetIII$_{DES}$ & $ 1.19^{+0.22}_{-0.19} $ & $ 20^{+4(14)}_{-3(9)} $ & $ 0.77^{+0.15}_{-0.12} $ & $ 20^{+4(15)}_{-3(10)} $ & $ 0.37^{+0.11}_{-0.12} $ & $ 46.34^{+257.79}_{-15.02} $ & $ 0.40^{+0.10}_{-0.12} $ & $ 48.25^{+267.53}_{-16.14} $ & 0.10 & 0.04 & -0.27 & $ 0.33^{+0.16}_{-0.11} $\\[2 mm]
SgrII$_{PS}$ & $ 1.46^{+0.11}_{-0.11} $ & $ 77^{+8(29)}_{-7(25)} $ & $ 2.88^{+0.23}_{-0.21} $ & $ 76^{+8(30)}_{-8(26)} $ & $ 0.23^{+0.08}_{-0.08} $ & $ 75.99^{+195.78}_{-12.04} $ & $ 0.24^{+0.09}_{-0.09} $ & $ 75.04^{+13.62}_{-12.40} $ & 0.44 & 0.15 & -0.45 & $ 0.44^{+0.14}_{-0.10} $\\[2 mm]
SegI$_{SDSS}$ & $ 3.98^{+0.52}_{-0.39} $ & $ 88^{+11(57)}_{-9(28)} $ & $ 2.41^{+0.21}_{-0.19} $ & $ 80^{+8(31)}_{-7(23)} $ & $ 0.18^{+0.07}_{-0.07} $ & $ 287.96^{+15.20}_{-198.33} $ & $ 0.19^{+0.08}_{-0.07} $ & $ 292.15^{+14.31}_{-15.07} $ & -0.30 & -0.50 & -0.24 & $ 0.58^{+0.16}_{-0.13} $\\[2 mm]
SegI$_{DCLS}$ & $ 4.33^{+0.90}_{-0.55} $ & $ 213^{+34(126)}_{-26(66)} $ & $ 2.29^{+0.14}_{-0.12} $ & $ 168^{+11(41)}_{-10(34)} $ & $ 0.31^{+0.06}_{-0.06} $ & $ 61.94^{+6.43}_{-6.95} $ & $ 0.35^{+0.05}_{-0.06} $ & $ 62.62^{+5.96}_{-6.56} $ & -0.62 & -0.98 & 0.12 & $ 0.62^{+0.16}_{-0.13} $\\[2 mm]
SegII$_{SDSS}$ & $ 5.76^{+7.73}_{-2.14} $ & $ 122^{+58(149)}_{-39(61)} $ & $ 2.19^{+0.14}_{-0.13} $ & $ 74^{+6(23)}_{-5(19)} $ & $ 0.19^{+0.07}_{-0.06} $ & $ 324.25^{+13.81}_{-24.40} $ & $ 0.21^{+0.08}_{-0.07} $ & $ 322.00^{+15.21}_{-27.90} $ & -10.12 & -0.10 & -0.17 & $ 1.16^{+0.20}_{-0.19} $\\[2 mm]
SegII$_{PS}$ & $ 3.34^{+0.45}_{-0.40} $ & $ 28^{+5(20)}_{-4(15)} $ & $ 2.22^{+0.28}_{-0.26} $ & $ 28^{+5(20)}_{-4(15)} $ & $ 0.26^{+0.10}_{-0.09} $ & $ 291.70^{+21.41}_{-218.17} $ & $ 0.26^{+0.11}_{-0.10} $ & $ 292.40^{+22.14}_{-217.11} $ & 0.10 & 0.02 & -0.35 & $ 0.40^{+0.17}_{-0.12} $\\[2 mm]
SegII$_{DCLS}$ & $ 3.01^{+0.10}_{-0.09} $ & $ 253^{+10(37)}_{-10(32)} $ & $ 1.99^{+0.06}_{-0.06} $ & $ 249^{+9(33)}_{-9(31)} $ & $ 0.32^{+0.03}_{-0.03} $ & $ 343.82^{+3.49}_{-3.88} $ & $ 0.32^{+0.03}_{-0.04} $ & $ 343.75^{+3.98}_{-4.11} $ & -0.16 & -0.23 & -0.48 & $ 0.30^{+0.05}_{-0.03} $\\[2 mm]
SexI$_{PS}$ & $ 17.58^{+0.36}_{-0.34} $ & $ 1921^{+44(152)}_{-43(150)} $ & $ 8.39^{+0.14}_{-0.13} $ & $ 1820^{+33(123)}_{-34(119)} $ & $ 0.18^{+0.02}_{-0.02} $ & $ 69.13^{+3.57}_{-3.50} $ & $ 0.18^{+0.02}_{-0.02} $ & $ 67.60^{+3.50}_{-3.47} $ & -3.20 & -4.07 & 0.89 & $ 0.72^{+0.10}_{-0.10} $\\[2 mm]
TriaII$_{PS}$ & $ 1.38^{+0.23}_{-0.20} $ & $ 21^{+4(17)}_{-4(12)} $ & $ 1.40^{+0.23}_{-0.20} $ & $ 21^{+4(17)}_{-4(12)} $ & $ 0.26^{+0.11}_{-0.10} $ & $ 298.34^{+29.82}_{-224.30} $ & $ 0.29^{+0.13}_{-0.11} $ & $ 303.35^{+28.44}_{-217.67} $ & 0.09 & -0.02 & -0.36 & $ 0.40^{+0.19}_{-0.13} $\\[2 mm]
TucII$_{DES}$ & $ 8.75^{+0.50}_{-0.47} $ & $ 248^{+17(64)}_{-18(62)} $ & $ 9.01^{+0.58}_{-0.53} $ & $ 255^{+20(79)}_{-19(66)} $ & $ 0.35^{+0.06}_{-0.07} $ & $ 277.56^{+5.88}_{-6.56} $ & $ 0.36^{+0.06}_{-0.07} $ & $ 276.44^{+5.86}_{-186.64} $ & -0.21 & -0.04 & -1.33 & $ 0.29^{+0.08}_{-0.06} $\\[2 mm]
TucIII$_{DES}$ & $ 5.08^{+0.56}_{-0.41} $ & $ 148^{+17(105)}_{-13(41)} $ & $ 10.11^{+0.98}_{-1.77} $ & $ 282^{+35(108)}_{-54(163)} $ & $ 0.25^{+0.10}_{-0.09} $ & $ 88.33^{+202.69}_{-7.11} $ & $ 0.78^{+0.03}_{-0.08} $ & $ 83.20^{+1.81}_{-1.65} $ & -0.94 & 0.02 & -1.30 & $ 0.43^{+0.11}_{-0.10} $\\[2 mm]
TucIV$_{DES}$ & $ 6.01^{+0.45}_{-0.42} $ & $ 142^{+13(49)}_{-12(42)} $ & $ 8.81^{+0.73}_{-0.64} $ & $ 146^{+14(56)}_{-14(44)} $ & $ 0.34^{+0.07}_{-0.07} $ & $ 335.86^{+8.63}_{-37.96} $ & $ 0.36^{+0.07}_{-0.08} $ & $ 335.92^{+8.42}_{-17.85} $ & -0.41 & -0.35 & -0.88 & $ 0.43^{+0.19}_{-0.12} $\\[2 mm]
TucV$_{DES}$ & $ 1.30^{+0.18}_{-0.16} $ & $ 31^{+4(19)}_{-4(13)} $ & $ 1.25^{+0.16}_{-0.14} $ & $ 29^{+4(17)}_{-4(12)} $ & $ 0.38^{+0.11}_{-0.12} $ & $ 40.01^{+9.63}_{-8.02} $ & $ 0.40^{+0.11}_{-0.13} $ & $ 39.34^{+10.03}_{-7.65} $ & -0.16 & -0.27 & -0.24 & $ 0.40^{+0.23}_{-0.16} $\\[2 mm]
UMaI$_{PS}$ & $ 4.92^{+0.62}_{-0.53} $ & $ 56^{+7(29)}_{-7(23)} $ & $ 3.29^{+0.41}_{-0.37} $ & $ 55^{+7(30)}_{-6(22)} $ & $ 0.49^{+0.07}_{-0.09} $ & $ 87.43^{+188.73}_{-8.08} $ & $ 0.50^{+0.08}_{-0.10} $ & $ 87.84^{+187.52}_{-9.19} $ & -0.11 & -0.18 & -0.55 & $ 0.24^{+0.13}_{-0.08} $\\[2 mm]
UMaI$_{SDSS}$ & $ 5.30^{+0.57}_{-0.54} $ & $ 51^{+6(23)}_{-7(21)} $ & $ 3.57^{+0.39}_{-0.36} $ & $ 50^{+7(26)}_{-6(20)} $ & $ 0.58^{+0.06}_{-0.08} $ & $ 77.03^{+4.89}_{-4.50} $ & $ 0.58^{+0.07}_{-0.08} $ & $ 76.88^{+5.35}_{-4.81} $ & 0.03 & -0.02 & -0.59 & $ 0.19^{+0.15}_{-0.07} $\\[2 mm]
UMaII$_{SDSS}$ & $ 8.49^{+0.38}_{-0.38} $ & $ 240^{+14(53)}_{-14(49)} $ & $ 13.04^{+0.72}_{-0.66} $ & $ 258^{+18(66)}_{-16(56)} $ & $ 0.53^{+0.04}_{-0.04} $ & $ 283.75^{+2.29}_{-2.38} $ & $ 0.59^{+0.03}_{-0.04} $ & $ 284.21^{+2.19}_{-2.08} $ & -1.26 & -0.19 & -1.38 & $ 0.75^{+0.61}_{-0.40} $\\[2 mm]
UMi$_{PS}$ & $ 11.56^{+0.16}_{-0.16} $ & $ 1836^{+31(113)}_{-30(104)} $ & $ 5.70^{+0.07}_{-0.07} $ & $ 1798^{+26(93)}_{-25(90)} $ & $ 0.53^{+.01}_{-.01} $ & $ 52.46^{+0.64}_{-0.63} $ & $ 0.53^{+.01}_{-.01} $ & $ 52.64^{+0.69}_{-0.68} $ & -0.78 & -1.21 & -0.47 & $ 0.16^{+0.03}_{-0.02} $\\[2 mm]
WilI$_{PS}$ & $ 1.73^{+0.17}_{-0.14} $ & $ 42^{+4(17)}_{-4(13)} $ & $ 1.13^{+0.10}_{-0.09} $ & $ 40^{+4(15)}_{-4(12)} $ & $ 0.45^{+0.08}_{-0.11} $ & $ 78.02^{+4.70}_{-4.48} $ & $ 0.42^{+0.10}_{-0.11} $ & $ 78.07^{+5.41}_{-5.18} $ & -0.23 & -0.38 & -0.13 & $ 0.27^{+0.12}_{-0.08} $\\[2 mm]
WilI$_{SDSS}$ & $ 1.44^{+0.11}_{-0.10} $ & $ 48^{+4(17)}_{-3(12)} $ & $ 0.94^{+0.07}_{-0.06} $ & $ 47^{+4(14)}_{-4(12)} $ & $ 0.55^{+0.05}_{-0.07} $ & $ 76.99^{+2.90}_{-2.97} $ & $ 0.52^{+0.06}_{-0.07} $ & $ 75.67^{+3.40}_{-3.48} $ & -0.09 & -0.33 & 0.36 & $ 0.26^{+0.18}_{-0.08} $\\[2 mm]
WilI$_{M18}$ & $ 2.05^{+0.10}_{-0.09} $ & $ 406^{+15(56)}_{-15(50)} $ & $ 4.24^{+0.11}_{-0.12} $ & $ 336^{+10(38)}_{-10(34)} $ & $ 0.51^{+0.02}_{-0.02} $ & $ 77.21^{+1.66}_{-1.59} $ & $ 0.51^{+0.02}_{-0.02} $ & $ 77.28^{+1.54}_{-1.52} $ & -1.42 & -4.47 & 4.01 & $ 0.43^{+0.08}_{-0.06} $\\[2 mm]